\documentclass[aps, pra, floatfix, reprint, showpacs, amsmath, amssymb]{revtex4-1}

\usepackage{hyperref}		%for hyperlinks
\usepackage{graphicx}		% Include figure files
\usepackage{color}			%for colored text
\usepackage[normalem]{ulem}	%for strikeout
\usepackage{soul}

\begin{document}

\title{Cavity-aided non-demolition measurements for atom counting and spin squeezing}

\author{Zilong Chen}\email{chen.zilong@gmail.com}
\author{Justin G. Bohnet}
\author{Joshua M. Weiner}
\author{Kevin C. Cox}
\author{James K. Thompson}\email{jkt@jila.colorado.edu}
\affiliation{JILA, NIST and University of Colorado, Department of Physics, University of Colorado, Boulder, Colorado 80309, USA}

\date{\today}

\begin{abstract}
Probing the collective spin state of an ensemble of atoms may provide a means to reduce heating via the photon recoil associated with the measurement and provide a robust, scalable route for preparing highly entangled states with spectroscopic sensitivity below the standard quantum limit for coherent spin states.  The collective probing relies on obtaining a very large optical depth that can be effectively increased by placing the ensemble within an optical cavity such that the probe light passes many times through the ensemble.  Here we provide expressions for measurement resolution and spectroscopic enhancement in such cavity-aided non-demolition measurements as a function of cavity detuning. In particular, fundamental limits on spectroscopic enhancements in $^{87}$Rb are considered.
\end{abstract}

\pacs{42.50.Dv, 42.50.Pq, 06.20.-f, 42.50.Lc}

\maketitle

\newcommand*{\ket}[1]{\ensuremath{|#1\rangle}}
\newcommand*{\bra}[1]{\ensuremath{\langle#1|}}
\newcommand*{\ketup}{\ensuremath{\ket{\!\!\uparrow}}}
\newcommand*{\ketdown}{\ensuremath{\ket{\!\!\downarrow}}}
\newcommand*{\Rb}{\ensuremath{^{87}\mathrm{Rb}}}
\newcommand*{\FSR}{\ensuremath{f_\mathrm{FSR}}}

\newcommand*{\kv}[1]{\ensuremath{\vec{k}_{#1}}}
\newcommand*{\Nup}{\ensuremath{N_{\uparrow}}}
\newcommand*{\Ndown}{\ensuremath{N_{\downarrow}}}
\newcommand*{\Nupbar}{\ensuremath{\overline{N}_{\uparrow}}}
\newcommand*{\Ndownbar}{\ensuremath{\overline{N}_{\downarrow}}}
\newcommand*{\avg}[1]{\ensuremath{\left\langle#1\right\rangle}}
\newcommand*{\std}[1]{\ensuremath{\Delta\left(#1\right)}}
\newcommand*{\rabi}{\ensuremath{\Omega_\uparrow}}
\newcommand*{\op}{\ensuremath{\omega_+}}
\newcommand*{\om}{\ensuremath{\omega_-}}
\newcommand*{\opm}{\ensuremath{\omega_\pm}}
\newcommand*{\ohf}{\ensuremath{\omega_\mathrm{hf}}}
\newcommand*{\oehf}{\ensuremath{\omega_\mathrm{ehf}}}

\newcommand*{\detmethod}{\ensuremath{\eta_d}}
\newcommand*{\dc}{\ensuremath{\delta_c}}
\newcommand*{\dpr}{\ensuremath{\delta_p}}
\newcommand*{\ms}{\ensuremath{m_s}}
\newcommand*{\msp}{\ensuremath{m_s^\mathrm{proj}}}
\newcommand*{\msopt}{\ensuremath{m_s^\mathrm{opt}}}
\newcommand*{\msbar}{\ensuremath{\overline{m}_s}}

\newcommand*{\dwproj}{\ensuremath{\Delta\omega^\mathrm{proj}}}

\newcommand*{\prepnoise}{\ensuremath{\zeta_\mathrm{prep}}}
\newcommand*{\qpn}{\ensuremath{\Delta J_{z,\mathrm{CSS}}}}

\newcommand*{\J}{\ensuremath{\hat{\bf J}}}
\newcommand*{\R}{\ensuremath{{\cal R}}}
\newcommand*{\sq}{\ensuremath{\xi_m}}
\newcommand*{\sqopt}{\ensuremath{\sq^\mathrm{opt}}}

%******************************************************************************************************
%*******************************************Introduction****************************
%******************************************************************************************************
\section{Introduction}\label{sect1}

High-resolution measurements of the populations of two-level systems are key for realizing high precision atomic sensors such as atomic clocks, magnetometers, and atom-based electric field, rotation, and inertial sensors~\cite{Kitching2011}.  Further, developing non-demolition, high-resolution measurement techniques to create and/or detect entangled states is a promising route to enhanced sensors with improved accuracy, precision and/or bandwidth~\cite{Wineland01,Wineland04,Oberthaler08, Polzik09,Vuletic10,Oberthaler10,Treutlein10,CBS11,Klempt11,Mitchell12,Chapman12,BohnetSqueeze13}.

In trapped neutral atom ensembles, non-demolition measurements that do not cause atom loss from the trap could also lead to significant advances in the repetition rates of sensors~\cite{Lemonde09}, allowing them to operate closer to the regime of ion-based sensors in which the ions can be stored over many repeated measurement cycles~\cite{Hume07}.   Furthermore, a quantum non-demolition measurement that also preserves quantum coherence can prepare conditionally spin-squeezed states with spectroscopic sensitivity below the Standard Quantum Limit (SQL) $\Delta \theta_{\rm SQL} = 1/\sqrt{N}$ that arises from the quantum projection noise of $N$ independent atoms~\cite{Wineland94}.  

Recently, cavity-aided, non-demolition measurements were used to generate and observe the largest entanglement enhancement to date in an ensemble of spin-squeezed $^{87}$Rb atoms, improving the sensitivity of the ensemble by an order of magnitude~\cite{BohnetSqueeze13}. Cavity-aided non-demolition measurement techniques are compatible with accurate precision measurements, and in particular optical lattice clocks. Therefore, establishing a firm understanding of the fundamental limitations to cavity-based collective measurements is going to be crucial for advancing quantum metrology beyond proof-of-principle experiments.

\begin{figure}
\includegraphics[width=3.4in]{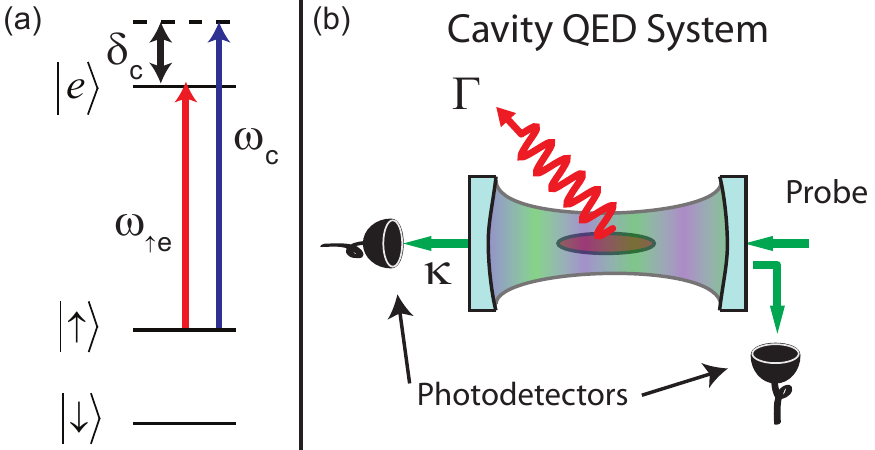}
\caption{(Color online) (a) Relevant energy levels. The pseudo-spin-1/2 system typically comprises two metastable states $\ketup, \ketdown$, which are utilized in atomic sensors and clocks.  The number of atoms $\Nup$ in $\ketup$ modifies the cavity resonance frequency.  Initially, we will assume the optically excited state $\ket{e}$ does not couple to $\ketdown$ because of dipole selection rules or  because the coupling is highly non-resonant $\dc \ll \ohf$, where $\ohf$ is the frequency separation between $\ketup$ and $\ketdown$. (b)  Transmitted and/or reflected probe light is monitored to determine the cavity resonance frequencies and hence the number of atoms in \ketup.  An applied NMR-like microwave rotation can swap the populations between ground states so that a subsequent measurement can also determine the  \ketdown~ population $\Ndown$.  The ensemble of atoms are trapped in an intracavity optical lattice (blue-green).   The total cavity power decay rate is $\kappa$.  Single-atom spontaneous decay  from $\ket{e}$ at rate $\Gamma$ leads to photon recoil heating and single-atom wavefunction collapse.  The goal then is to extract collective information from the probe mode more rapidly than the undesired single-atom photon scattering into free space.}
\label{BasicEnergyLevelsCavity}
\end{figure}

The results described in this work are relevant to recent approaches for generating entangled states in large ensembles using many diverse approaches, including quantum non-demolition measurements \cite{Polzik09,Vuletic10,CBS11,BohnetSqueeze13},  one-axis twisting arising from probe-mediated atom-atom interactions~\cite{Vuletic10b, Vuletic10c, Vuletic12b, Takahashi05, Deutsch10, Deutsch12}, and direct collisional interactions that generate one-axis squeezing~\cite{Oberthaler08, Oberthaler10,Treutlein10} or parametric pair generation~\cite{Klempt11, Chapman12}.  In all of these cases,  a low noise readout such as the approach described in this paper is always required to actually exploit the enhanced phase-sensing properties of these states.

\begin{figure}
\includegraphics[width=3.4in]{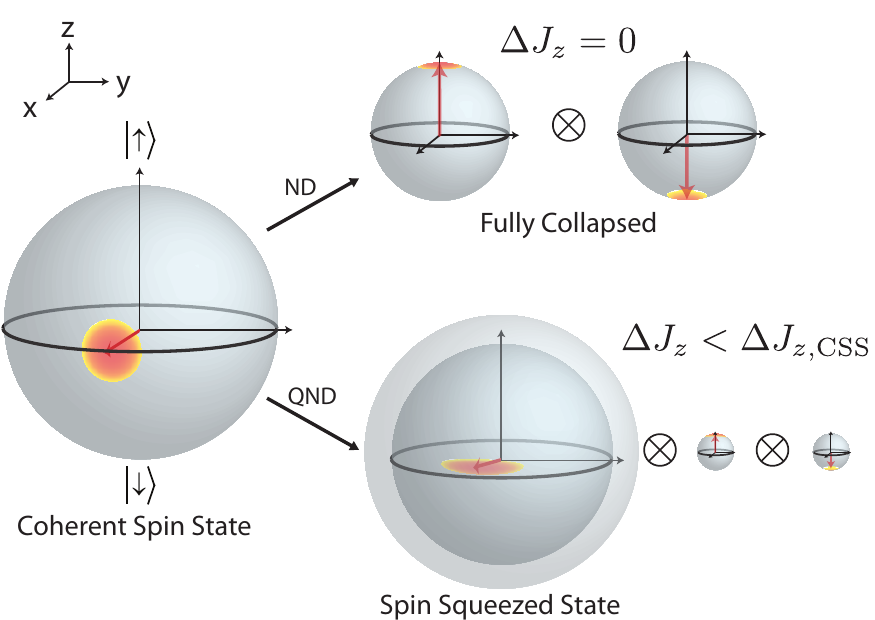}
\caption{(Color online)  Visualization of a non-demolition (ND) (upper right) and a quantum non-demolition (QND) (lower right) measurement.   A coherent spin state (CSS) for an ensemble of $N$ spin-1/2 atoms prior to measurement is represented as a Bloch vector (red arrow) of length $N/2$.  The quantum noise in the orientation of the Bloch vector is visualized as a quasi-probability distribution (red/yellow region) perpendicular to the Bloch vector, with rms opening angle at the Standard Quantum Limit (SQL) $\Delta \theta_{\rm SQL}=1/\sqrt{N}$.    A non-demolition measurement in this paper will refer to a measurement of the state's $\hat{z}$ spin projection $J_z= (\Nup-\Ndown)/2$ with an rms imprecision  $\Delta J_z < \qpn = \sqrt{N}/2$, and with the majority of the atoms remaining trapped after the measurement.   For example, in the ND measurement visualized on the right, the measurement imprecision is  $\Delta J_z =0$, but after the measurement the atoms are described by a product state of atoms in spin up and down due to single-atom state information gained by the environment via free space scattering of light.     We define a quantum non-demolition measurement via the additional requirement that  a sufficiently large number of atoms remain in a coherent superposition of spin up and down such that the resulting state, conditioned on the measurement outcome, has a polar angle uncertainty $\Delta \theta<\Delta \theta_{\rm SQL}$.  The definition of quantum non-demolition employed here is related to, but less restrictive than that of Ref.~\cite{Grangier98}.  The lower right QND example visualizes the conditional state as a product state of a squeezed state and the atoms that have been collapsed into spin up and down.}
\label{QND_ND_Definitions}
\end{figure}

In particular, there has been substantial recent interest in developing non-demolition readout schemes for large laser-cooled and quantum-degenerate neutral atomic ensembles consisting of roughly $10^{3}$ to $10{^7}$ atoms~\cite{Jessen06, Lemonde09, Takahashi99, Takano09,Polzik09, Vuletic10, CBS11,Mitchell12}.   It is well known that significant improvements in readout sensitivity can be achieved by optically probing ensembles in free space along directions of large resonant optical depth. This approach has been extensively analyzed theoretically~\cite{Mandel98,Wiseman02,Mabuchi03,Molmer04a,Molmer04b,Kennedy04,Auzinsh04,Hope04,Mabuchi04,Hope05,Polzik05,Deutsch10,Baragiola13}, and experimentally studied~\cite{Kuzmich99,Bigelow00,Jessen06,Polzik08,Polzik09,Polzik10b,Mitchell10a,Mitchell10b,Romalis10,Romalis11}.  

More recently, the technique of free space probing of large optical depth samples has been extended to using optical cavities to effectively increase the optical depth of the atomic ensemble (Fig.~ \ref{BasicEnergyLevelsCavity})~\cite{Ye98,Vuletic06,Molmer08,Kasevich08,Vuletic10,CBS11,Bouyer11,Vuletic12}. While free space ensembles of large optical depth have been realized, a cavity can enhance the already-large optical depth to a regime difficult to achieve using free space techniques alone.  It is crucial to develop techniques that are compatible with current cold atom technology to go beyond proof of principle experiments.  For reference, cold atom precision measurement experiments including optical lattice clocks, microwave fountain clocks, and matter wave interferometers, operate with of order $10^3$ to $10^7$ atoms. Optical cavities are amenable to the geometry in these kinds of experiments, and in fact, some optical lattice clocks are already incorporating optical cavities to build up power in the lattice trap.

From a metrology perspective, cavity probing achieves the same optical depth as free space probing using atomic densities lower by of order the cavity finesse, reducing atomic-density dependent atom loss, dephasing, and systematic errors.   In most of these experiments and proposals, the cavity is far-detuned from the optical transition that was probed. Probing in the resonant regime~\cite{CBS11} is an exception rather than the norm. In principle, the cavity detuning $\dc$ can be chosen almost arbitrarily. Therefore, a natural question to ask is: How does cavity detuning affect both the fundamental and technical atomic population measurement resolution or the degree of spin-squeezing for a given cavity geometry and cavity finesse?

To answer the question posed above, we provide detailed expressions for the fundamental scalings for probing an atomic ensemble using an optical cavity that smoothly connects the resonant to the far-detuned probing  regime. We apply our results to first estimate the amount of photon-recoil heating of the ensemble when the cavity-aided measurement has an imprecision at the quantum projection noise level.  The average number of photon recoils per atom sets the degree to which the measurement can be considered non-demolition. This analysis is then extended to estimate bounds on the degree of conditional spin squeezing (see Fig.~\ref{QND_ND_Definitions} for details) that can be obtained using $\Rb$ atoms in a cavity~(see Fig.~ \ref{BasicEnergyLevelsCavity}a) \cite{Vuletic10,CBS11}. We show that the fundamental limitations are set by the collective cooperativity parameter $NC$, the probe detection quantum efficiency $q$,and atomic properties alone.  The collective cooperativity parameter $NC$ plays a similar role to the resonant optical depth of atoms in free space, where $N$ is the number of atoms in the probe volume and $C$ is the single-atom  cooperativity parameter~\cite{Vuletic11}. 

This paper is organized as follows.  In Sec.~\ref{sect2}, we begin with a review of the properties of the coupled atoms-cavity system including dissipation. This review also provides precise definitions and notation used throughout the paper.

In Sec.~\ref{sect3}, we derive the quantum-limited signal-to-noise ratio for non-demolition measurements of atomic populations considered as a function of the cavity detuning from atomic resonance ($\delta_c$ in Fig.~\ref{BasicEnergyLevelsCavity}a). We identify three different probing regimes, a resonant regime and two detuned regimes separated by a critical detuning $\dc^\circ$.  

In Sec.~\ref{sect5}, we address quantum back-action effects due to probe-induced spin flips on estimates of atomic populations in a simple three-level model. We show how the optimal measurement resolution can be achieved by balancing between noise added by spin flips and averaging down the probe's vacuum or photon shot noise.

In Sec.~\ref{sect6}, we consider the limits set on coherence preservation.  Coherence is lost due to wavefunction collapse into spin up or down driven by the same probe induced free space scattering that also causes photon recoil heating.  We then obtain the optimal spectroscopic enhancement, the figure of merit for a quantum  non-demolition measurement, as a function of the spin flip probability $p$.

In Sec.~\ref{sect7}, we apply the results of Sec.~\ref{sect6} to two concrete examples in \Rb: generating conditional spin squeezing first using a non-cycling optical transition, and then a cycling optical transition.  Here, we demonstrate the key role of the ratio of the ground state hyperfine splitting $ \omega_\mathrm{hf}$ to the optical transition width $\Gamma$ for determining scalings and fundamental limits on conditional spin squeezing.

%******************************************************************************************************
%*************************************Coupled Atom Cavity Modes****************************
%******************************************************************************************************
\section{Coupled Atom Cavity Modes}\label{sect2}

To begin, we provide a brief review of the open coupled atoms-cavity system with the goal of providing a framework for understanding the experimental work and to explicitly enumerate the assumptions made to reduce this system to a classical two-mode system~\cite{Molmer09,Auffeves11,Albert12}. The dynamics of the system under a classical drive and dissipation are then studied with the goal of obtaining the full complex response of the reflected and transmitted cavity field. Finally, a discussion of the probe signal-to-noise sets the stage for addressing measurement resolution at the projection noise level in Sec.~\ref{sect3}.

%******************************************************************************************************
%*************************************Linearized Response***********************************
%******************************************************************************************************
\subsection{System Hamiltonian}\label{sect2a}

\begin{figure}
\includegraphics[width=3.4in]{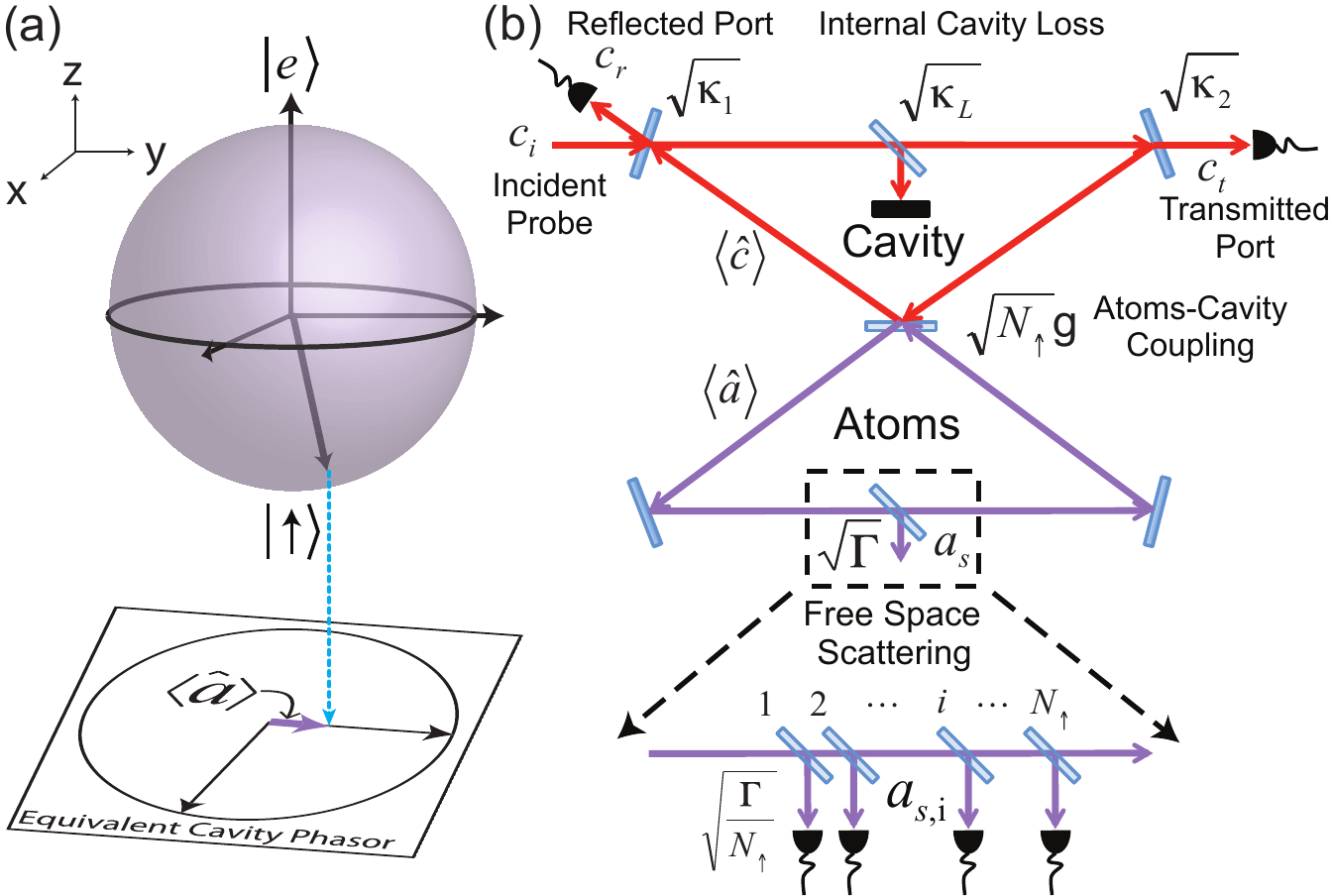}
\caption{(Color online) (a) Graphical representation of the linearization achieved via the Holstein-Primakoff approximation.   The atomic sub-system $\ketup$ and $\ket{e}$ is described by a Bloch vector of length $(\Nup +N_\mathrm{e})/2$.  In the regime in which the probe light only weakly excites the atoms such that $N_\mathrm{e} \ll \Nup$, the two-level system can be described by its projection onto a 2D plane equivalent to that describing a light field.  (b) With this approximation,  the collective atomic mode may be treated as an equivalent cavity mode (lower purple mode) whose coupling to the actual cavity (upper red mode) is governed by a partially reflecting mirror (center) described by a field coupling rate constant $(\sqrt{\Nup}g)$ that depends on the number of atoms in state $\ketup$.  The physical mirrors have transmission coefficients described by the field coupling rates $\sqrt{\kappa_1}$ and $\sqrt{\kappa_2}$, and internal cavity losses are described by $\sqrt{\kappa_L}$, such that the total power decay rate is $\kappa = \kappa_1+\kappa_2+\kappa_L$.  Decay of the atoms by emission of a photon into free space is described by the field transmission coefficient $\sqrt{\Gamma}$ that is the sum of the field scattering into free space modes by the atoms in $\ketup$.  Because the scattered modes are distinguishable (expanded view of free space scattering port), it is possible to tell which atoms are in $\ketup$ from the free-space-scattered photons, destroying any coherent superposition between $\ketup$ and another state $\ketdown$ (not shown) that may have been prepared for sensing a quantum phase.}
\label{CoupledCavityAtomModel}
\end{figure}

We consider an ensemble of $N$ atoms with two ground states $\ketup$ and $\ketdown$ whose populations we wish to estimate precisely.  The ensemble is confined and collectively coupled to a cavity mode (see Fig.~\ref{BasicEnergyLevelsCavity}(b)).  Atoms in $\ketup$ interact with the cavity mode by absorbing a cavity photon and being promoted to an optically excited state $\ket{e}$.  On the other hand, atoms in $\ketdown$ are assumed to not interact with the cavity mode because of dipole selection rules, a large energy splitting between the ground states, or otherwise.  A quantum phase may be encoded in the coherence between $\ketup$ and $\ketdown$, but is otherwise not important in this section. 

The Tavis-Cummings Hamiltonian that describes the coupled atoms-cavity system is 
\begin{eqnarray}
H =  \hbar \dc \hat{c}^\dagger \hat{c} + \hbar g \left(\hat{J}_- \hat{c}^{\dagger} +\hat{J}_+  \hat{c}  \right)  \, .
\end{eqnarray}

\noindent The Hamiltonian is written in a frame rotating at the $\ketup \rightarrow \ket{e}$ atomic transition angular frequency $\omega_{\uparrow e}$.   In this paper, we assume every atom couples to the cavity mode with the same coupling strength parametrized by the coupling angular frequency $g$. Uniform coupling can be implemented with ring cavities for example.   In the case of non-uniform coupling, an effective $g$ and $N$ may be defined~\cite{CBS11}. The cavity field is described by the photon annihilation operator $\hat{c}$, with cavity photon number $\hat{M}_c =  \hat{c}^\dagger \hat{c}$. The cavity detuning is $\dc = \omega_c -\omega_{\uparrow e}$, where $\omega_c$ is the empty cavity frequency. The collective raising and lowering operators $\hat{J}_\pm = \sum_{i} \hat{\sigma}_{i\pm}$ are written in terms of the single-atom raising and lowering operators $\hat{\sigma}_{i+} = \ket{e_i}\bra{\uparrow_i}$ and $\hat{\sigma}_{i-} = \ket{\!\!\uparrow_i}\bra{e_i}$.   The atomic populations are given by the collective projection operators $\hat{N}_\uparrow = \sum_{i}  \ket{\!\!\uparrow_i}\bra{\uparrow_i\!\!}$, $\hat{N}_\downarrow = \sum_{i}  \ket{\!\!\downarrow_i}\bra{\downarrow_i\!\!}$, and $\hat{N}_e = \sum_{i}  \ket{e_i}\bra{e_i}$. For brevity, we use the following abbreviations throughout this paper: $\Nup \equiv \avg{\hat{N}_\uparrow}$, $\Ndown \equiv \avg{\hat{N}_\downarrow}$ and $N_e \equiv \avg{\hat{N}_e}$.

Although the atoms may in general exist in a superposition of $\ketup$ and $\ketdown$, in the following analysis we will consider the atoms to be in a definite eigenstate $\Nup$ of the $\hat{N}_{\uparrow}$ operator.  We will then reintroduce the fluctuations in the operator $\hat{N}_{\uparrow}$  for atoms in a superposition of $\ketup$ and $\ketdown$ using the rms projection noise about the mean value $ \Delta\Nup = \sqrt{\left\langle  (\hat{N}_\uparrow - \langle \hat{N}_\uparrow \rangle  )^2 \right\rangle}$.

To gain information about the atoms, the effect of the atoms on an incident cavity probe field is measured in transmission and/or reflection. We assume the system is driven weakly by the probe such that the mean number of atoms in the optically excited state $\ket{e}$ is a small fraction of the total number of atoms in $\ketup$, i.e. $N_e / \Nup \ll 1$.   In the weak excitation limit, the Holstein-Primakoff approximation~\cite{HP40} may be employed, replacing the atomic raising and lowering operators with effective creation and annihilation operators as $\hat{a}^\dagger \approx \hat{J}_+ / \sqrt{\Nup}$ and $\hat{a} \approx \hat{J}_- / \sqrt{\Nup}$  that satisfy the usual commutation relation $[\hat{a}, \hat{a}^\dagger] =1$.  The resulting Hamiltonian in the Holstein-Primakoff approximation can be described by two coupled cavities, visualized in Fig.~\ref{CoupledCavityAtomModel}(b),
\begin{eqnarray}
H =  \hbar \dc \hat{c}^\dagger \hat{c} + \hbar  \sqrt{N_\uparrow} g \left( \hat{a} \hat{c}^{\dagger} +  \hat{a}^{\dagger} \hat{c}  \right) \, .
\end{eqnarray}

%******************************************************************************************************
%******************************Driven and Damped Dynamics*******************************
%******************************************************************************************************
\subsection{Driven and damped dynamics}\label{sect2b}

Using input-output theory~\cite{Gardiner85}, the Heisenberg-Langevin equations of motion for the cavity and atomic operators that include driving and damping are as follows:
\begin{eqnarray}
\frac{d\left<\hat{c}\right>}{d t} &=& - \left(\imath \dc + \frac{\kappa}{2} \right) \left<\hat{c}\right> - \imath \sqrt{\Nup}g \left<\hat{a}\right> + \sqrt{\kappa_1} c_{i} \, , \nonumber \\
\frac{d \left<\hat{a}\right>}{d t}  &=&  - \frac{\Gamma}{2} \left<\hat{a}\right> -  \imath \sqrt{\Nup}g \left<\hat{c}\right> \, .
\label{eqn:Langevin}
\end{eqnarray}

\noindent The complex amplitude $c_i$, with units of $\sqrt{\mathrm{photons/sec}}$, describes the incident cavity driving field at frequency $\omega_p$ in the lab frame. As the above equation is written in a rotating frame at the atomic frequency $\omega_{\uparrow e}$, the incident cavity field $c_i$ in Eq.~\ref{eqn:Langevin} is $c_i = |c_i| e^{-i \delta_{pe} t}$ where $\delta_{pe} = \omega_p - \omega_{\uparrow e}$ is the drive detuning from the optically excited state $\ket{e}$. The non-unitary damping and drive terms are shown schematically in Fig. \ref{CoupledCavityAtomModel}(b). The eigenfrequencies $\omega_{\pm}$ and linewidths $\kappa'_\pm$ of the normal modes described by the coupled equations are given by \begin{eqnarray}
\omega_\pm = \frac{\dc\pm \sqrt{\dc^2+\rabi^2 }}{2} \, ,
\label{eqn:modefreq}
\end{eqnarray}

\begin{eqnarray}
\kappa'_\pm =\frac{  \kappa + \left( \frac{\rabi}{2\omega_\pm} \right)^2 \Gamma}{1+ \left( \frac{\rabi}{2\omega_\pm} \right)^2} \, .
\label{eqn:Dressedkappa}
\end{eqnarray}

\noindent where 

\begin{eqnarray}
\rabi \equiv \sqrt{\Nup }  2g \, ,
\end{eqnarray}

\noindent is the collective vacuum Rabi frequency, and $\rabi^2 \gg \Gamma \kappa$ is assumed.  The collective vacuum Rabi splitting $\rabi$ sets the difference in the normal mode frequencies $\op - \om$ at zero detuning $\dc=0$.  For atom number counting via cavity probing, the normal mode that is farthest from atomic resonance is most useful because this normal mode is predominantly cavity-like in character. For brevity, we refer to this mode's linewidth and frequency as simply $\kappa'$ and $\omega$ such that $\kappa' = \kappa'_\pm$ and $\omega = \omega_\pm$ when $\left | \omega_\pm \right | \ge  \left | \omega_\mp \right | $.

%******************************************************************************************************
%*************************Cavity Damping and Input-Output Fields****************************
%******************************************************************************************************
\subsubsection{Cavity damping and input-output fields}\label{sect2b1}

As shown in Fig. \ref{CoupledCavityAtomModel}(b), the damping of the cavity field at rate $\kappa/2 = (\kappa_1 + \kappa_2+\kappa_L$)/2 is set by the mirror power transmission coefficients $T_{1,2}$ such that $\kappa_{1,2}= T_{1,2}\times \FSR$. The cavity free spectral range is $\FSR = c/2l$, with $2l$ being the round-trip cavity length, and $c$ the speed of light.  The total round-trip scattering and absorption fractional power losses at the mirrors $L$ can be modeled by an additional beam splitter with field decay rate $\kappa_L = L \times \FSR$.

The reflected and transmitted complex field amplitudes, $c_r$ and $c_t$ respectively, will be detected to infer the number of atoms in $\ketup$.  The external field normalizations are chosen such that $\left |c_{i, r, t} \right |^2$ is the flux of incident, reflected, and transmitted probe photons in units of photons/second.  The average number of incident, reflected, and transmitted photons $M_{i,r,t}$ in a measurement time interval $T_m$ is then
\begin{eqnarray}
M_{i,r,t} = \int^{T_m}_{0} \left |c_{i, r, t}(t') \right |^2 \mathrm{d} t' .
\end{eqnarray}

\noindent In our experiments~\cite{CBS11}, it is convenient to express the number of probe photons coupled into the atoms-cavity system in terms of the measured ``missing"  photons in the reflected mode compared to the incident beam $M_m \equiv M_i-M_r$.

The reflected and transmitted fields can be found by first solving the coupled-driven Eq.~(\ref{eqn:Langevin}) for $ \left<\hat{c}\right>$ and then using the results in the approximate relationships
\begin{eqnarray}
 c_r &= & \sqrt{\kappa_1} \left<\hat{c}\right> - c_i \, , \nonumber \\
 c_t  &=& \sqrt{\kappa_2} \left<\hat{c}\right> \, ,
\label{InputOutputRelations}
\end{eqnarray}

\noindent that hold in the limit of a high finesse cavity $T_{1,2}, L \ll1$.

%******************************************************************************************************
%****************************Atomic Damping via Free Space Decay****************************
%******************************************************************************************************
\subsubsection{Atomic damping via free space decay}\label{sect2b2}

The atomic damping via scattering of light into free space (i.e. not into the cavity mode) is described by an effective amplitude damping rate $\Gamma/2$. To good approximation, the probability decay rate $\Gamma$ is simply the single-particle excited state \ket{e} decay rate in free space~\footnote{The approximation that the single-particle decay rate $\Gamma$ into all modes other than the cavity mode holds true in the limit that the cavity subtends a small fraction of the total solid angle as seen by the atom~\cite{Kimble98}.}.  The rate of scattering into free space is described by the field amplitude $a_s = \sqrt{\Gamma} \left<\hat{a}\right>$, normalized such that the rate of photons scattered into free space is simply $\dot{M}_s=\left | a_s \right |^2$.   

The above picture of atomic damping can be further refined as shown in Fig. \ref{CoupledCavityAtomModel}(b).  While the decay of excitation from the cavity mirrors is single-mode in nature, the atoms scatter light into many free space modes. This multimode scattering can be envisioned by replacing the single decay process via a single mirror with a weak beam splitter for each atom in $\ketup$.  If the ensemble is optically thin along all directions except the cavity mode, then one can approximate that each atom decays into its own bath of states with an amplitude $a_{s,i} = \sqrt\Gamma \left( \left< \hat{a} \right>/\sqrt{\Nup} \right)$.   The total scattering rate is the incoherent sum of the decay rates, reproducing the previous decay rate $\dot{M}_s=\Gamma \left | \avg{\hat{a}} \right |^2$.  However, this refinement importantly emphasizes that the multimode free space scattering leads to in-principle information gain as to which particular atoms are in $\ketup$, causing single particle collapse of the atomic wavefunction from a coherent superposition into an energy eigenstate, for example $\left(\ketup + \ketdown\right)/\sqrt2 \to \ketup$, thus destroying coherence.  In contrast, the decay of light through the cavity mirrors leads to only collective information as to how many atoms total are in spin up and therefore preserves coherence.   Thus information gained through the cavity will be useful for preparing conditionally spin-squeezed states,  while the free space scattering is a competing decoherence mechanism that serves to reduce the attainable degree of spin squeezing.

%******************************************************************************************************
%*******************Full Complex Field Response to Probing***************************
%******************************************************************************************************
\subsection{Full complex field response to probing}\label{sect2c}

\begin{figure*}
\includegraphics[width=6.5in]{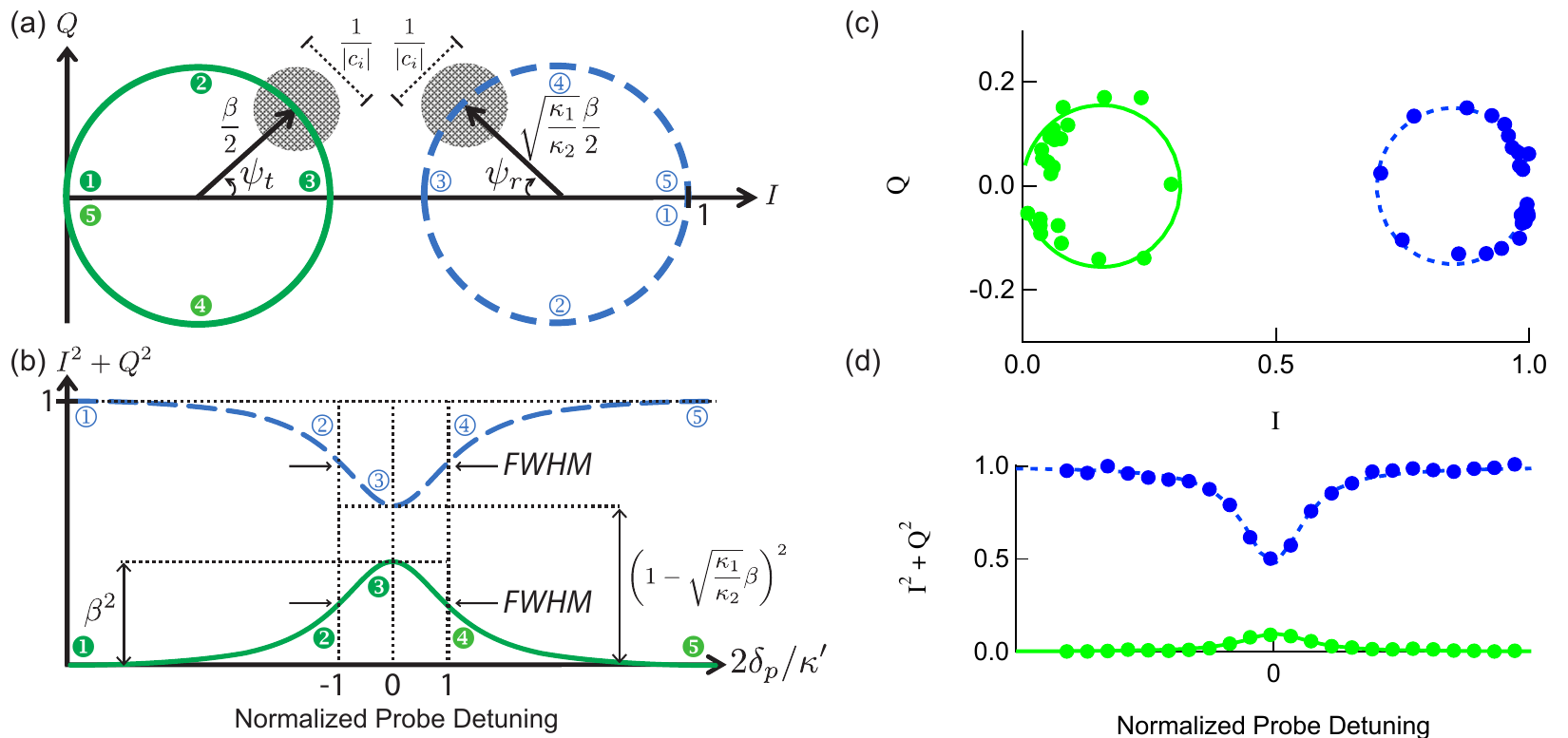}
\caption{(Color online) Transmitted and reflected probe electric fields from driving the atoms-cavity system through the cavity mode. (a) The electric field phasors trace out circles in the $I$, $Q$-quadrature plane as the probe detuning $\delta_{p}$ from the dressed cavity resonance varies from far-below to far-above resonance. The normalization is chosen such that the reflected electric field goes to 1 when far off resonance. The quantum noise of the probe normalized to the incident electric field is represented as a fuzzy blob with rms diameter $1/|c_i|$. In this illustration, a symmetric cavity  $\kappa_1 = \kappa_2$ is assumed, so that the circles have the same diameter $\beta$. (b) Corresponding power transmission and reflection signals.  (c, d) Typical experimental data (points) and least square fits to the $I$, $Q$-quadrature data  (solid and dashed circles) from Ref.~\cite{CBS11}. }
\label{IQ_SignalToNoise}
\end{figure*}

The reflected and transmitted fields relative to the incident field $c_{r,t}/c_i = I_{r, t} +  \imath Q_{r,t}$ can be described in the complex plane by the real amplitudes  $I_{r,t}$ and $Q_{r,t}$.  We consider a single incident probe frequency close to resonance such that the probe frequency $\omega_p$ is detuned by $\delta_{p} = (\omega_p - \omega_{\uparrow e} - \omega)$.   We assume the probe is near resonance $|\delta_{p}| \ll \op -\om$ and the modes are well resolved  $\omega_+ - \omega_- \gg \kappa_\pm'$ so that interference effects between normal modes can be ignored.  The normalized transmitted electric field through the cavity is then, to a good approximation, given by
\begin{eqnarray} 
I_t=  \frac{\beta}{1+\left(2 \delta_{p}/\kappa' \right)^2} \, , \\
Q_t =  \frac{\beta \left( 2 \delta_{p}/\kappa' \right)}{1+\left(2 \delta_{p}/\kappa' \right)^2} \, ,
\end{eqnarray}

\noindent  where the dimension-less amplitude $\beta$ is given by 
\begin{eqnarray}
\beta = \frac{2 \sqrt{\kappa_1 \kappa_2}}{\kappa + \Gamma \left(\frac{\rabi}{2 \omega}\right)^2} \, .
\label{eqn:beta}
\end{eqnarray}

From Eq.~(\ref{InputOutputRelations}), the reflected field $c_r$ is just the sum of the transmitted field (rescaled for relative transmission coefficients) and the largely reflected field such that $I_r =1 - \sqrt{\kappa_1/\kappa_2}I_t$, and $Q_r= \sqrt{\kappa_1/\kappa_2}Q_t$. 

As shown in Fig.~\ref{IQ_SignalToNoise}(a), the phasor $c_t$ traces out a circle of radius $\beta/2$ in the complex plane as $\delta_{p}$ varies from $\ll \kappa'/2$ to $\gg \kappa'/2$.   The translation $I'_t = I_t - \beta/2$ centers the circle traced out by the phasor $I'_t + \imath Q_t$ at the origin. One then sees that the angle with respect to the real axis is given by $\psi_{t} = \arctan(Q_t/I'_t)= \arctan(2 \delta_{p}/\kappa')$. Similarly, the translation $I'_r = I_r - (1 - \sqrt{\kappa_1/\kappa_2}\beta/2)$ centers the circle traced out by the phasor $I'_r + \imath Q_r$ at the origin with the angle $\psi_{r}$ defined with respect to the real axis such that $\psi_{r} = \arctan(-Q_r/I'_r)$.   The angles $\psi_r$ and $\psi_t$ are the same, but the quantum-limited estimation of the phases may be different if $\kappa_1 \neq \kappa_2$.

%******************************************************************************************************
%*******************Probe Vacuum Noise and Measurement Resolution***************************
%******************************************************************************************************
\subsection{Probe vacuum noise and measurement resolution}\label{sect2d}

The size of the quantum vacuum noise that contributes uncertainty to measuring the position of the phasor is not changed by a linear transformation of coordinates in the complex plane. For our purposes, the noise can be described as a Gaussian probability distribution with equal and uncorrelated real and imaginary rms fluctuations of magnitude $\sigma_v = 1/2$.  The rms quantum vacuum uncertainty $\Delta \psi_t$ on the angle $\psi_t$ is then
 independent of the average value $\psi_t$ and is set only by the average number of detected photons in transmission $M_d = q_d M_t$ as
 \begin{eqnarray}
 \Delta\psi_t = \frac{1}{2 \sqrt{M_d}} \, .
 \label{DeltaPsi}
\end{eqnarray}

\noindent The detection quantum efficiency $q_d$  includes any light loss and any excess technical or thermal noise of the detector relative to vacuum noise.   The uncertainty  $\Delta\psi_t$ maps onto an uncertainty on the estimation of $\delta_{p}$ through $\Delta \delta_{p} = \left | d \delta_{p}/d\psi_t \right|\Delta \psi_t = \kappa' \Delta\psi_t /2 \detmethod$.  The detection sensitivity $\detmethod$ is given by
\begin{eqnarray}
\detmethod = \frac{1}{1+\left(2 \delta_{p}/\kappa' \right)^2} \, .
\end{eqnarray}

\noindent  Probing near resonance $\delta_{p} =0$, one finds $\detmethod =1$.  For side-of-fringe probing $\delta_{p}=\kappa'/2$, one finds $\detmethod =1/2$.  If the probe frequency is linearly and adiabatically scanned  from $\delta_{p} \ll \kappa'$ to $\delta_{p}\gg  \kappa'$ such that the total number of detected photons is fixed to the same $M_d$ as in the two previous scenarios, one finds $\detmethod =1/2$.   The optimal readout assumes that as $\delta_{p}$ is changed, an adaptive homodyne readout is employed to maximize the measurement sensitivity to small changes in $\psi_t$.  In Ref.~\cite{CBS11}, heterodyne detection is employed so that adaptive detection is not required. However, the effective quantum efficiency $q_d$ was reduced by 1/2 as a result of the heterodyne detection.

It is straightforward to extend the analysis to a probe signal detected in reflection.  However, one must parameterize in terms of the measurable  average number of missing photons in the reflection port $M_m$ and the average number of incident photons $M_i$ such that in Eq.~(\ref{DeltaPsi}), one substitutes $M_d \rightarrow (\kappa_2 /\kappa_1) M_i q_d (1 \mp \sqrt{1-M_m/M_i} )^2$ when $\beta \sqrt{\kappa_1/\kappa_2} \lessgtr 1$.

%******************************************************************************************************
%*******************Quantum-Limited Signal to Noise and Free Space Scattering*****************
%******************************************************************************************************
\section{Quantum-Limited Signal-to-Noise and Free Space Scattering}\label{sect3}

The measurement of the atomic population $\Nup$ in $\ketup$ is achieved by precisely measuring the dressed mode frequency $\op$ or $\om$ or some combination of the two.  In essence, the approach used here  converts the problem of measuring an atomic population into a frequency measurement.  For atoms in a coherent superposition of $\ketup$ and $\ketdown$, quantum projection noise in the atomic population $\Nup$ causes the dressed mode frequency to fluctuate from one trial to the next. 

In this section, we first derive the trial to trial fluctuations on the dressed mode frequency due to quantum projection noise as a function of cavity detuning $\dc$. We then use the results of Sec.~\ref{sect2} to obtain the average number of free-space-scattered photons per atom $\msp$, when the measurement imprecision on the probe field is sufficient to resolve the projection noise fluctuations of the mode frequency $\dwproj$.  The quantity $\msp$ is the key figure of merit that characterizes the degree to which a measurement is non-demolition. Three limits of cavity probing are identified, and a summary table of various key quantities in different regimes is presented.

%******************************************************************************************************
%*******************Projection-Noise-Driven Fluctuations of Mode Frequencies*******************
%******************************************************************************************************
\subsection{Projection-noise-driven fluctuations of mode frequencies}\label{sect3a}

\begin{figure}
\includegraphics[width=3.4in]{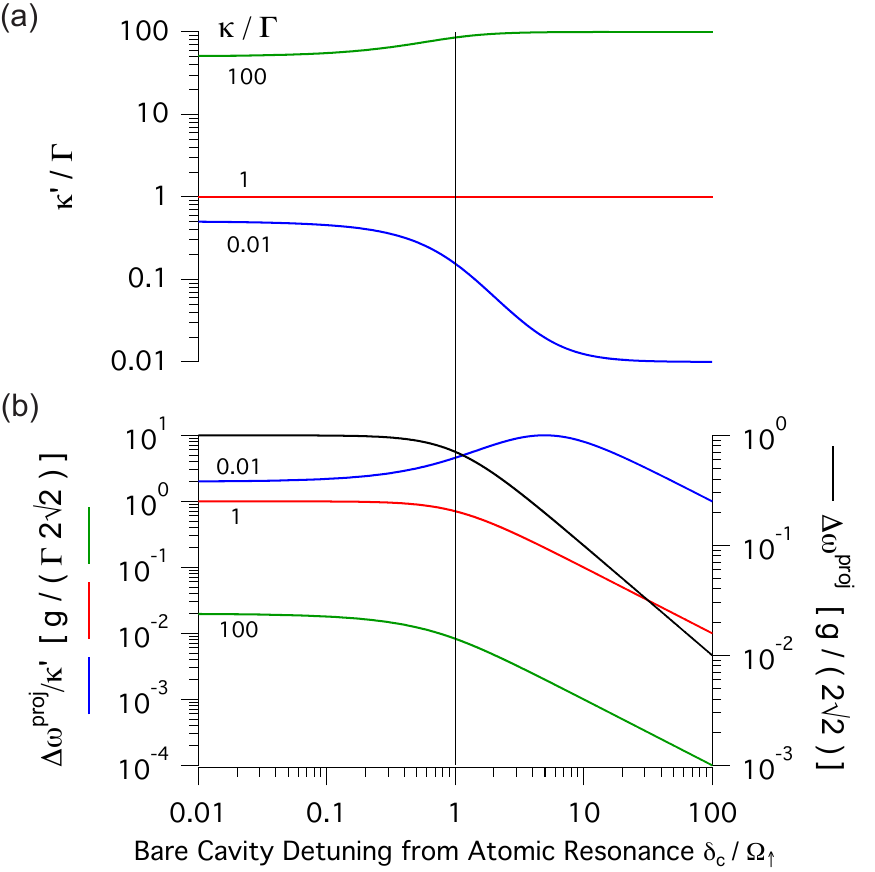}
\caption{(Color online) Theoretical scaling of key  quantities with cavity detuning $\dc$ expressed in units of the the collective vacuum Rabi frequency $\rabi$.  (a) The dressed cavity linewidth $\kappa'$ in units of the atomic excited state linewidth $\Gamma$.  In (a) and (b), the scaling for different cavity finesses is shown for  $\kappa/\Gamma = $ 0.01, 1, and 100 (blue, red, and green, respectively).  (b, right, black curve)  The rms fluctuation of the dressed cavity mode frequency due to projection noise fluctuations $\Delta \op^\mathrm{proj}$ decreases as $1/\dc$ above $\dc/\rabi=1$.  The normalization is chosen such that one should multiply by $g/2\sqrt{2}$. (b, left, red, blue, green curves).  The ratio of the projection noise fluctuation of the cavity mode to the dressed cavity linewidth $\dwproj/\kappa'$ is shown normalized such that the plotted values should be multiplied by $g/(2\sqrt{2}\Gamma)$.  A large ratio is desirable because technical noise may limit the ability to split the probed resonance by more than a fractional amount.}
\label{ModeScaling}
\end{figure}

\begin{figure}
\includegraphics[width=3.4in]{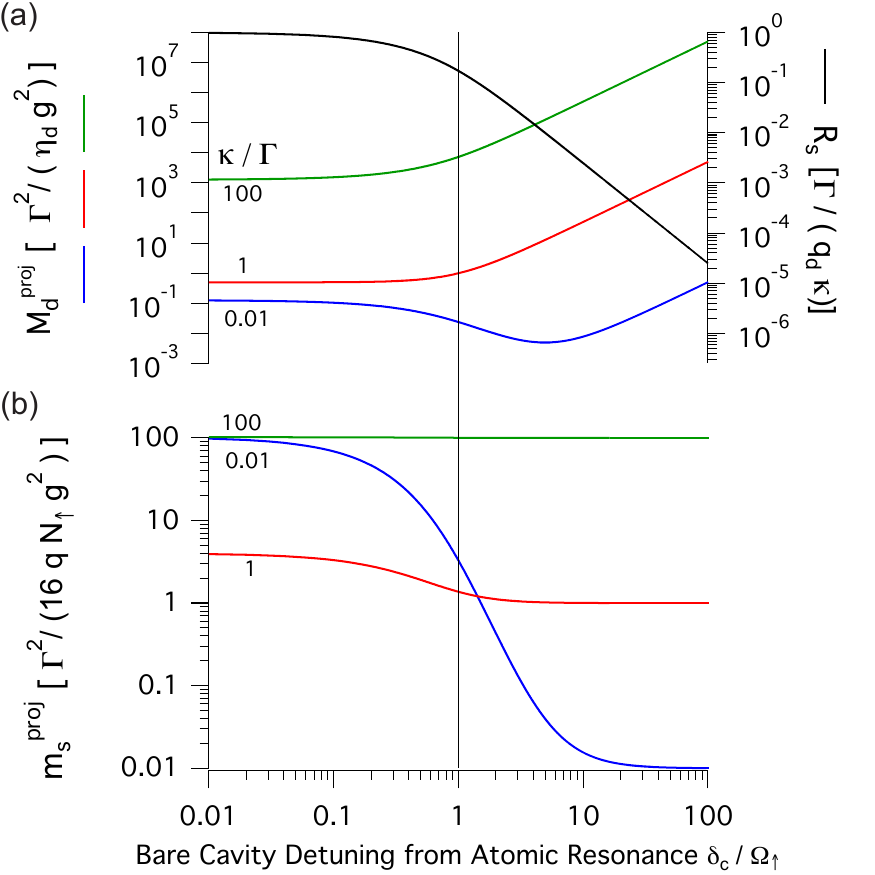}
\caption{(Color online) Theoretical scaling of key  quantities with cavity detuning $\dc$ expressed in units of the the collective vacuum Rabi frequency $\rabi$.  In (a) and (b),  the scaling for different cavity finesses is shown giving $\kappa/\Gamma = $ 0.01, 1, and 100 (blue, red, and green, respectively).  (a, left, blue, red, green )  The average number of detected photons needed to resolve the projection noise fluctuations $M_d^\mathrm{proj}$ normalized such that the plotted values should be multiplied by $\Gamma^2/(\detmethod g^2)$. (a, right, black)  The ratio of the number of free-space-scattered photons for every detected photon $R_s$, normalized such that the plotted values should be multiplied by $\Gamma/(q_d\kappa)$.  (b)  The crucial average number of scattered photons into free space per atom $\msp$ when the atomic population measurement precision  is equal to the projection noise fluctuations.  The normalization is such that the plotted values should be multiplied by $\Gamma^2/(16 q \Nup g^2)$.  In the bad-cavity limit of $\kappa\gg \Gamma$ (green curves), there is little fundamental advantage to operating away from resonance $\dc =0$. The technical requirements are simply increased as a result of detuning.  As the finesse of the cavity $F$ is increased, the amount of free space scattering falls roughly as $1/F$ until the good cavity regime is reached when  $\kappa \ll \Gamma$ (blue curves).  Here, one must detune by roughly the critical detuning $\dc^\circ$ in order to realize the full advantage of having increased the cavity finesse.  Importantly, note that $\msp$ does not significantly decrease above $\dc^\circ$ owing to cancellation in this regime of the scaling of $R_s\sim 1/\dc^2$ with the scaling of $M_d^\mathrm{proj}\sim \dc^2$.}
\label{TheoryScalingQuantumNoiseLimit}
\end{figure}

\begin{figure}
\includegraphics[width=3.4in]{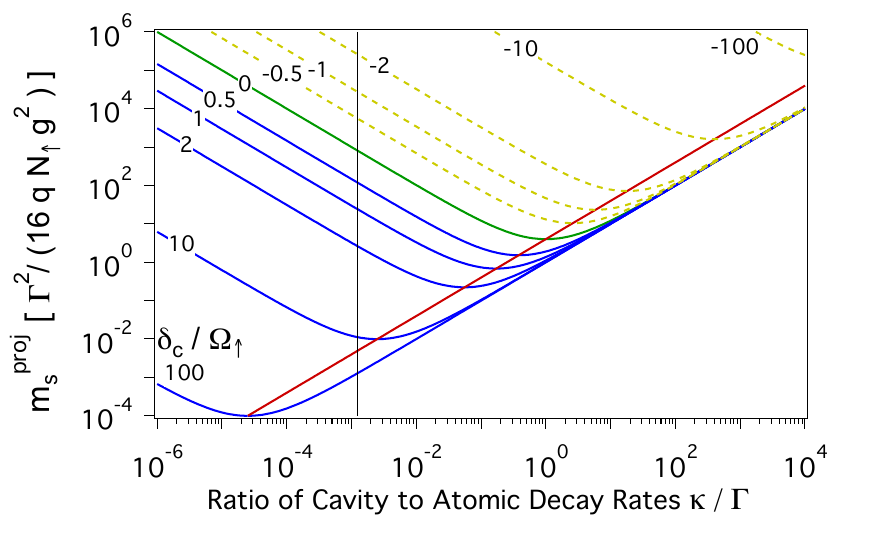}
\caption{(Color online) The crucial average number of scattered photons into free space per atom such that the detected probe light allows resolution of projection noise fluctuations $\msp$  versus the ratio of the cavity power to atomic population decay rates $\kappa/\Gamma$.  The normalization is chosen such that the plotted values should be multiplied by $\Gamma^2/(16 q \Nup g^2)$.  Here, we assume that $\kappa$ is varied by changing the cavity finesse while holding the cavity length fixed (such that $g$ is constant).  Each trace represents a fixed ratio of the bare cavity detuning to the collective vacuum Rabi frequency $\dc/\Omega_\uparrow$ , with values labeled on the traces and blue denoting $\dc/\Omega_\uparrow>0$, green $\dc/\Omega_\uparrow=0$ and dashed yellow $\dc/\Omega_\uparrow<0$ .  Here we consider probing the dressed mode at frequency $\op$.  At a fixed detuning (and therefore fixed projection-noise-driven frequency fluctuation size) a minimum is reached below which a better cavity is detrimental due to the dressed cavity linewidth $ \kappa'$ becoming clamped while the ratio of free space scattering to detected photon number $R_s$ continues to rise.  In the units above, the locus of the minimum is just $4 \kappa/\Gamma$ (red line). Including the normalization, the locus of the minimum reduces to  $\msp = 1/(q \Nup C)$.  When including the dependence of $g^2$, $\kappa \sim 1/l$ on cavity length $l$, the minimum value of $\msp$ is not changed by just shortening the cavity at fixed finesse.  Only shortening the cavity length while simultaneously increasing the cavity finesse (such that $\kappa$ is constant) leads to a net fundamental reduction of $\msp$.  Finally, the vertical black line indicates the cavity linewidth for a finesse $F=10^6$ cavity, near the highest currently achievable, for $\Rb$ and $l= 2$~cm.}
\label{Fig7}
\end{figure}

As stated earlier, the atom number $\Nup$ can be determined by precisely measuring one or both the dressed mode frequencies $\op$, $\om$ with a cavity probe.  The collective enhancement of the Rabi splitting by $\sqrt{\Nup}$ produces an important enhancement of the measurement sensitivity that is key to resolving projection noise. To concretely analyze the signal to noise of the probing, we consider a measurement procedure most relevant to spectroscopy:  we assume that for each experimental trial, all of the $N \gg 1$ total atoms are initially prepared in spin down via optical pumping or otherwise.  Each atom is then rotated into an equal superposition of spin up and down, preparing the ensemble in a coherent spin state (CSS).  The population in spin up and down  fluctuate about the average $\Nup  = \Ndown = N/2$ with equal magnitude but perfectly anti-correlated projection noise fluctuations  $\Delta \Nup = \Delta \Ndown = \sqrt{N}/2$.

The rms fluctuation $\dwproj$ of the individual mode frequencies $\opm$~caused by the projection-noise-driven fluctuations in $\Nup$  is  found by linear expansion as $\dwproj= \left | d \opm / d \Nup \right | \Delta \Nup$ evaluated at $\Nup = N/2$. Making use of Eq.~(\ref{eqn:modefreq}), one finds
\begin{eqnarray}
\dwproj= \frac{g}{2\sqrt{2}}\frac{\rabi }{\sqrt{\rabi^2 + \dc^2}}  \, .
\label{eqn:ProjNoise_vs_Detune}
\end{eqnarray}

\noindent Note that $\dwproj$ carries an $N$-dependence from the Rabi splitting $\rabi$. The fluctuations of the two mode frequencies are equal in magnitude but opposite in sign such that the rms differential fluctuation is $\Delta(\op - \om)^\mathrm{proj} = 2 \dwproj$.

The projection noise variance $(\dwproj)^2$ decreases as a Lorentzian versus the bare cavity detuning $\dc$ with half width at half maximum (HWHM) $\rabi$.  Figure \ref{ModeScaling}(b) shows this scaling with detuning (black, left curve).  The technical requirements on the experiment for resolving $\dwproj$ are increased with detuning.  Other experimental imprecision and inaccuracies scale relative to the mode linewidth $\kappa'$ that one must split to the level of  $\dwproj$, therefore the ratio $\dwproj/\kappa'$ is shown in Fig.~\ref{ModeScaling}(b, left) for three different bare cavity linewidths $\kappa/\Gamma = 0.01, 1, 100$ (blue, red, green).  Note that in the good cavity limit $\kappa/\Gamma\ll1$ (blue), the experimental requirement on splitting the mode line can be somewhat reduced at larger detuning owing to the rapid fall off of $\kappa'$ as $1/\dc^2$ in the approximate region  $\dc/\rabi \in \left\{1, 10\right\}$.

%******************************************************************************************************
%****Fundamental Measurement Noise and Free Space Scattering at Arbitrary Detuning $\dc$}****
%******************************************************************************************************
\subsection{Fundamental measurement noise and free space scattering at arbitrary detuning $\dc$}\label{sect3b}

The resonance frequency of the farther detuned of the two dressed modes $\op$ or $\om$ is measured relative to the known frequency of a coherent (and un-squeezed) laser probe.  The rms uncertainty on the probe detuning $\Delta \dpr$ is equal to the projection noise fluctuation level $\dwproj$ at an average detected photon number of 
 \begin{eqnarray}
M_{d}^\mathrm{proj} = \frac{1}{2 \eta_d} \left( \frac{\kappa' }{g}\right)^2\left(1+\frac{\dc^2}{\rabi^2}\right) \, .
\label{eq:GenMd}
\end{eqnarray}

The passage of light through the cavity also leads to the scattering of $M_s = |a_s|^2 T_m$ probe photons into free space modes by the atoms in spin up.  The ratio of free space scattered to detected photons $R_s=M_s/M_d$  is given by a weighted ratio of the two damping rates as
 \begin{eqnarray}
R_{s} =  \frac{1}{q_d \eta_s}\frac{\Gamma}{\kappa} \frac{\rabi^2}{4 \omega^2} \, .
\label{eq:GenRs}
\end{eqnarray}

\noindent  The factor $\eta_s$ plays an equivalent role to a quantum efficiency and separately accounts for photons exiting the cavity via an undetected port.   In the symmetric cavity example we consider here, only the transmission port 2 is measured, and $\eta_s= \kappa_2/\kappa$ (see Fig.~\ref{CoupledCavityAtomModel}(b) for an illustration).

The {\it key} number of scattered photons into free space normalized to the total number of atoms $N$, denoted $\msp$, may then be found from Eqs.~(\ref{eq:GenRs}) and (\ref{eq:GenMd}) as
\begin{align}
\msp &=& &\frac{R_{s} M_{d}^\mathrm{proj}}{N} \\
&=& &\frac{1}{4 q \Nup C}\left(\frac{\kappa'}{\kappa}\right)^2\left( 1+\frac{\dc^2}{\rabi^2}\right) \frac{\rabi^2}{\omega^2} \, ,
\label{eqn:Genms}
\end{align}

\noindent where $C$ is the single-atom cooperativity parameter  
\begin{eqnarray}
C = \frac{(2 g)^2}{\kappa\Gamma} \, ,
\end{eqnarray}

\noindent and the total effective quantum efficiency is  
\begin{equation}
q= q_d \eta_d \eta_s \, .
\end{equation}

\noindent  For any arbitrary measurement imprecision $\Delta \dpr = \alpha \dwproj$ relative to the projection noise level, the required  average number of detected photons is  simply $M_{d} = M_{d}^\mathrm{proj}/\alpha^2$, and the average number of scattered photons normalized to the total atom number $N$ is $\ms =\msp/\alpha^2$.

A key result is that $\msp$ saturates to a finite value in the far-detuned limit

\begin{eqnarray}
\msp \rightarrow \frac{1}{4q \Nup C} \; \mathrm{as} \; |\dc | \rightarrow \infty.
\end{eqnarray}

\noindent The reason for this saturation is because in Eq.~(\ref{eqn:Genms}), the ratio of free space to detected photons asymptotically decreases as $1/\dc^2$,  but the required number of detected photons increases asymptotically as $\dc^2$.  The non-demolition character of the measurement is ultimately set by the collective cooperativity parameter and quantum efficiency $q \Nup C$.  This quantity physically sets the maximum rate at which collective information can be extracted from the ensemble compared to the rate at which single-particle information is gained by the environment via multimode scattering of light into the many modes of free space.

In the good cavity limit $\kappa \ll \Gamma$, the frequency dependence $\dc$ of Eq.~(\ref{eqn:Genms}) can be understood in three regimes:  the far-detuned dispersive regime $\left| \dc \right | > \dc^\circ$, the near-detuned dispersive regime $\left| \dc \right | < \dc^\circ$, and the resonant regime $\dc =0$.  The critical cavity detuning $\dc^\circ$ is given by
\begin{eqnarray}
\dc^\circ = \sqrt{\frac{\Gamma}{\kappa}}\frac{\rabi}{2} \frac{1}{\sqrt{q}} = \frac{\Gamma}{2} \sqrt{\frac{N_\uparrow C}{q}} \, .
 \label{eqn:DispersiveDeltaCrit}
\end{eqnarray}

\noindent  The critical detuning is the cavity detuning at which the dressed cavity linewidth is $\kappa'  = 2  \kappa$, possible only in the good cavity limit $\kappa \ll \Gamma$.  Expressions for the number of photons scattered into free space per atom $\msp$, the absolute size of the projection noise fluctuations of the mode frequency $\dwproj$, and the dressed cavity linewidth $\kappa'$ are summarized in these different regimes in Table \ref{Table:Scalings}.  Again, the quantity $\msp$ is critical for understanding the fundamental limits on both probe-induced heating of the sample and potential improvements on measurement sensitivity beyond the SQL. Collective information gained from the cavity results from a forward scattering process that leaves the momentum state of the atom unmodified and therefore does not cause recoil heating. In contrast, the probing-induced free space scattering always cause recoil heating on average, even if the atoms are tightly confined in the Lamb-Dicke regime in all three dimensions.
 
\begin{table*}[t!]
\caption{Regimes of cavity probing.  The regime name and assumptions used to define the regime are provided in the first and last columns.  The quantity $\msp$ is the average number of photons scattered into free space normalized to the total atom number $N$, required to resolve an rms  fluctuation $\sqrt{N}/2$ in the spin up population equal to the projection noise level.   The quantity $\dwproj$ is the rms angular frequency fluctuation of a single coupled atoms-cavity mode $\opm$ due to projection noise.  The quantity $\kappa'$ is the dressed cavity power decay linewidth, here taken for the mode detuned farthest from atomic resonance.  The single-particle cooperativity $C$, the number of atoms in spin up $\Nup= N/2$, the single-particle cavity coupling $g$, the empty cavity power and atomic population decay rates $\kappa$ and $\Gamma$ respectively, and the collective vacuum Rabi frequency $\rabi$ are related by the following:  $\rabi = \sqrt{\Nup} 2 g$, $\Nup C = \Nup (2 g)^2/\kappa \Gamma$.  The detuning of the empty cavity resonance frequency from the atomic transition frequency is $\dc$, and the critical detuning at which $\kappa' = 2 \kappa$ is $(\dc^{\circ})^2=\rabi^2 \Gamma/ 4 \kappa$, assuming the good cavity limit $\kappa \ll \Gamma$. The maximally detuned regime assumes that the quantity $\Nup C$ is chosen to minimize $\msp$ in the presence of the constraint that the cavity detuning cannot be made larger than some maximum value $\delta_\mathrm{max}$ set by technical constraints on resolving the projection noise fluctuations or fundamental constraints set by the internal energy level structure of the atoms being probed (for instance the ground state hyperfine splitting in $\Rb$).  }
\renewcommand{\arraystretch}{2}
    \begin{tabular}{| l | c | c| c | c  | }
    \hline
  Regime Name  				&  $\msp$ 									& $\Delta \opm^\mathrm{proj}$ 											& $\kappa'$  		&	 Assumptions 																	\\ 
    							&  $\left[\times \frac{1}{4 q}\right]$ 			& 															& $\left[\times \kappa\right]$ 		&																				 \\ \hline
   Resonant 					&  $\frac{1}{ \Nup C}\left(1+\frac{\Gamma}{\kappa} \right)^2$ 	& $\frac{g}{2\sqrt{2}}$ 											& $\frac{1}{2}\left(1+\frac{\Gamma}{\kappa}\right)$					& $\dc=0$  																		\\ \hline
  Detuned  					& $\frac{1}{\Nup C}\left(\frac{\kappa'}{\kappa}\right)^2$ 		& $\sqrt{\frac{\Nup}{2}} \frac{g^2}{ \left | \dc \right |}$	 		& $1 + \frac{\Nup g^2}{\dc^2}\left(\frac{\Gamma}{\kappa} -1\right)$						& $\dc\gg \rabi$  																	\\ \hline
  Near Detuned, Good Cavity   				& $  \Nup C   \frac{\Gamma^4}{ \left(2 \dc\right)^4}$ 			& $\sqrt{\frac{\Nup}{2}} \frac{g^2}{ \left | \dc \right |}$ 	& $ \frac{\Nup g^2}{\dc^2}\frac{\Gamma}{\kappa}$							& $ \dc^\circ\gg\dc\gg \rabi; \Gamma\gg \kappa$  											\\ \hline
  Critically Detuned, Good Cavity			&  $ \frac{4}{\Nup C}$ 								& $ \frac{g}{\sqrt{2}}  \sqrt{\frac{\kappa}{\Gamma}}$  				& $2$						& $ \dc = \dc^\circ \gg  \rabi ; \Gamma\gg \kappa$  											\\ \hline
   Far Detuned  					& $ \frac{1}{\Nup C}$ 								& $\sqrt{\frac{\Nup}{2}} \frac{g^2}{ \left | \dc \right |}$ 			& $1 + \frac{\Nup g^2}{\dc^2}\frac{\Gamma}{\kappa}$						& $ \dc \gg \dc^\circ,  \rabi ; \Gamma\gg \kappa$  											\\ \hline
    Maximally Detuned, Good Cavity, 	& $ \left(\frac{\Gamma}{2 \delta_\mathrm{max}}\right)^2$ 			&$ \frac{g}{\sqrt{2}} \sqrt{\frac{\kappa}{\Gamma}}$	&  $2$				& $ \dc =\delta_\mathrm{max}\gg  \rabi ; \Gamma\gg \kappa;$  \\ 

    Optimized	&	&	&	& $\Nup C = \left(\frac{2 \delta_\mathrm{max}}{\Gamma}\right)^2$  \\ \hline

 \end{tabular}
  \label{Table:Scalings}
\end{table*}

%******************************************************************************************************
%*****************Minimizing $\msp$ at Fixed Maximum Detuning***************************
%******************************************************************************************************
\subsection{Minimizing $\msp$ at fixed maximum detuning}\label{sect3c}

In some experimental situations, a maximum probe detuning $|\dc|\leq \delta_\mathrm{max}$ is set by the energy structure of the atom.  For instance, the ground state hyperfine splitting in $\Rb$ imposes $ \delta_\mathrm{max} \approx 6.8/2$~GHz~\cite{Vuletic10}.   An optimum value of $\Nup C$ can be found that minimizes $\msp$ when $|\dc|=\delta_\mathrm{max}$.  The scaling for this case is shown in the last line of  Table \ref{Table:Scalings}.  Physically, the optimum value of $\Nup C$ is reached (at a fixed detuning) when the dressed cavity linewidth is related to the bare cavity and atomic linewidths by $\kappa'/\kappa = 2 \Gamma/(\Gamma+ \kappa)$.  In the resonant limit, $\dc =0$, one finds an optimum $\kappa' /\kappa= 1$, while in the detuned limit one finds $\kappa' \approx 2 \kappa$, i.e., the detuned cavity resonance is broadened by a factor of 2 at optimum.  In this same limit, the ratio of rms fluctuation size to dressed cavity HWHM is given by $2\dwproj / \kappa'/2 = \sqrt{C/8}$.  Larger single-atom cooperativity $C$ reduces the technical requirements on resolving the projection noise fluctuations of the cavity mode.

%******************************************************************************************************
%*******************************Fundamental Scalings related to Projection Noise ***********************************************
%******************************************************************************************************
\subsection{Probing dressed modes in the resonant cavity limit, $\dc =0$}\label{sect4}

Here we consider the special case of probing in the resonant cavity limit, $\dc=0$, utilized in the experiment of Ref.~\cite{CBS11}.   On resonance $\dc=0$, the absolute size of the projection noise fluctuations is maximized, i.e., 
\begin{eqnarray}
\Delta \opm^\mathrm{proj}= \frac{g}{2\sqrt{2}} \, .
\end{eqnarray}

\noindent  Note that the rms fluctuation is independent of $N$. The same is true for the  FWHM linewidth which is simply equal to the average linewidth~\cite{ZGM90} due to the equal photonic and atomic contributions to the normal modes:
\begin{eqnarray}
\kappa' = (\kappa+\Gamma)/2 \, .
\end{eqnarray}

\noindent To be able to resolve projection noise, one must detect, on average, a number of probe photons in transmission given by
 \begin{eqnarray}
M_{d}^\mathrm{proj} = \frac{1}{2 \detmethod} \left( \frac{\kappa' }{g}\right)^2 \, .
\label{eqn:ResMd}
\end{eqnarray}

\noindent The ratio of free-space-scattered photons to detected probe photons in transmission is
 \begin{eqnarray}
R_{s} = \frac{1}{q_d \eta_s}\frac{\Gamma}{\kappa} \, .
\label{eqn:ResRs}
\end{eqnarray}

\noindent Finally, the number of scattered photons into free space for measurement uncertainty at the projection noise level normalized to the total number of atoms $N$ is
\begin{eqnarray}
m_{s}^\mathrm{proj} = \frac{1}{ 2 q N C}\left(1+\frac{\Gamma}{\kappa} \right)^2 \, .
\label{eqn:ResMs}
\end{eqnarray}

\noindent  If the cavity length and mode volume are fixed by experimental constraints, then one is, in principle, free to minimize Eq.~(\ref{eqn:ResMs}) by varying the finesse of the cavity mirrors, until a minimum value of $\msp = 2/q N C$ is reached when $\kappa = \Gamma$. The minimization with respect to cavity finesse accounts for the fact that the cooperativity $C$ scales as $1/(T_1+T_2+L)$.

%******************************************************************************************************
%****************************Quantum Back-Action on determining $J_z$************************
%******************************************************************************************************
\section{Quantum Back-Action Limits on determining $J_z$}\label{sect5}

In this section, we study the limitations on measurement resolution on the spin projection  $J_z$ arising from Raman spin flips caused by free space scattering.  We will begin by considering a simple three level model that will be used in Sec.~\ref{sect6} to calculate spectroscopic enhancements relative to the SQL. The simple model will be extended in Sec.~\ref{sect7} to describe probing of the clock and cycling transitions in $\Rb$.

%******************************************************************************************************
%********************Average Impact on $J_z$ due to Free Space Scattering**********************
%******************************************************************************************************
\subsection{Definitions}
This section defines symbols that are relevant to the later discussions on measurement resolution and spectroscopic enhancement.  We define collective spin operators $\hat{J}_{x, y, z} = \sum^N_{i=1} \hat{J}^i_{x, y, z}$, where $\hat{J}^i_{x, y, z}$ is the single-atom spin operator for the $i$th atom such that $\hat{J}^i_z\ket{\!\uparrow_i} = \frac{1}{2} \ket{\!\uparrow_i}$, $\hat{J}^i_z\ket{\!\downarrow_i} = -\frac{1}{2} \ket{\!\downarrow_i}$, etc. The collective spin operator is $\hat{J}_z = (\hat{\Nup} - \hat{\Ndown})/2$ where $\hat{\Nup}$ and $\hat{\Ndown}$ are the atomic population operators defined in Sec.~\ref{sect2a}.   Expectation values of the collective spin operators are denoted by $J_{x, y, z} \equiv \avg{\hat{J}_{x, y, z}}$. The collective Bloch vector is $\J \equiv (\hat{J}_x, \, \hat{J}_y, \, \hat{J}_z )$. The radius of the collective Bloch sphere is $\R \equiv \sqrt{\avg{\J^2}}$.

%******************************************************************************************************
%****************************Simple Three-level Model************************
%******************************************************************************************************
\subsection{Simple model for $J_z$ diffusion}\label{sect5a1}

In this subsection, we consider how the free space scattering changes atomic population in the spin up and spin down two-level manifold through Raman or spin flip events.   For arriving at the results presented in this section, only the atomic populations matter, and coherences are irrelevant.

We consider here the simplest model that captures the essential physics. In this toy model, the only states in the problem are the two-level system $\ketup$, $\ketdown$ and the optically excited state $\ket{e}$, as described in Fig.~\ref{BasicEnergyLevelsCavity}(b). We assume that a free space scattering event causes an atom to spin flip from $\ketup$ to $\ketdown$ via the intermediate state $\ket{e}$ with probability $p$.  This simple model may be straightforwardly extended to provide accurate predictions for a multi-level atom by accounting for all possible Raman scattering processes.

%******************************************************************************************************
%*****************************************Diffusion of $J_z$ *****************************************
%******************************************************************************************************
%\subsubsection{Diffusion of $J_z$}\label{sect5b1}

Free space scattering causes $J_z$ to change on average by a certain amount, while the random nature of the spin flip process leads to a random walk or diffusion of the collective Bloch vector's spin projection $J_z$. Provided multiple scattering can be neglected, i.e. $p \ms \ll 1$, the diffusion process can be described by the relation
\begin{eqnarray}
\frac{\avg{\left( J_z(\ms) - J_z(0) \right)^2}}{(\qpn)^2} = 4\, p\, \ms \, ,
\label{eqn:RamanVar}
\end{eqnarray}

\noindent with the spin flip probability setting the diffusion constant $4p$, and the random variable $J_z(\ms)$ describing the $z$-component of the Bloch vector after $\ms$ scattering events per atom.  The diffusion is normalized to the projection noise level for a CSS $(\qpn)^2 = N/4$.

%******************************************************************************************************
%*********************Measurement Imprecision due to Photon Shot Noise************************
%******************************************************************************************************
\subsection{Measurement imprecision due to photon shot noise}\label{sect5b2}

The measurement imprecision $\Delta J_z^\mathrm{meas}$ is due to probe vacuum noise alone is
\begin{eqnarray}
\left( \frac{\Delta J_z^\mathrm{meas}}{\qpn} \right)^2 =  \frac{\msp}{\ms} \, .
\label{eqn:MeasVar}
\end{eqnarray}

%******************************************************************************************************
%***************Balancing Measurement Imprecision against Diffusion of $J_z$*******************
%******************************************************************************************************
\subsection{Balancing measurement imprecision against $J_z$ diffusion}\label{sect5b3}

\begin{figure}
\includegraphics[width=3.4in]{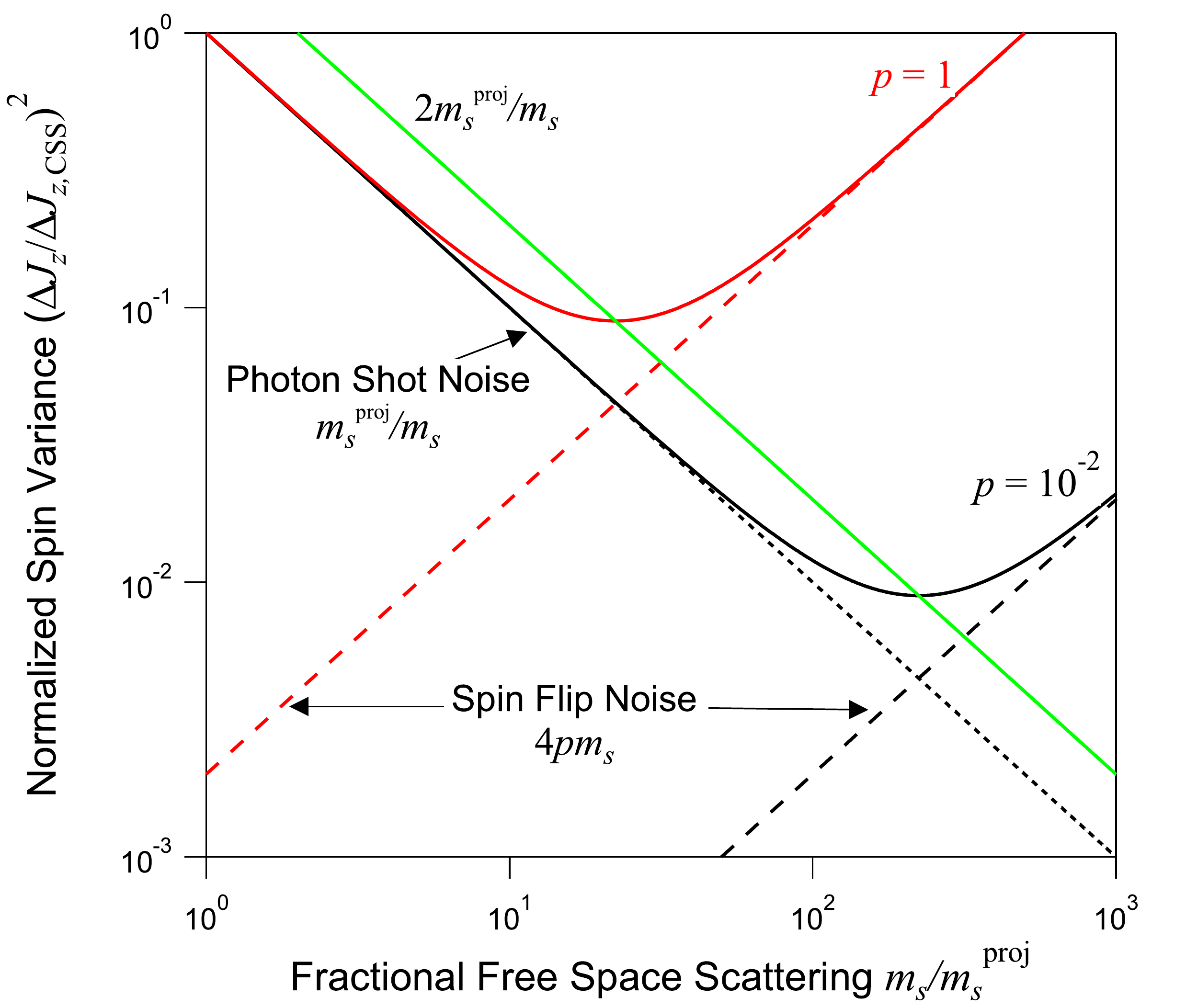}
\caption{(Color online) Normalized spin noise variance versus fractional free space scattering $\ms/\msp$. The curves assume no prior knowledge, i.e., $\prepnoise \to \infty$, and $NC = 10^3$. The measurement variance decreases as $1/\ms$ from averaging down photon shot noise (dotted black line) until noise due to random spin flips (dashed red and black lines) takes over at large $\ms$. The minimum noise variance is $4\sqrt{p \msp} \to \sqrt{8p/NC}$ assuming perfect quantum efficiency and probing in the far-detuned limit. This minimum is reached when two noise contributions are equal at $\msopt = \sqrt{\msp/4p} \to 1/ \sqrt{8 p N C}$ assuming perfect quantum efficiency and probing in the far-detuned limit. A larger collective cooperativity $NC$ or lower spin flip probability $p$ allows photon shot noise to be averaged down more before spin flip noise takes over, as illustrated by the two curves for $p/NC = 10^{-5}$ (black) and $10^{-3}$ (red). The locus of the minimum variance is $2\ms/\msp$ (green). }
\label{SpinNoise_vs_ms}
\end{figure}

Given the diffusion in $J_z$, and the photon shot noise in the probe, we must ask: How well can one determine the value of $J_z$ or equivalently the atomic population $\Nup$ prior to the measurement (which disturbs $J_z$ through spin flips)? As a first pass, the total variance $(\Delta J'_z)^2$ in the estimate of $J_z$ is found by adding the measurement imprecision [Eq.~(\ref{eqn:MeasVar})] and spin-flip-induced diffusion of $J_z$ [Eq.~(\ref{eqn:RamanVar})]:
\begin{eqnarray}
\left( \frac{\Delta J'_z}{\qpn} \right)^2 = \frac{\msp}{\ms}  + 4\, p\, m_s \, .
\label{eqn:NormTotalVar}
\end{eqnarray}

\noindent  Averaging down photon shot noise determines the $z$-projection of the Bloch vector more and more precisely. Eventually, however, scattering-induced diffusion of $J_z$  causes the value of $J_z$ measured at earlier times to become less correlated with the value of $J_z$ measured at later times, adding noise to the estimate of $J_z$, as shown in Fig.~\ref{SpinNoise_vs_ms}. The optimal resolution $\Delta J_z^\mathrm{opt}$ occurs at an optimal scattering $\msopt$ where the noise contributions from measurement imprecision and diffusion due to Raman spin flips are equal:
\begin{align}
\msopt = \frac{1}{\sqrt{8 p q N C h(\dc)} }  \, , \\
\left( \frac{\Delta J^\mathrm{opt}_z}{\qpn} \right)^2 =  \sqrt{ \frac{8p}{q  N C h(\dc)}}  \, , \label{eqn:noise_opt}
\end{align}

\noindent The detuning dependence has been lumped into the factor $h(\dc)$
\begin{eqnarray}
h(\dc) = \left( \frac{\kappa}{\kappa'(\dc)} \frac{\omega(\dc)}{\sqrt{ \dc^2 + \rabi^2 }} \right)^2 \, ,
\end{eqnarray}

\noindent  where $\kappa'(\dc)$ is the dressed cavity mode linewidth introduced in Eq.~(\ref{eqn:Dressedkappa}), $\omega(\dc)$ is the dressed cavity mode frequency introduced in Eq.~(\ref{eqn:modefreq}), and $\rabi = \sqrt{\Nup} 2g$ is the collective vacuum Rabi splitting. The parameter $h(\dc^\circ) = 1/4$ at the critical detuning, and $h(\dc) \rightarrow 1$ as $|\dc|\rightarrow \infty$.  In the far-detuned limit, Raman spin flip diffusion limits the achievable resolution to $\left( \frac{\Delta J_z^\mathrm{opt}}{\qpn} \right)^2 = \sqrt{8p/qNC}$.

%******************************************************************************************************
%***************************Single Spin Measurement Resolution**********************************
%******************************************************************************************************
\subsection{Single spin measurement resolution}\label{sect5c}

Single spin resolution $\Delta J_z^\mathrm{opt} \le 1/2$  is required for  conditionally preparing states with spectroscopic sensitivity at the Heisenberg limit as well as the parity measurements needed for reading out NOON states, or Dicke states~\cite{Bollinger96}.  Single spin resolution in ensembles of $\sim100~\Rb$ atoms has recently been demonstrated using cavity-aided nondemolition measurements~\cite{Vuletic12}.  We find that in the far-detuned limit, single spin resolution is reached for $p \le  \frac{ q C}{8 N}$, quantifying how ideal a cycling probe transition needs to be in order to resolve single spins. For $ q C \sim 1$ and $N \sim 10^6$, one would need $p\le10^{-7}$, which is highly unrealistic for real multi-level alkali atoms due to off-resonant scattering from other hyperfine states, as will be discussed in Sec.~\ref{sect7} for the case of $\Rb$.   Alternatively, with high single-atom cooperativity, $q C \ge 8 N p$, single spin resolution could be attained without a cycling transition. For example, with $q C \sim 100$ and a worst-case open transition with $p=1/2$, single spin resolution would be reached for  $N\le25$ atoms.

%******************************************************************************************************
%*************************************Spectroscopic Enhancement**********************************
%******************************************************************************************************
\section{Spectroscopic Enhancement}\label{sect6}

Spectroscopic sensitivity refers to the ability to resolve the angle
through which a Bloch vector or a Dicke state is rotated. To first approximation, the polar angular resolution is set by the conditional spin noise, discussed in Sec.~\ref{sect5}, and the radius $\R$ of the collective Bloch sphere on which the Bloch vector or Dicke state lives. In this section, we discuss how the radius $\R$ of the collective Bloch sphere is reduced by the measurement due to free space scattering, and derive the fundamental limits to the spectroscopic sensitivity. The radius $\R$ is proportional to the Ramsey contrast $\cal C$ for a CSS or a
slightly spin-squeezed state. The radius is used here because it is possible to consider a conditional measurement with imprecision below a single spin. If all atoms remain in a superposition, the resulting state would be a fully symmetric Dicke state or eigenstate of the operator $\hat{J}_z$, with $\R=N/2$, but with $\avg{\J}, \, {\cal C} =0$. Nonetheless, Dicke states have near-Heisenberg limited spectroscopic sensitivity~\cite{Holland93}.

Enhanced sensitivity in one degree of angular resolution, say the polar angle $\theta$, can be gained at the expense of enhanced uncertainty in an orthogonal degree of freedom, namely the azimuthal angle $\phi$.  For concreteness,  the Bloch vector is initially prepared in a CSS along $\hat{x}$ with $\avg{\J} = \hat{x}N/2$.   The angular resolution of the polar angle for the CSS defines the SQL $\Delta \theta_\mathrm{SQL} = \Delta J_{z,\mathrm{CSS}}  / \R = 1/\sqrt N$.  If the actual angular resolution is $\Delta \theta$, then the metrologically relevant squeezing parameter is $\sq\equiv \left( \Delta \theta_\mathrm{SQL}/ \Delta \theta \right)^2$ with $\sq\ge1 $ representing a spectroscopic enhancement in sensitivity that must arise from entanglement.

Angular resolution is reduced if the radius $\R$ of the collective Bloch sphere is reduced below its initial value without a corresponding decrease in the spin-noise.  In the simplest model, each free-space-scattered photon from an atom in a superposition state leads to the collapse of its spin into spin up or down, leading to an average reduction in $\R$ by 1/2.  If the free space scattering rate for each spin is unchanged by the scattering process, then the  collective Bloch sphere radius normalized to its initial value $\widetilde{\R}$ as a function of the number of scattered photons per atom $m_s$ is given by
\begin{eqnarray}
\widetilde{\R} = e^{-m_s} \, .
\label{eqn:LengthDecay}
\end{eqnarray}

\noindent Note that both Rayleigh and Raman scattering lead to a reduction in the collective Bloch sphere radius.   In certain cases, free space Rayleigh scattering does not create wavefunction collapse~\cite{Uys2010}, but this requires indistinguishability in the scattering process which reduces the information that can be extracted from the probe mode.

Putting Eq.~\ref{eqn:NormTotalVar} and Eq.~\ref{eqn:LengthDecay} together, the spectroscopic enhancement is given by
\begin{equation}
\sq = \left( \frac{\Delta J'_z}{\qpn} \right)^{-2} e^{-2\ms} \, .
\end{equation}

\noindent   If Raman spin-flips dominate over Rayleigh scattering events, the optimal spectroscopic enhancement $\sq^\mathrm{opt} \sim \sqrt{q N C/p}$ is limited by spin-flip diffusion noise as considered in sub-section~\ref{sect6a}. If Rayleigh scattering dominates, then the optimal spectroscopic enhancement $\sq^\mathrm{opt} \sim q N C$ is limited by shrinkage of the collective Bloch sphere radius $\R$, discussed in Sec.~\ref{sect6b}. The change in the scaling of $\sqopt$ from $\sqrt{NC}$ in the Raman spin-flip limit to $NC$ in the cycling transition limit allows far greater amounts of squeezing on a cycling transition. However, the loss of quantum efficiency $q$ degrades squeezing as $\sqopt \sim q$ on a cycling transition, compared to the more favorable scaling $\sqopt \sim \sqrt q$ in the Raman spin-flip limited regime.

%******************************************************************************************************
%**************************Small Decoherence/Spin Flip Limit*************************
%******************************************************************************************************
\subsection{Small decoherence/spin flip limit}\label{sect6a}

Here we consider the case where the reduction in the radius $\R$ of the collective Bloch sphere may be ignored.  This is justified if the optimal scattering $\msopt$ that optimizes the measurement resolution of $J_z$ is small, $\msopt \ll 1$, equivalently $p q N C \left(\kappa / \Gamma \right)^2 \gg 1$ for probing in the far-detuned limit. In this regime, the radius $\widetilde{\R}$ remains approximately 1, so that the spectroscopic enhancement is primarily set by the reduction in the spin noise, discussed previously in Sec.~\ref{sect5}
\begin{align}
\sq^\mathrm{opt} \approx \left( \frac{\Delta J_z^\mathrm{opt}}{\qpn} \right)^{-2} \, , 
\label{eqn:squeeze_opt}
\end{align}

\noindent where $\left( \frac{\Delta J_z^\mathrm{opt}}{\qpn} \right)^2$ has been introduced in Eq.~\ref{eqn:noise_opt}.  In the far-detuned limit, Raman spin flip noise limits the achievable squeezing to $\sq^\mathrm{opt} = \sqrt{q NC/8p}$. 
 
\begin{figure}
\includegraphics[width=3.4in]{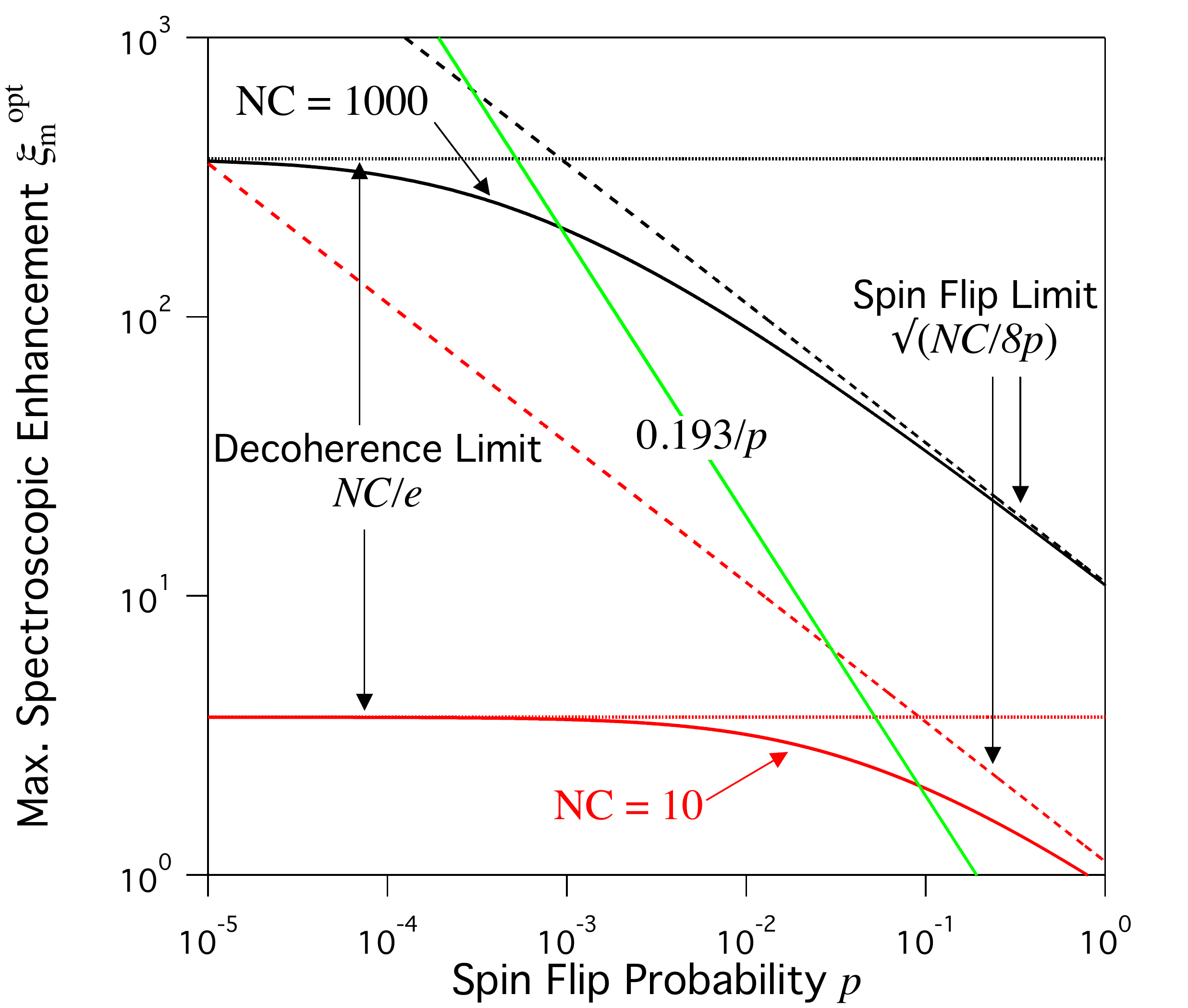}
\caption{(Color online) Optimal squeezing $\sq^\mathrm{opt}$ versus spin flip probability $p$ for $NC = 1000$ (black curve) and $NC = 10$ (red curve) for probing in the far-detuned limit. Both curves take into account spin flip noise and decoherence. The curves assume perfect quantum efficiency $q=1$ and no prior knowledge, i.e., $\prepnoise \to \infty$. The decoherence-limited squeezing $\sq^\mathrm{opt} = NC/e$ (dotted lines) is approached for $p \ll e^2/8 NC$, and the spin-flip-limited squeezing $\sq^\mathrm{opt} = \sqrt{NC/8p}$ (dashed lines) is approached for $p \gg e^2/8 NC$. The locus of optimal squeezing at the crossover point $p = e^2/8 NC$ is $0.193/p$ (green line).}
\label{MaxSqueeze_vs_SpinFlipProbability}
\end{figure}

%******************************************************************************************************
%********************Large Decoherence/Cycling Transition Limit*********************
%******************************************************************************************************
\subsection{Large decoherence/cycling transition limit}\label{sect6b}

Here we consider the case where the reduction in the radius $\R$ of the collective Bloch sphere plays an important role in determining the angular resolution. If the probing is performed on a nominally closed transition where Raman scattering spin-flips due to probe polarization imperfections and off-resonant scattering are very improbable, spin-flip diffusion noise is negligible and the only limit to spectroscopic enhancement is the shrinking of the radius $\R$ due to free space (Rayleigh) scattering.  The probing is performed in the large decoherence or cycling transition regime when the optimal scattering $\msopt$ that optimizes the measurement resolution of $J_z$ is not small. Formally,  this regime occurs when the optimum spectroscopic enhancement calculated from Eq.~(\ref{eqn:squeeze_opt}) is $\sq^\mathrm{opt} \geq 0.193/p$.

In the large decoherence regime, the spectroscopic enhancement is  given by $\sq = \frac{\ms}{\msp} e^{-2\ms}$, ignoring the small improvement that may result from prior knowledge.  An optimum is reached when the radius $\widetilde{\R} = e^{-1/2}$  or equivalently $\msopt = 1/2$, yielding an optimum $\sq^\mathrm{opt} = 1/(2e \, \msp)$.   In the far-detuned limit, the optimal squeezing $\sq^\mathrm{opt} \to q N C/e$.  We caution that the simple model presented here is not valid near the Heisenberg limit because while measurement resolution below a single spin may be achieved, the effective spin noise variance must be clamped to 1/4 in order to enforce the Heisenberg limit on spectroscopic sensitivity.

%******************************************************************************************************
%**************************Optimal Squeezing for  $\Rb$**********************
%******************************************************************************************************
\section{Optimal Squeezing for $\Rb$}\label{sect7}

Using the framework above, we now analyze the limits of two separate probing schemes. From Sec.~\ref{sect3}, we have shown that the measurement resolution at fixed free space scattering improves with cavity detuning $\dc$, but ultimately saturates to a value set by the collective cooperativity parameter $N C$. For squeezing on a clock transition comprised of two hyperfine ground states, the maximal detuning is approximately half of the ground state hyperfine splitting $\dc = \ohf/2$. We are interested in the ultimate limits of squeezing in such a system, taking into account Raman spin flips and decoherence. Motivated by the fact that Raman spin flips limit the achievable squeezing on a hyperfine clock transition, we then analyze squeezing via probing on a cycling transition, where Raman spin flips are greatly reduced, but introduce an additional scaling with $\dc$, resulting in a region of saturation of the spectroscopic enhancement as $N C$ is increased.  In both scenarios, the ratio $\ohf /\Gamma$ of the hyperfine splitting to the excited state decay linewidth plays a critical role.

%******************************************************************************************************
%**************************Optimal Squeezing via Differential Measurement of $\Rb$ Clock Transition**********************
%******************************************************************************************************
\subsection{Optimal squeezing via differential measurement of $\Rb$ clock transition}\label{sect7a}

We now consider the measurement scheme demonstrated by the MIT group~\cite{Vuletic10} in which the pseudospin states were as in Ref.~\cite{CBS11}, namely the clock states of $\Rb$ $\ketup \equiv \ket{5 ^{2}S_{1/2}, F=2, m_F=0}$ and $\ketdown \equiv \ket{5 ^{2}S_{1/2}, F=1, m_F=0}$.   The bare cavity frequency is tuned to the average of the two ground states to optically excited state transitions near 780~nm.  This tuning ensures that an atom in $\ketdown$ shifts the dressed cavity resonance frequency by an equal, but opposite amount as an atom in $\ketup$.  The excited state hyperfine splitting  $\sim 500$~MHz  is much less than the ground hyperfine splitting and is taken to be zero for the following analysis.

The problem is analyzed by extending the linearized two-mode model of Eq.~(\ref{eqn:Langevin}) to a linearized three-mode model in which the atomic operator $\hat{a}$ is generalized to operators $\hat{a}_\downarrow$ and $\hat{a}_\uparrow$ to yield three coupled differential equations, along with the same input-output relations given by Eq.~\ref{InputOutputRelations}:
\begin{eqnarray}
\frac{d\left<\hat{c}\right>}{d t} &=& -  \frac{1}{2} \kappa \left<\hat{c}\right> -\imath g \left(\sqrt{\Nup}\left<\hat{a}_\uparrow\right> - \sqrt{\Ndown}\left<\hat{a}_\downarrow \right> \right )+ \sqrt{\kappa_1} c_{i} \nonumber \, , \\
\frac{d \left<\hat{a}_\uparrow\right>}{d t}  &=&   -\frac{1}{2}\left( \Gamma +\imath\, \ohf\right)\left<\hat{a}_\uparrow\right> -  \imath \sqrt{\Nup}g \left<\hat{c}\right> \nonumber \, , \\
\frac{d \left<\hat{a}_\downarrow\right>}{d t}  &=&   -\frac{1}{2}\left( \Gamma -\imath\, \ohf  \right)\left<\hat{a}_\downarrow\right> -  \imath \sqrt{\Ndown}g \left<\hat{c}\right> \, .
\label{eqn:Langevin3mode}
\end{eqnarray}

\noindent The equations are now written in a rotating frame at the bare cavity resonance frequency that is chosen such that the two optical atomic transitions are detuned by $\pm \ohf/2$.  The rate of scattering into free space is described by the two field amplitudes $a_{s, \uparrow,\downarrow} = \sqrt{\Gamma} \left<\hat{a}_{\uparrow, \downarrow}\right>$ and normalized such that the rate of photons scattered into free space is simply $\dot{M}_s=\left | a_{s, \uparrow} \right |^2+ \left | a_{s,\downarrow} \right |^2$.   

From the coupled set of Eq.~(\ref{eqn:Langevin3mode}), we find that the rms phase shift of the transmitted light field caused by the rms projection noise level fluctuation in the population difference is:
\begin{equation}
\Delta\phi^\mathrm{proj} =  \sqrt{N}C \left( \frac{\Gamma}{\ohf}\right)\left(\frac{1}{1+ N C \Gamma^2/\ohf^2}\right) \, .
\end{equation}

\noindent This expression assumes the damping rates are small $\kappa, \Gamma \ll \ohf, 2g \sqrt{N/2}$.  The phase shift initially climbs with increasing atom number as $\sqrt{N}$, but saturates to a maximum value $\Delta\phi^\mathrm{proj}= \sqrt{C}/2$ at a critical atom number given by $N C = (\ohf/\Gamma)^2$, after which the phase shift decreases as $1/\sqrt{N}$.  The physical interpretation for this decrease is that above the critical atom number, the dressed cavity mode linewidth $\kappa'$ rapidly starts to broaden with increasing atom number.  The number of free-space-scattered photons required to resolve the projection noise level phase shift of the probe is
\begin{equation}
\msp = \frac{1}{4 q N C }\left(1 +  N C \left(\frac{\Gamma}{\ohf}\right)^2\right) \, .
\end{equation}

The diffusion of the difference between the estimate of $J_z$  and the actual value of $J_z$ is driven by Raman transitions that move atoms from $\ketup$ to $\ket{F=1}$ or $\ketdown$ to $\ket{F=2}$.   Raman transitions between states of the same $F$ (i.e., $\Delta F =0$) lead to loss of coherence, but do not change the coupling of the atom to the cavity mode in the limit where the excited state hyperfine splitting is neglected, as we do here. Hyperfine changing transitions $\Delta F\neq 0$ cause the detuning to change sign, but not magnitude, making such a process equivalent to a spin flip.  Accounting for transition branching ratios, we find that to a good approximation, we can apply Eq.~(\ref{eqn:NormTotalVar}), with an effective spin flip probability $p = 1/6$.  Assuming the loss of coherence is small, then the optimal spectroscopic enhancement with respect to average probe photon number is
\begin{equation}
\sq^\mathrm{opt} = \frac{ \sqrt{6 q N C }}{1+  4 N C \Gamma^2 /\ohf^2} \, .
\end{equation}

\noindent At small $N$, the spectroscopic enhancement scales as $\sqrt{6 q N C}$, reaching a peak value of $\sq^\mathrm{opt} = \sqrt{3q/8}\, \ohf\,/\Gamma$ at a value $N C = \frac{1}{4}(\ohf/\Gamma)^2$, slightly before the maximum phase shift is reached.  At larger $N C$, the spectroscopic enhancement scales as  $\sqopt = \sqrt{3 q /8 N C}(\ohf/\Gamma)^2$.   

Taking the quantum efficiency to be $q=1$, the maximum spectroscopic enhancement for $\Rb$ is quite large at 28~dB.  The exact details of the full measurement sequence  (i.e., whether rotations such as $\pi$-pulses are used to cancel sources of technical noise as was done in Refs.~\cite{Vuletic10, CBS11}) are needed to construct an optimal estimator of $J_z$, but, at best, a 3~dB further improvement may result.

\begin{figure}
\includegraphics[width=3.4in]{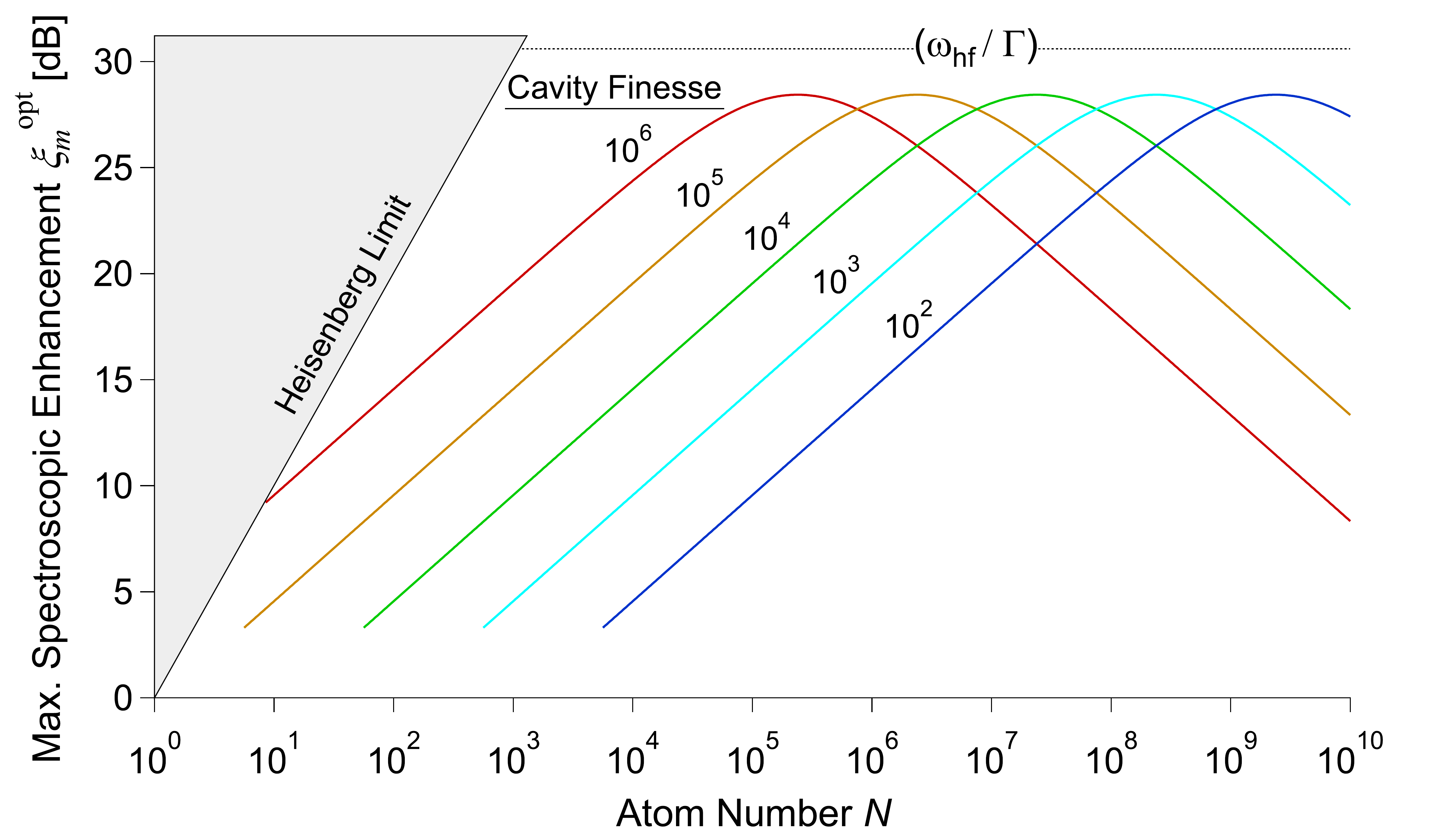}
\caption{(Color online) The fundamental optimum spectroscopic enhancement for differential probing of the $\Rb$ clock transition, as performed in Ref.~\cite{Vuletic10}.  The calculations assume a net quantum efficiency of $q=1$, a cavity mode waist $w_0 = 71~\mu$m and cavity length $l = 1.91$~cm used for our experiments~\cite{CBS11}.  The purple, light blue, green, orange, and red curves correspond to cavity finesses of $F=10^2,10^3,10^4,10^5$, and $10^6$, respectively. All these finesses are experimentally feasible. The atom number $N$ at which the spectroscopic enhancement is maximized scales with the cavity mode waist $w_0$ and cavity finesse $F$ as $w_0^2/F$.}
\label{dB_clock}
\end{figure}

Because $C$ does not depend on the cavity length, the optimum $N$  for peak spectroscopic enhancement scales as $(w_0^2 /F )(\ohf/\Gamma)^2$, where $w_0$ is the cavity mode waist, and $F$ is the cavity finesse. More fundamentally, no change in the cavity geometry ($w_0$ and $l$) or finesse $F$ changes the maximum obtainable enhancement in spectroscopic sensitivity. This enhancement is determined solely by the atomic properties. Figure~\ref{dB_clock} shows the spectroscopic enhancement versus atom number for a range of technologically feasible cavity finesses.

\begin{figure}
\includegraphics[width=3.4in]{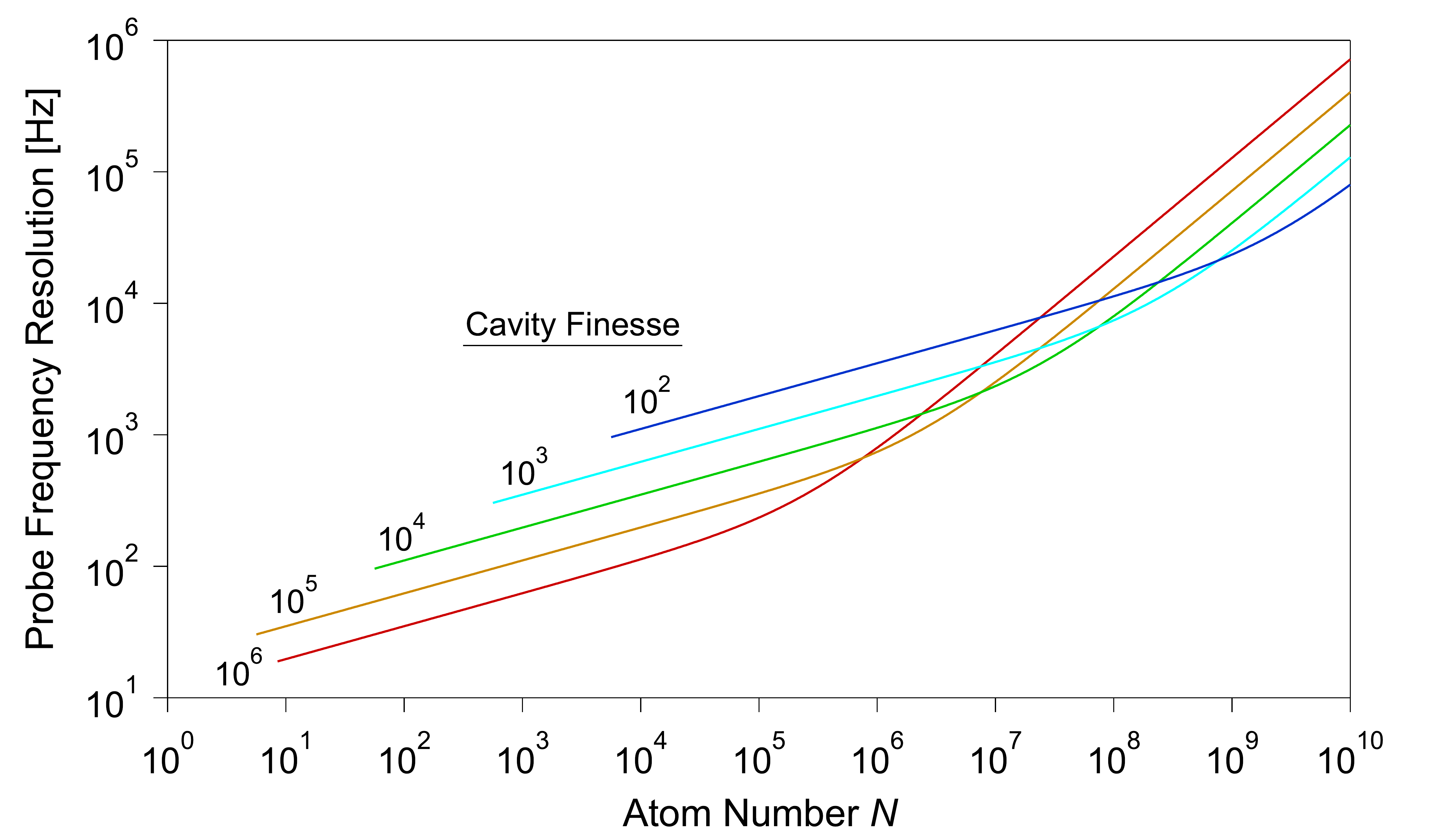}
\caption{(Color online) Frequency resolution with which the relative frequency of the probe and dressed cavity mode must be measured to obtain the spectroscopic enhancements shown in (and under the same conditions as) Fig.~\ref{dB_clock} .  The purple, light blue, green, orange, and red curves correspond to cavity finesses of $F=10^2,10^3,10^4,10^5$, and $10^6$, respectively. }
\label{Hz_clock}
\end{figure}

\begin{figure}
\includegraphics[width=3.4in]{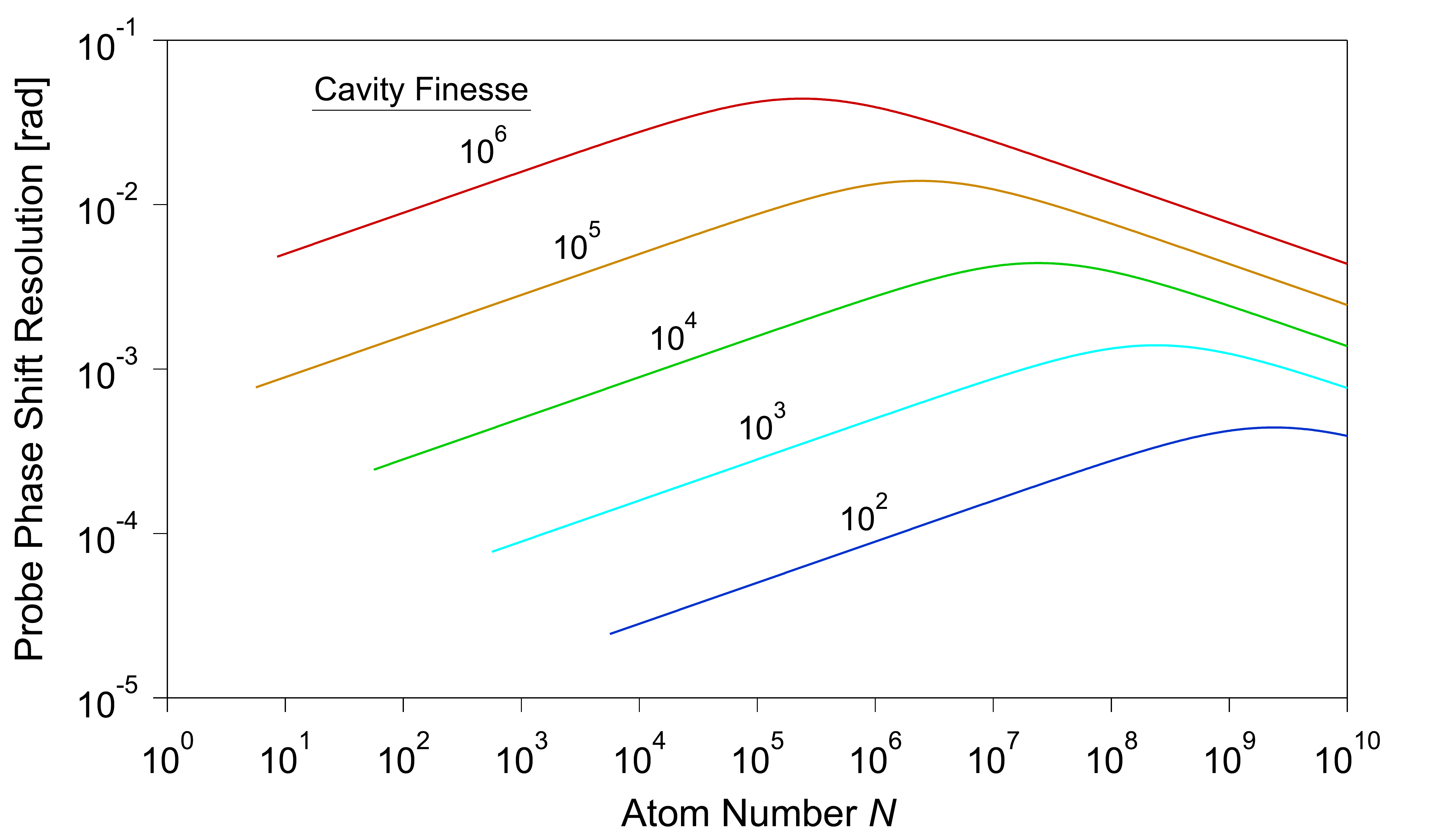}
\caption{(Color online) Phase resolution with which the transmitted probe light must be measured to obtain the spectroscopic enhancements shown in (and under the same conditions as)  Fig.~\ref{dB_clock}.  The purple, light blue, green, orange, and red curves correspond to cavity finesses of $F=10^2,10^3,10^4,10^5$, and $10^6$, respectively. }
\label{rad_clock}
\end{figure}

Resolving very small phase deviations or small frequency shifts imposes technical challenges that are modified with cavity geometry or finesse, as shown by the probe frequency (Fig.~\ref{Hz_clock}) and probe phase shift (Fig.~\ref{rad_clock})  resolutions required to obtain the spectroscopic sensitivities shown in Fig.~\ref{dB_clock}.  All three figures assume the cavity geometry of Ref.~\cite{CBS11}, $l =1.91$~cm and $w_0 = 71~\mu$m.   Finally, we note that Fig.~\ref{Hz_clock} shows the technical requirement on probe frequency resolution is more relaxed above the optimum $N$ compared to achieving the same spectroscopic enhancement at a value below the optimum $N$.

%******************************************************************************************************
%**************************Optimal Squeezing on $\Rb$ Cycling Transition*******************
%******************************************************************************************************
\subsection{Optimal squeezing via probing on $\Rb$ cycling transition}\label{7b}

Having seen that the optimal squeezing on a clock transition is fundamentally limited by Raman spin flips, we consider a situation in which Raman spin flips are reduced, namely probing on a cycling transition, and show that larger amounts of squeezing are possible in this configuration than on the clock transition~\cite{Saffman09, Vuletic12}. 

\begin{figure}
\includegraphics[width=3.4in]{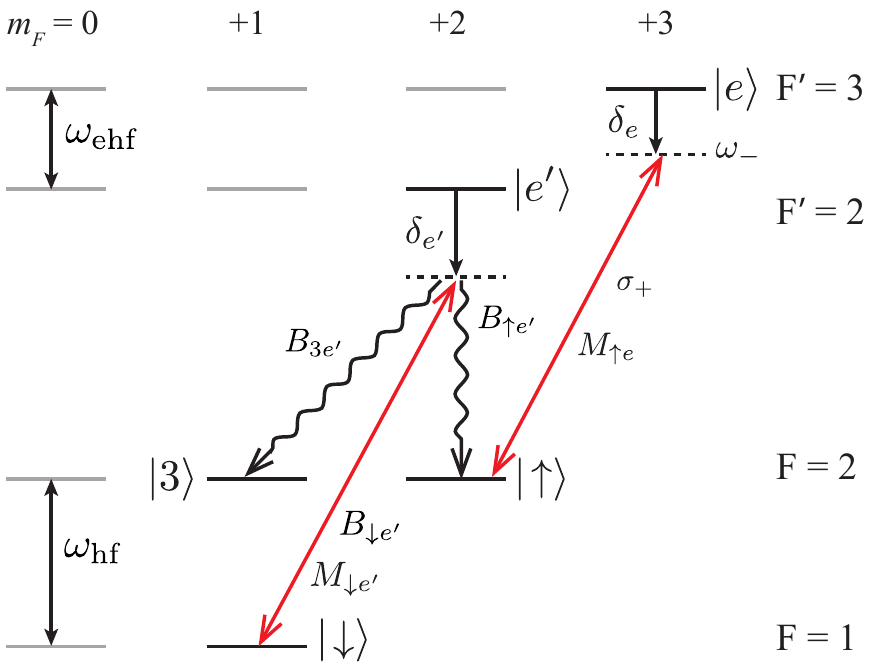}
\caption{(Color online) Probing scheme for the cycling transition in $\Rb$.  The ground hyperfine states $F=2$ and $F=1$ are split by $\ohf = 2 \pi \times (6834~\mathrm{MHz})$.  The values of $m_F$ are labeled across the top.  The relevant $F'=3$ and $F'=2$ excited D2 transitions at wavelength 780~nm are also shown with the excited state splitting $\oehf = 2 \pi \times (267~\mathrm{MHz})$. The pseudo spin-1/2 system is here composed of $\ketup = \ket{F=2, m_F = 2}$  and $\ketdown = \ket{F=1, m_F =1 }$. Ideally, the $\sigma^+$-polarized probing laser couples only $\ketup$ to the optically excited state $\ket{e}$ with dipole matrix element $M_{\uparrow e}$ at a frequency detuning $\delta_e$ that is approximately equal to the dressed-cavity mode frequency $\om$.  In this illustration, $\delta_e$ is negative. By dipole selection rules, $\ket{e}$ can only decay  back to $\ketup$.  However, the same probe laser also couples $\ketdown$ to the single excited state $\ket{e'}$ with dipole matrix element $M_{\downarrow e'}$ and larger detuning from resonance $\delta_{e'}=\delta_e- \ohf +\oehf$.  The ratio of matrix elements is $M_{\downarrow e'}/M_{\uparrow e}=1/\sqrt{2}$.  Finally, the state $\ket{e'}$ decays to states $\ketdown$, $\ketup$, and $\ket{3}$ with fractional branching ratios $B_{\downarrow e'} = 1/2$, $B_{\uparrow e'} = 1/3$, and $B_{3 e'} = 1/6$ respectively.}
\label{Rb87Cycling_EnergyLevels}
\end{figure}

As a concrete example of how a cycling transition can be used to enhance probing, we consider the cycling transition in $\Rb$, $\ketup \equiv \ket{F=2, m_F = 2}$ to $\ket{e}\equiv \ket{F=3', m_F = 3}$ at wavelength 780~nm.   The spin down state is  chosen as $\ketdown \equiv \ket{F=1, m_F = 1}$. For the following, the probing scheme with relevant energy levels, dipole matrix elements, decay branching ratios, dressed mode frequencies, and probe laser detunings are shown and defined in Fig.~\ref{Rb87Cycling_EnergyLevels}.  Here we will extend the previous models of the precision of the estimation of $J_z$ and the loss of signal due to wavefunction collapse to capture the essential physics for this system.  Key results are that there exists a region of saturation, or universal spectroscopic enhancement, set only by atomic properties and  in which varying atom number and cavity finesse can have little impact.  However, unlike in the previous section, the asymmetry in the cavity coupling to $\ketup$ and $\ketdown$ allows this saturation region to be surpassed at large values of $N C$.

We must first consider what limits the rate of Raman scattering processes that can lead to diffusion of the spin projection $J_z$.  The probe polarization can be set to pure $\sigma^+$ to better than $10^{-4}$, so that Raman scattering from $\ketup$ is suppressed to at least this level or greater.   The more fundamental Raman scattering limitation arises from the finite hyperfine splitting $\ohf = 2 \pi \times (6834$~MHz).  Specifically, atoms in $\ketdown$ can non-resonantly Raman scatter probe photons from $\ket{e'} \equiv \ket{F=2', m_F= 2 }$.  

In the following discussion, the quantity $m_s$ is importantly defined as the average number of probe photons (normalized to the total atom number)  Rayleigh scattered into free space by atoms in $\ketup$.  All other scattering processes will be scaled from this quantity using the quantities defined in Fig.~\ref{Rb87Cycling_EnergyLevels}.   The key parameters for rescaling are the ratio of the dipole matrix elements $r = M_{\downarrow e'}/M_{\uparrow e} = 1\sqrt{2}$, the decay branching ratio $B_{3 e'}= 1/6$ from $\ket{e'}$ to $\ket{3}\equiv \ket{F= 2, m_F=1}$, and the detunings of the probe light $\delta_e$ and $\delta_{e'} = \delta_{e}-\ohf + \oehf$ from resonance with the transitions $\ketup \rightarrow \ket{e}$ and $\ketdown \rightarrow \ket{e'}$ respectively. As in the previous section, we neglect the excited state hyperfine splittings so that $\delta_{e'} \approx \delta_e - \ohf$.

The rms imprecision in the estimate of $J_z$ relative to the projection noise level can be approximately modeled as 
\begin{eqnarray}
\left(\frac{\Delta J'_z}{\qpn}\right)^2 = 4 p_\uparrow m_s + p_3 m_s + \frac{\widetilde{m}_s^\mathrm{proj} }{m_s} \, .
\label{eqn:CyclingNoise}
\end{eqnarray}

\noindent  Starting in order of physical significance, the first and second terms arise from diffusion of $J_z$ caused by Raman transitions from $\ketdown \rightarrow \ketup$ and $\ket{3}$ with effective probabilities $p_\uparrow$ and $p_3$  given approximately by
\begin{eqnarray}
 p_{\uparrow,3}=  B_{\uparrow,3\,\, e'}  r^2 \left(\frac{\delta_e}{\delta_{e'}}\right)^2 \, .
\label{eqn:pflip}
\end{eqnarray}

\noindent  In this simple treatment, Raman decays to $\ket{3}$ are treated as loss, as reflected in the smaller numerical pre-factor in front of the second term of Eq.~(\ref{eqn:CyclingNoise}).

The third term in Eq.~(\ref{eqn:CyclingNoise}) is modified to reflect that both states can interact with the probe at large detunings such that the dressed cavity mode frequency is less sensitive to quantum projection noise in $J_z$, and thus more probe photons must be used to resolve $J_z$ at the projection noise level, i.e., 
\begin{eqnarray}
\widetilde{m}_s^\mathrm{proj} = \frac{\msp}{R_\mathrm{ray}} \,\, .
\label{eqn:mscycling}
\end{eqnarray}

\noindent Here, $\msp$ is defined by Eqs.~(\ref{eqn:Genms}), (\ref{eqn:Dressedkappa}), and (\ref{eqn:modefreq}).  Indistinguishability is accounted for by 
\begin{eqnarray}
R_\mathrm{ray} = \left( 1 -r  \left| \frac{\delta_{e}}{\delta_{e'}} \right|  \right)^2 \,\, .
\label{eqn:LengthDecayRay} 
\end{eqnarray}

\noindent Note that $R_\mathrm{ray} \leq 1$ with an asymptotic value of $R_\mathrm{ray} \to 1$  at large detunings.

\begin{figure}
\includegraphics[width=3.4in]{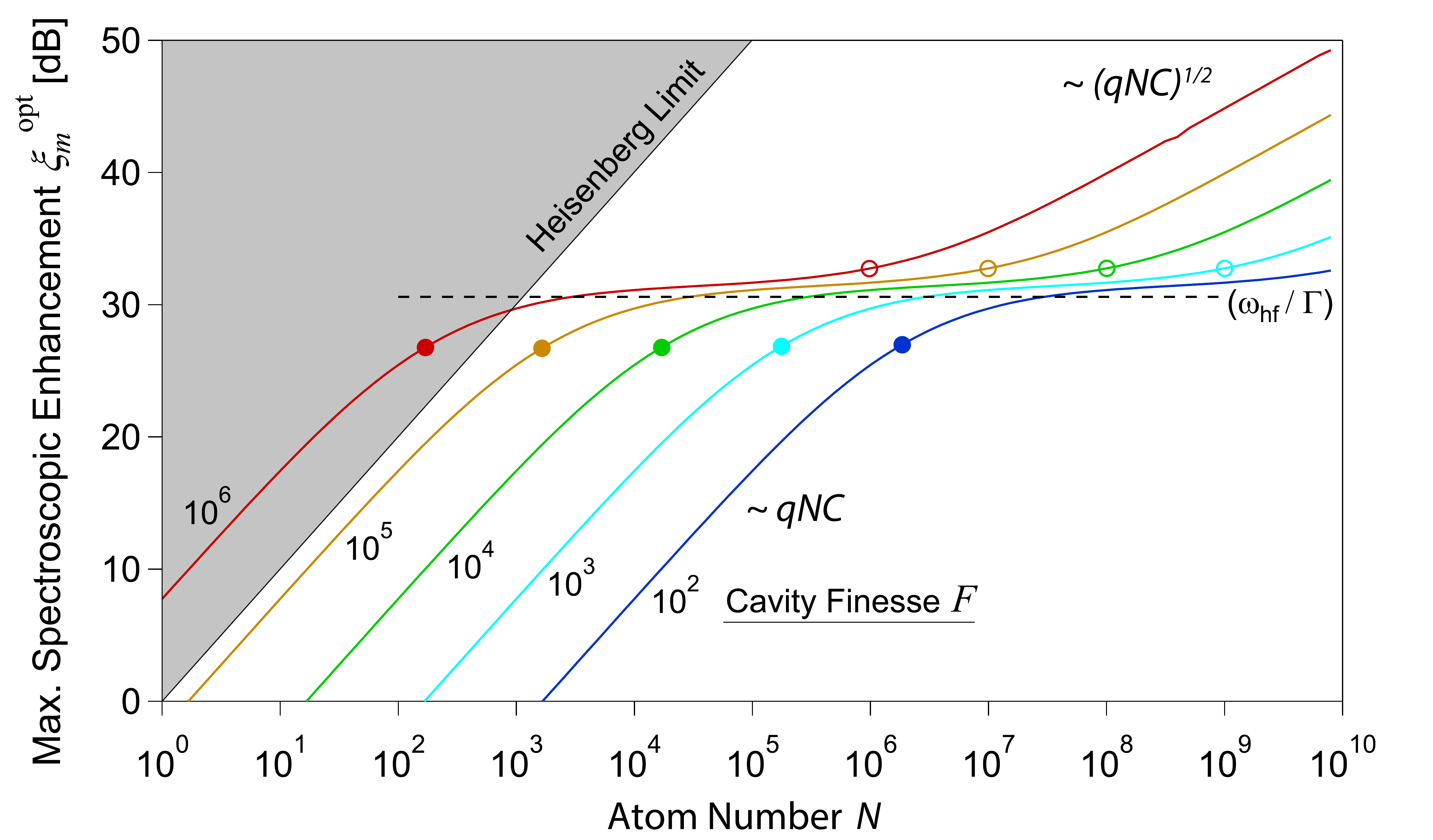}
\caption{(Color online) Theoretical optimal spectroscopic enhancement $\sq^\mathrm{opt}$ (solid curves) in dB relative to the SQL versus effective $^{87}$Rb atom number $N$ when probing near the D2 cycling transition as shown in Fig.~\ref{Rb87Cycling_EnergyLevels}.   The calculations are for the cavity geometry of Ref.~\cite{CBS11} (waist $w_0 = 71~\mu$m and cavity length $l=1.91$~cm), and assuming a net probe quantum efficiency $q=1$.  The optimization of the spectroscopic enhancement is performed for a range of technically realizable cavity finesses $F= 10^2, 10^3, 10^4, 10^5,$ and $10^6$ (blue, cyan, green, yellow, and red curves respectively).    Note that the regime below the Heisenberg limit is unphysical (disallowed grey region).   A region of saturation of $\sq^\mathrm{opt}$ versus $N$ occurs near $\sq^\mathrm{sat}\approx \ohf/\Gamma = 30.5$~dB shown by the horizontal dashed line.   The physical origin for the saturation region arises from competition between the scaling of the off-resonance Raman scattering probabilities and the dressed cavity mode broadening, as described in the text.  The points of 3~dB deviation from $qNC\propto q N F/ w_0^2$ scaling at low effective atom number (solid circles) and $\sqrt{qNC}\propto  \sqrt{q N F}/ w_0$ scaling at large effective atom number (open circles) occur when the solid curves cross  $\sq^\mathrm{opt} =27$~dB and 33~dB, respectively.  For different cavity waist size  $w_0'$ and finite quantum efficiency $q$, the saturation level is given by $\sq^\mathrm{sat}\approx \sqrt{q}\ohf/\Gamma$.   The lower saturation atom number  $N_{lower}$ (value of $N$ at the solid circles) scales $N_{lower}\propto q^{-1/2} \times \left(w_0'/w_0\right)^{-2} \times F^{-1}$, and the upper saturation atom number $N_{upper}$ (value of $N$ at the open circles) scales as $N_{upper} \propto q^0 \times \left(w_0'/w_0\right)^{-2} \times F^{-1}$ .  Note that the cavity length $l$ is largely irrelevant here (although important for technical reasons) so long as  one operates in the good-cavity limit $\kappa \ll \Gamma$.  This is the case for all curves except for the finesse $F=10^2$ curve where $\kappa \lesssim \Gamma$. }
\label{CyclingLimit_AtomNum_Finesse}
\end{figure}

There are two effects that are neglected in Eq.~\ref{eqn:mscycling} by first assuming they are small, and then verifying this to be the case after the calculations.  First, in applying the dressed cavity linewidth result for $\kappa'$ from  Eq.~(\ref{eqn:Dressedkappa}),  we assume that the cavity mode is negligibly further broadened by atoms in state $\ketdown$.  By estimating the additional broadening evaluated at the optimal cavity detuning and average number of scattered photons, we find that the optimal spectroscopic enhancements calculated in Fig.~\ref{CyclingLimit_AtomNum_Finesse} are reduced by $<0.3$~dB due to the neglected  mode broadening.  Second, we assume that the dressed mode frequency $\om$ calculated from Eq.~(\ref{eqn:modefreq}) is only modified by a small fraction by atoms in state $\ketdown$.  Again, this assumption is verified to be the case at the optimal cavity detuning and the average number of scattered photons, with the exception of the case where $N>10^8$ and cavity finesse $F= 100$, as shown in Fig.~\ref{CyclingLimit_AtomNum_Finesse} where several dB of deviations are possible due to this effect.

Next, we consider how collapse due to free space scattering reduces the radius of the collective Bloch sphere, specifically 
\begin{eqnarray}
\widetilde{\R} = e^{-m_s (R_\mathrm{ray} +R_\mathrm{ram})} \, ,
\label{eqn:LengthDecayRayRam}
\end{eqnarray}

\noindent where the partial cancellation of wavefunction collapse due to indistinguishable Rayleigh scattering  off of both $\ketup$ and $\ketdown$ (see~\cite{Uys2010}) is accounted for by $R_\mathrm{ray}$.

The term $R_\mathrm{ram}$ accounts for  Raman scattering from $\ketdown$ to $\ket{3}$.
\begin{eqnarray}
R_\mathrm{ram} = B_{3 e'} r^2 \left( \frac{\delta_{e}}{\delta_{e'}} \right)^2 \, .
\label{eqn:LengthDecayRam}
\end{eqnarray}

\noindent As before, we assume that Raman scattering to state $\ket{3}$  is  equivalent to atom loss. Note also that $R_\mathrm{ram} \leq B_{3 e'} r^2$.

\begin{figure}
\includegraphics[width=3.4in]{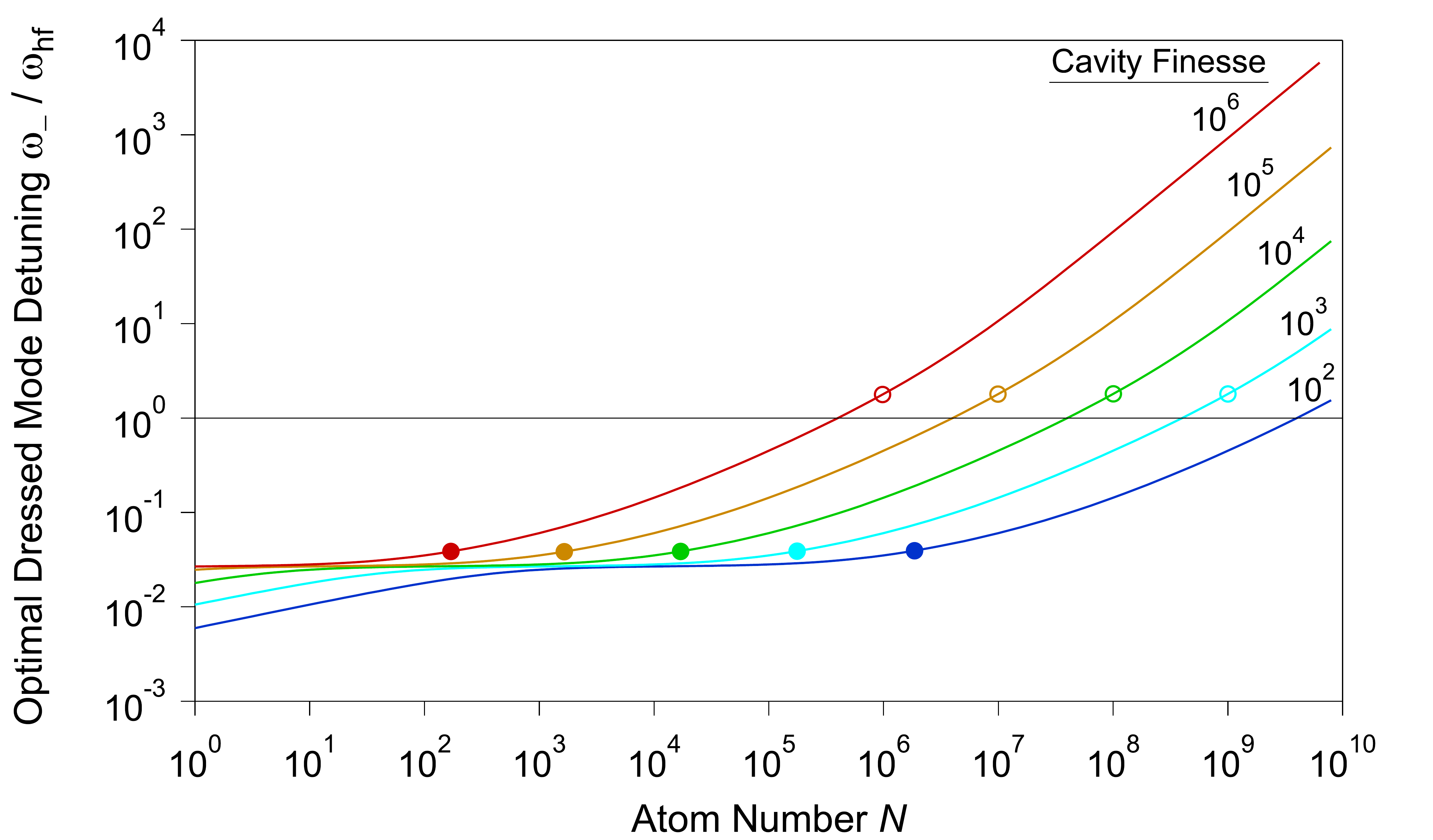}
\caption{(Color online) The dressed mode frequency $\om$ that optimizes the spectroscopic enhancement $\sq$ for the conditions described in Fig.~\ref{CyclingLimit_AtomNum_Finesse}, where again the cavity finesses $F= 10^2, 10^3, 10^4, 10^5,$ and $10^6$ correspond to  blue, cyan, green, yellow, and red curves, respectively.  The open and closed circles indicate the locations of the 3 dB points in Fig.~\ref{CyclingLimit_AtomNum_Finesse}.  The mode frequency is normalized to the hyperfine splitting $\ohf/2\pi = 6.834$~GHz.  Note that $\om \approx \Gamma$ at $\om /\ohf\approx 10^{-3}$.  Comparing to Fig.~\ref{CyclingLimit_AtomNum_Finesse},  one finds that the transition to spectroscopic enhancement scaling as $\sq^\mathrm{opt} \propto \sqrt{q N C}$ occurs near $\om \sim 1.8\times \ohf$. Above this point, the spin flip probabilities $p_{\uparrow, 3}$ change little with $\om$.}
\label{CyclingLimit_AtomNum_fp}
\end{figure}

\begin{figure}
\includegraphics[width=3.4in]{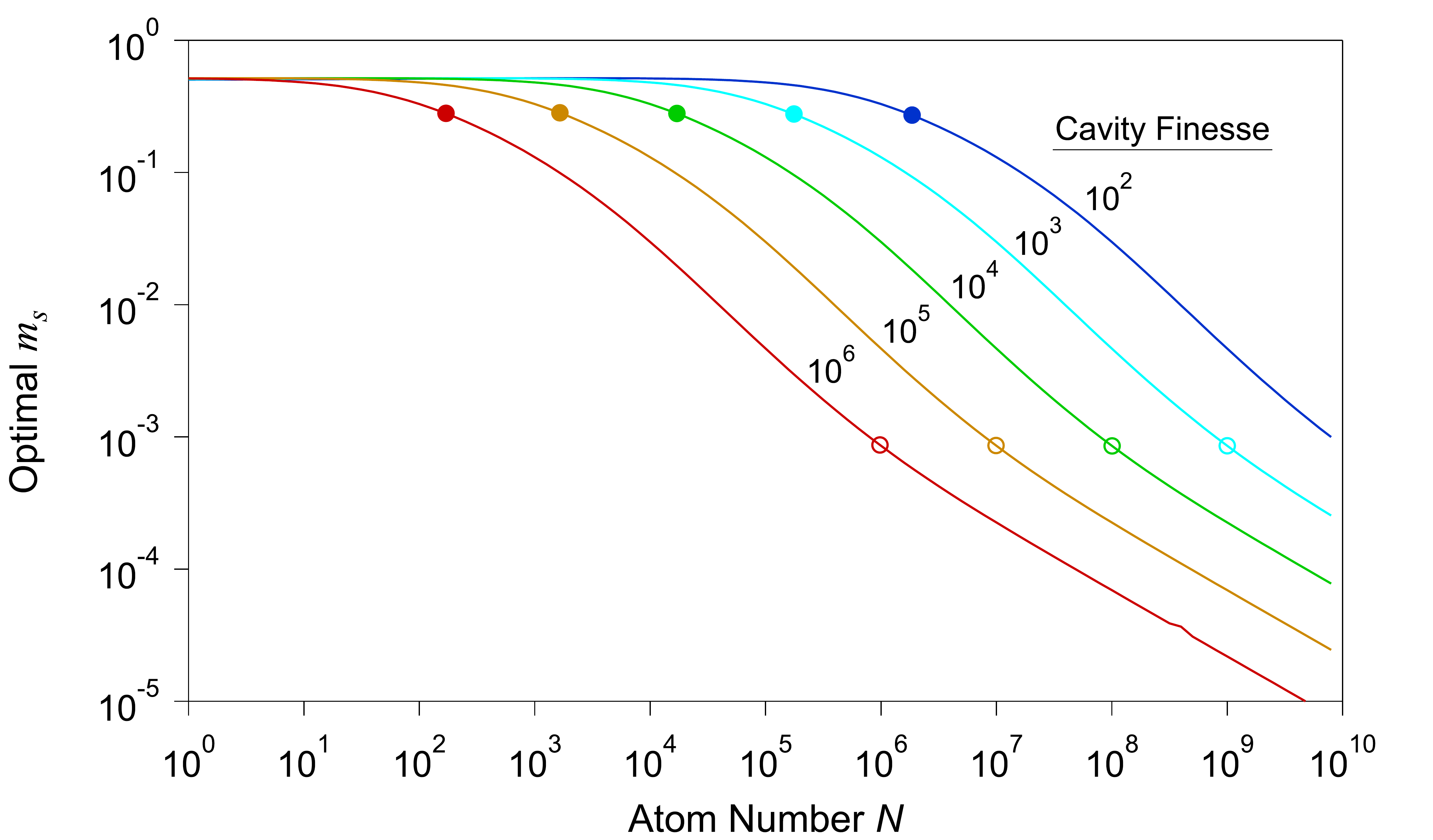}
\caption{(Color online) The scattering $m_s$ that optimizes the spectroscopic enhancement $\sq$ shown in Fig.~\ref{CyclingLimit_AtomNum_Finesse}.   Again, the cavity finesses $F= 10^2, 10^3, 10^4, 10^5,$ and $10^6$ correspond to the solid blue, cyan, green, yellow, and red curves, respectively.  The open and closed circles indicate the locations of the 3 dB points in Fig.~\ref{CyclingLimit_AtomNum_Finesse}.  Comparing to Fig.~\ref{CyclingLimit_AtomNum_Finesse},  one finds that the transition to spectroscopic enhancement scaling as $\sq^\mathrm{opt} \propto {q N C}$ occurs when $m_s \approx 0.25$, indicating that at larger $N$, far off-resonance Raman scattering begins to create significant diffusion of $J_z$ that limits the precision of the estimation of $J_z$ as discussed in the main text.}
\label{CyclingLimit_AtomNum_LossofContrast}
\end{figure}

Equations (47)-(52) are used to numerically estimate the optimal spectroscopic enhancement $\sq^\mathrm{opt}$ shown versus atom number $N$ in Fig. \ref{CyclingLimit_AtomNum_Finesse} for a range of technologically reasonable cavity finesses and assuming the cavity geometry of Ref.~\cite{CBS11} (cavity length $l = 1.91$~cm, mode waist size $w_0=71~\mu$m), and perfect quantum efficiency $q=1$.  The optimization is done with respect to both $m_s$ and the dressed cavity mode frequency $\om$ (tuned by changing the bare cavity frequency).  The mode frequency $\om$ at the optimum is shown in Fig. \ref{CyclingLimit_AtomNum_fp}. The loss of signal due to wavefunction collapse and scattering to $\ket{3}$ at the optimum is shown in Fig. \ref{CyclingLimit_AtomNum_LossofContrast}.  

At low atom number, the spectroscopic enhancement scales as $\sq^\mathrm{opt} \sim q N C$.  At high atom number, the spectroscopic enhancement scales as  $\sq^\mathrm{opt} \sim \sqrt{q N C}$.  There is an intermediate region of atom number for which the spectroscopic enhancement is relatively flat versus atom number with $\sq^\mathrm{opt} \sim \ohf / \Gamma$. 

The physical origin of this plateau arises from the form of the critical detuning of Eq.~(\ref{eqn:DispersiveDeltaCrit}).     In the good cavity limit, the scattering necessary to reach projection noise level sensitivity $\msp$ falls as $1/\dc^4$ for $\dc<\dc^\circ$, making it beneficial to operate with   $|\om| \sim \dc \ge \Gamma  \sqrt {N C/ q}$.  However, the Raman transition probabilities $p_{\uparrow, 3}$ continue to grow quadratically with the detuning, while detuning farther no longer rapidly reduces $\msp$.   

Assuming the critical detuning $\dc^\circ$ is optimal for the reasons above, and in the limit of $|\om|< \ohf$,  the Raman transition probabilities scale as $p_{\uparrow, 3}\sim  (N C/ q) (\Gamma/  \ohf)^2$, while $\msp\sim 1/(q N C)$.  Optimizing the total noise in our estimate of $J_z$ using Eq.~(\ref{eqn:CyclingNoise}) with respect to $m_s$ reproduces the observed plateau value $\sq^\mathrm{opt} \sim \ohf/ \Gamma \sim 10^3$.  The plateau region is exited at low atom number when loss of signal [described by Eq.~(\ref{eqn:LengthDecayRayRam})] dominates the reduction in spectroscopic enhancement, as illustrated by the loss of signal due to wavefunction collapse shown in Fig.~\ref{CyclingLimit_AtomNum_fp}.  At high atom number, the plateau region is exited when the optimum mode frequency becomes large compared to the hyperfine splitting $|\om| > \ohf$, as shown in Fig. \ref{CyclingLimit_AtomNum_LossofContrast}.

Importantly, this analysis shows that there is a range in which increasing either finesse or atom number can have little effect on the optimal spectroscopic enhancement achieved.  Also, note that the value of the plateau does not depend on the cavity geometry, and therefore represents a universal value that depends only on the atomic properties and the quantum efficiency.  See the caption  of Fig.~\ref{CyclingLimit_AtomNum_Finesse} for various scalings with physical parameters.  Finally, for atom numbers below $10^3$, it appears possible to both prepare and readout states near the Heisenberg limit using this approach and technologically feasible cavity finesses.  Indeed, Ref.~\cite{Vuletic12} recently demonstrated single atom measurement resolution for $N \sim 100$ using the approach described here.

%******************************************************************************************************
%*****************************************Conclusions*************************************************
%******************************************************************************************************
\section{Conclusions}\label{sect9}
In conclusion, we have presented detailed expressions for how cavity-aided, nondemolition measurements of atomic populations scale with key experimental parameters: cavity linewidth, cavity geometry, collective cooperativity, and Raman transition probabilities. We have analyzed two different probing schemes in $\Rb$ and estimated fundamental limits on conditional spin-squeezing in ensembles of $\Rb$ atoms.

This in depth look at the fundamental limits for cavity-aided measurements will be an important part of moving beyond   proof-of-principle experiments to achieve large amounts of observed squeezing for advancing precision measurements with cold atoms.  The present analysis was particularly important for guiding recent work in $\Rb$ where we have observed 10.2(6) dB of spectroscopic enhancement~\cite{BohnetSqueeze13}. To the best of our knowledge, this represents the largest entanglement enhancement ever observed in matter systems and is ideal for implementing in state of the art precision measurement experiments such as optical lattice clocks~\cite{Hinkley2013,Bloom2013}. The analysis in this manuscript enabled our cavity-aided non-demolition measurement to greatly improve upon the previous best observation of $\le 1.5$~dB of spectroscopic enhancement using quantum non-demolition techniques~\cite{Polzik10b}. Our analysis can be adapted to other atoms with ground state hyperfine structure such as Cs and other alkali elements used in spectroscopy and quantum information experiments, and further points the way to implementing the techniques presented here in alkali-earth elements in optical lattice clocks.

\section{Acknowledgements}\label{sect10}
This work was supported in parts by NIST, the NSF PFC, ARO, and DARPA QUASAR.  Z.C. was supported in part by A*STAR Singapore. J.G.B. acknowledges support from NSF GRF, and K.C.C. acknowledges support from NDSEG.

%\bibliography{SpinSqueezingPRA}

\begin{thebibliography}{67}%
\makeatletter
\providecommand \@ifxundefined [1]{%
 \@ifx{#1\undefined}
}%
\providecommand \@ifnum [1]{%
 \ifnum #1\expandafter \@firstoftwo
 \else \expandafter \@secondoftwo
 \fi
}%
\providecommand \@ifx [1]{%
 \ifx #1\expandafter \@firstoftwo
 \else \expandafter \@secondoftwo
 \fi
}%
\providecommand \natexlab [1]{#1}%
\providecommand \enquote  [1]{``#1''}%
\providecommand \bibnamefont  [1]{#1}%
\providecommand \bibfnamefont [1]{#1}%
\providecommand \citenamefont [1]{#1}%
\providecommand \href@noop [0]{\@secondoftwo}%
\providecommand \href [0]{\begingroup \@sanitize@url \@href}%
\providecommand \@href[1]{\@@startlink{#1}\@@href}%
\providecommand \@@href[1]{\endgroup#1\@@endlink}%
\providecommand \@sanitize@url [0]{\catcode `\\12\catcode `\$12\catcode
  `\&12\catcode `\#12\catcode `\^12\catcode `\_12\catcode `\%12\relax}%
\providecommand \@@startlink[1]{}%
\providecommand \@@endlink[0]{}%
\providecommand \url  [0]{\begingroup\@sanitize@url \@url }%
\providecommand \@url [1]{\endgroup\@href {#1}{\urlprefix }}%
\providecommand \urlprefix  [0]{URL }%
\providecommand \Eprint [0]{\href }%
\providecommand \doibase [0]{http://dx.doi.org/}%
\providecommand \selectlanguage [0]{\@gobble}%
\providecommand \bibinfo  [0]{\@secondoftwo}%
\providecommand \bibfield  [0]{\@secondoftwo}%
\providecommand \translation [1]{[#1]}%
\providecommand \BibitemOpen [0]{}%
\providecommand \bibitemStop [0]{}%
\providecommand \bibitemNoStop [0]{.\EOS\space}%
\providecommand \EOS [0]{\spacefactor3000\relax}%
\providecommand \BibitemShut  [1]{\csname bibitem#1\endcsname}%
\let\auto@bib@innerbib\@empty
%</preamble>
\bibitem [{\citenamefont {Kitching}\ \emph {et~al.}(2011)\citenamefont
  {Kitching}, \citenamefont {Knappe},\ and\ \citenamefont
  {Donley}}]{Kitching2011}%
  \BibitemOpen
  \bibfield  {author} {\bibinfo {author} {\bibfnamefont {J.}~\bibnamefont
  {Kitching}}, \bibinfo {author} {\bibfnamefont {S.}~\bibnamefont {Knappe}}, \
  and\ \bibinfo {author} {\bibfnamefont {E.~A.}\ \bibnamefont {Donley}},\
  }\href {\doibase 10.1109/JSEN.2011.2157679} {\bibfield  {journal} {\bibinfo
  {journal} {IEEE Sensors Journal}\ }\textbf {\bibinfo {volume} {11}},\
  \bibinfo {pages} {1749} (\bibinfo {year} {2011})}\BibitemShut {NoStop}%
\bibitem [{\citenamefont {Meyer}\ \emph {et~al.}(2001)\citenamefont {Meyer},
  \citenamefont {Rowe}, \citenamefont {Kielpinski}, \citenamefont {Sackett},
  \citenamefont {Itano}, \citenamefont {Monroe},\ and\ \citenamefont
  {Wineland}}]{Wineland01}%
  \BibitemOpen
  \bibfield  {author} {\bibinfo {author} {\bibfnamefont {V.}~\bibnamefont
  {Meyer}}, \bibinfo {author} {\bibfnamefont {M.~A.}\ \bibnamefont {Rowe}},
  \bibinfo {author} {\bibfnamefont {D.}~\bibnamefont {Kielpinski}}, \bibinfo
  {author} {\bibfnamefont {C.~A.}\ \bibnamefont {Sackett}}, \bibinfo {author}
  {\bibfnamefont {W.~M.}\ \bibnamefont {Itano}}, \bibinfo {author}
  {\bibfnamefont {C.}~\bibnamefont {Monroe}}, \ and\ \bibinfo {author}
  {\bibfnamefont {D.~J.}\ \bibnamefont {Wineland}},\ }\href {\doibase
  10.1103/PhysRevLett.86.5870} {\bibfield  {journal} {\bibinfo  {journal}
  {Phys. Rev. Lett.}\ }\textbf {\bibinfo {volume} {86}},\ \bibinfo {pages}
  {5870} (\bibinfo {year} {2001})}\BibitemShut {NoStop}%
\bibitem [{\citenamefont {Leibfried}\ \emph {et~al.}(2004)\citenamefont
  {Leibfried}, \citenamefont {Barrett}, \citenamefont {Schaetz}, \citenamefont
  {Britton}, \citenamefont {Chiaverini}, \citenamefont {Itano}, \citenamefont
  {Jost}, \citenamefont {Langer},\ and\ \citenamefont {Wineland}}]{Wineland04}%
  \BibitemOpen
  \bibfield  {author} {\bibinfo {author} {\bibfnamefont {D.}~\bibnamefont
  {Leibfried}}, \bibinfo {author} {\bibfnamefont {M.~D.}\ \bibnamefont
  {Barrett}}, \bibinfo {author} {\bibfnamefont {T.}~\bibnamefont {Schaetz}},
  \bibinfo {author} {\bibfnamefont {J.}~\bibnamefont {Britton}}, \bibinfo
  {author} {\bibfnamefont {J.}~\bibnamefont {Chiaverini}}, \bibinfo {author}
  {\bibfnamefont {W.~M.}\ \bibnamefont {Itano}}, \bibinfo {author}
  {\bibfnamefont {J.~D.}\ \bibnamefont {Jost}}, \bibinfo {author}
  {\bibfnamefont {C.}~\bibnamefont {Langer}}, \ and\ \bibinfo {author}
  {\bibfnamefont {D.~J.}\ \bibnamefont {Wineland}},\ }\href {\doibase
  10.1126/science.1097576} {\bibfield  {journal} {\bibinfo  {journal}
  {Science}\ }\textbf {\bibinfo {volume} {304}},\ \bibinfo {pages} {1476}
  (\bibinfo {year} {2004})}\BibitemShut {NoStop}%
\bibitem [{\citenamefont {Est\`{e}ve}\ \emph {et~al.}(2008)\citenamefont
  {Est\`{e}ve}, \citenamefont {Gross}, \citenamefont {Weller}, \citenamefont
  {Giovanazzi},\ and\ \citenamefont {Oberthaler}}]{Oberthaler08}%
  \BibitemOpen
  \bibfield  {author} {\bibinfo {author} {\bibfnamefont {J.}~\bibnamefont
  {Est\`{e}ve}}, \bibinfo {author} {\bibfnamefont {C.}~\bibnamefont {Gross}},
  \bibinfo {author} {\bibfnamefont {A.}~\bibnamefont {Weller}}, \bibinfo
  {author} {\bibfnamefont {S.}~\bibnamefont {Giovanazzi}}, \ and\ \bibinfo
  {author} {\bibfnamefont {M.~K.}\ \bibnamefont {Oberthaler}},\ }\href
  {\doibase 10.1038/nature07332} {\bibfield  {journal} {\bibinfo  {journal}
  {Nature}\ }\textbf {\bibinfo {volume} {455}},\ \bibinfo {pages} {1216}
  (\bibinfo {year} {2008})}\BibitemShut {NoStop}%
\bibitem [{\citenamefont {Appel}\ \emph {et~al.}(2009)\citenamefont {Appel},
  \citenamefont {Windpassinger}, \citenamefont {Oblak}, \citenamefont {Hoff},
  \citenamefont {Kj¾rgaard},\ and\ \citenamefont {Polzik}}]{Polzik09}%
  \BibitemOpen
  \bibfield  {author} {\bibinfo {author} {\bibfnamefont {J.}~\bibnamefont
  {Appel}}, \bibinfo {author} {\bibfnamefont {P.~J.}\ \bibnamefont
  {Windpassinger}}, \bibinfo {author} {\bibfnamefont {D.}~\bibnamefont
  {Oblak}}, \bibinfo {author} {\bibfnamefont {U.~B.}\ \bibnamefont {Hoff}},
  \bibinfo {author} {\bibfnamefont {N.}~\bibnamefont {Kj¾rgaard}}, \ and\
  \bibinfo {author} {\bibfnamefont {E.~S.}\ \bibnamefont {Polzik}},\ }\href
  {\doibase 10.1073/pnas.0901550106} {\bibfield  {journal} {\bibinfo  {journal}
  {Proc. Nat. Acad. Sci.}\ }\textbf {\bibinfo {volume} {106}},\ \bibinfo
  {pages} {10960} (\bibinfo {year} {2009})}\BibitemShut {NoStop}%
\bibitem [{\citenamefont {Schleier-Smith}\ \emph
  {et~al.}(2010{\natexlab{a}})\citenamefont {Schleier-Smith}, \citenamefont
  {Leroux},\ and\ \citenamefont {Vuleti\ifmmode~\acute{c}\else
  \'{c}\fi{}}}]{Vuletic10}%
  \BibitemOpen
  \bibfield  {author} {\bibinfo {author} {\bibfnamefont {M.~H.}\ \bibnamefont
  {Schleier-Smith}}, \bibinfo {author} {\bibfnamefont {I.~D.}\ \bibnamefont
  {Leroux}}, \ and\ \bibinfo {author} {\bibfnamefont {V.}~\bibnamefont
  {Vuleti\ifmmode~\acute{c}\else \'{c}\fi{}}},\ }\href {\doibase
  10.1103/PhysRevLett.104.073604} {\bibfield  {journal} {\bibinfo  {journal}
  {Phys. Rev. Lett.}\ }\textbf {\bibinfo {volume} {104}},\ \bibinfo {pages}
  {073604} (\bibinfo {year} {2010}{\natexlab{a}})}\BibitemShut {NoStop}%
\bibitem [{\citenamefont {Gross}\ \emph {et~al.}(2010)\citenamefont {Gross},
  \citenamefont {Zibold}, \citenamefont {Nicklas}, \citenamefont {Est\`{e}ve},\
  and\ \citenamefont {Oberthaler}}]{Oberthaler10}%
  \BibitemOpen
  \bibfield  {author} {\bibinfo {author} {\bibfnamefont {C.}~\bibnamefont
  {Gross}}, \bibinfo {author} {\bibfnamefont {T.}~\bibnamefont {Zibold}},
  \bibinfo {author} {\bibfnamefont {E.}~\bibnamefont {Nicklas}}, \bibinfo
  {author} {\bibfnamefont {J.}~\bibnamefont {Est\`{e}ve}}, \ and\ \bibinfo
  {author} {\bibfnamefont {M.~K.}\ \bibnamefont {Oberthaler}},\ }\href
  {\doibase 10.1038/nature08919} {\bibfield  {journal} {\bibinfo  {journal}
  {Nature}\ }\textbf {\bibinfo {volume} {464}},\ \bibinfo {pages} {1165}
  (\bibinfo {year} {2010})}\BibitemShut {NoStop}%
\bibitem [{\citenamefont {Riedel}\ \emph {et~al.}(2010)\citenamefont {Riedel},
  \citenamefont {B\"{o}hi}, \citenamefont {Li}, \citenamefont {H\"{a}nsch},
  \citenamefont {Sinatra},\ and\ \citenamefont {Treutlein}}]{Treutlein10}%
  \BibitemOpen
  \bibfield  {author} {\bibinfo {author} {\bibfnamefont {M.~F.}\ \bibnamefont
  {Riedel}}, \bibinfo {author} {\bibfnamefont {P.}~\bibnamefont {B\"{o}hi}},
  \bibinfo {author} {\bibfnamefont {Y.}~\bibnamefont {Li}}, \bibinfo {author}
  {\bibfnamefont {T.~W.}\ \bibnamefont {H\"{a}nsch}}, \bibinfo {author}
  {\bibfnamefont {A.}~\bibnamefont {Sinatra}}, \ and\ \bibinfo {author}
  {\bibfnamefont {P.}~\bibnamefont {Treutlein}},\ }\href {\doibase
  10.1038/nature08988} {\bibfield  {journal} {\bibinfo  {journal} {Nature}\
  }\textbf {\bibinfo {volume} {464}},\ \bibinfo {pages} {1170} (\bibinfo {year}
  {2010})}\BibitemShut {NoStop}%
\bibitem [{\citenamefont {Chen}\ \emph {et~al.}(2011)\citenamefont {Chen},
  \citenamefont {Bohnet}, \citenamefont {Sankar}, \citenamefont {Dai},\ and\
  \citenamefont {Thompson}}]{CBS11}%
  \BibitemOpen
  \bibfield  {author} {\bibinfo {author} {\bibfnamefont {Z.}~\bibnamefont
  {Chen}}, \bibinfo {author} {\bibfnamefont {J.~G.}\ \bibnamefont {Bohnet}},
  \bibinfo {author} {\bibfnamefont {S.~R.}\ \bibnamefont {Sankar}}, \bibinfo
  {author} {\bibfnamefont {J.}~\bibnamefont {Dai}}, \ and\ \bibinfo {author}
  {\bibfnamefont {J.~K.}\ \bibnamefont {Thompson}},\ }\href {\doibase
  10.1103/PhysRevLett.106.133601} {\bibfield  {journal} {\bibinfo  {journal}
  {Phys. Rev. Lett.}\ }\textbf {\bibinfo {volume} {106}},\ \bibinfo {pages}
  {133601} (\bibinfo {year} {2011})}\BibitemShut {NoStop}%
\bibitem [{\citenamefont {L\"{u}cke}\ \emph {et~al.}(2011)\citenamefont
  {L\"{u}cke}, \citenamefont {Scherer}, \citenamefont {Kruse}, \citenamefont
  {Pezz\'{e}}, \citenamefont {Deuretzbacher}, \citenamefont {Hyllus},
  \citenamefont {Topic}, \citenamefont {Peise}, \citenamefont {Ertmer},
  \citenamefont {Arlt}, \citenamefont {Santos}, \citenamefont {Smerzi},\ and\
  \citenamefont {Klempt}}]{Klempt11}%
  \BibitemOpen
  \bibfield  {author} {\bibinfo {author} {\bibfnamefont {B.}~\bibnamefont
  {L\"{u}cke}}, \bibinfo {author} {\bibfnamefont {M.}~\bibnamefont {Scherer}},
  \bibinfo {author} {\bibfnamefont {J.}~\bibnamefont {Kruse}}, \bibinfo
  {author} {\bibfnamefont {L.}~\bibnamefont {Pezz\'{e}}}, \bibinfo {author}
  {\bibfnamefont {F.}~\bibnamefont {Deuretzbacher}}, \bibinfo {author}
  {\bibfnamefont {P.}~\bibnamefont {Hyllus}}, \bibinfo {author} {\bibfnamefont
  {O.}~\bibnamefont {Topic}}, \bibinfo {author} {\bibfnamefont
  {J.}~\bibnamefont {Peise}}, \bibinfo {author} {\bibfnamefont
  {W.}~\bibnamefont {Ertmer}}, \bibinfo {author} {\bibfnamefont
  {J.}~\bibnamefont {Arlt}}, \bibinfo {author} {\bibfnamefont {L.}~\bibnamefont
  {Santos}}, \bibinfo {author} {\bibfnamefont {A.}~\bibnamefont {Smerzi}}, \
  and\ \bibinfo {author} {\bibfnamefont {C.}~\bibnamefont {Klempt}},\ }\href
  {\doibase 10.1126/science.1208798} {\bibfield  {journal} {\bibinfo  {journal}
  {Science}\ }\textbf {\bibinfo {volume} {334}},\ \bibinfo {pages} {773}
  (\bibinfo {year} {2011})}\BibitemShut {NoStop}%
\bibitem [{\citenamefont {Sewell}\ \emph {et~al.}(2012)\citenamefont {Sewell},
  \citenamefont {Koschorreck}, \citenamefont {Napolitano}, \citenamefont
  {Dubost}, \citenamefont {Behbood},\ and\ \citenamefont
  {Mitchell}}]{Mitchell12}%
  \BibitemOpen
  \bibfield  {author} {\bibinfo {author} {\bibfnamefont {R.~J.}\ \bibnamefont
  {Sewell}}, \bibinfo {author} {\bibfnamefont {M.}~\bibnamefont {Koschorreck}},
  \bibinfo {author} {\bibfnamefont {M.}~\bibnamefont {Napolitano}}, \bibinfo
  {author} {\bibfnamefont {B.}~\bibnamefont {Dubost}}, \bibinfo {author}
  {\bibfnamefont {N.}~\bibnamefont {Behbood}}, \ and\ \bibinfo {author}
  {\bibfnamefont {M.~W.}\ \bibnamefont {Mitchell}},\ }\href {\doibase
  10.1103/PhysRevLett.109.253605} {\bibfield  {journal} {\bibinfo  {journal}
  {Phys. Rev. Lett.}\ }\textbf {\bibinfo {volume} {109}},\ \bibinfo {pages}
  {253605} (\bibinfo {year} {2012})}\BibitemShut {NoStop}%
\bibitem [{\citenamefont {Hamley}\ \emph {et~al.}(2012)\citenamefont {Hamley},
  \citenamefont {Gerving}, \citenamefont {Hoang}, \citenamefont {Bookjans},\
  and\ \citenamefont {Chapman}}]{Chapman12}%
  \BibitemOpen
  \bibfield  {author} {\bibinfo {author} {\bibfnamefont {C.~D.}\ \bibnamefont
  {Hamley}}, \bibinfo {author} {\bibfnamefont {C.~S.}\ \bibnamefont {Gerving}},
  \bibinfo {author} {\bibfnamefont {T.~M.}\ \bibnamefont {Hoang}}, \bibinfo
  {author} {\bibfnamefont {E.~M.}\ \bibnamefont {Bookjans}}, \ and\ \bibinfo
  {author} {\bibfnamefont {M.~S.}\ \bibnamefont {Chapman}},\ }\href {\doibase
  10.1038/nphys2245} {\bibfield  {journal} {\bibinfo  {journal} {Nature Phys.}\
  }\textbf {\bibinfo {volume} {8}},\ \bibinfo {pages} {305} (\bibinfo {year}
  {2012})}\BibitemShut {NoStop}%
\bibitem [{\citenamefont {Bohnet}\ \emph {et~al.}(2013)\citenamefont {Bohnet},
  \citenamefont {Cox}, \citenamefont {Norcia}, \citenamefont {Weiner},
  \citenamefont {Chen},\ and\ \citenamefont {Thompson}}]{BohnetSqueeze13}%
  \BibitemOpen
  \bibfield  {author} {\bibinfo {author} {\bibfnamefont {J.~G.}\ \bibnamefont
  {Bohnet}}, \bibinfo {author} {\bibfnamefont {K.~C.}\ \bibnamefont {Cox}},
  \bibinfo {author} {\bibfnamefont {M.~A.}\ \bibnamefont {Norcia}}, \bibinfo
  {author} {\bibfnamefont {J.~M.}\ \bibnamefont {Weiner}}, \bibinfo {author}
  {\bibfnamefont {Z.}~\bibnamefont {Chen}}, \ and\ \bibinfo {author}
  {\bibfnamefont {J.~K.}\ \bibnamefont {Thompson}},\ }\href@noop {} {\bibfield
  {journal} {\bibinfo  {journal} {ArXiv e-prints}\ } (\bibinfo {year}
  {2013})},\ \Eprint {http://arxiv.org/abs/1310.3177} {arXiv:1310.3177
  [quant-ph]} \BibitemShut {NoStop}%
\bibitem [{\citenamefont {Lodewyck}\ \emph {et~al.}(2009)\citenamefont
  {Lodewyck}, \citenamefont {Westergaard},\ and\ \citenamefont
  {Lemonde}}]{Lemonde09}%
  \BibitemOpen
  \bibfield  {author} {\bibinfo {author} {\bibfnamefont {J.}~\bibnamefont
  {Lodewyck}}, \bibinfo {author} {\bibfnamefont {P.~G.}\ \bibnamefont
  {Westergaard}}, \ and\ \bibinfo {author} {\bibfnamefont {P.}~\bibnamefont
  {Lemonde}},\ }\href {\doibase 10.1103/PhysRevA.79.061401} {\bibfield
  {journal} {\bibinfo  {journal} {Phys. Rev. A}\ }\textbf {\bibinfo {volume}
  {79}},\ \bibinfo {pages} {061401} (\bibinfo {year} {2009})}\BibitemShut
  {NoStop}%
\bibitem [{\citenamefont {Hume}\ \emph {et~al.}(2007)\citenamefont {Hume},
  \citenamefont {Rosenband},\ and\ \citenamefont {Wineland}}]{Hume07}%
  \BibitemOpen
  \bibfield  {author} {\bibinfo {author} {\bibfnamefont {D.~B.}\ \bibnamefont
  {Hume}}, \bibinfo {author} {\bibfnamefont {T.}~\bibnamefont {Rosenband}}, \
  and\ \bibinfo {author} {\bibfnamefont {D.~J.}\ \bibnamefont {Wineland}},\
  }\href {\doibase 10.1103/PhysRevLett.99.120502} {\bibfield  {journal}
  {\bibinfo  {journal} {Phys. Rev. Lett.}\ }\textbf {\bibinfo {volume} {99}},\
  \bibinfo {pages} {120502} (\bibinfo {year} {2007})}\BibitemShut {NoStop}%
\bibitem [{\citenamefont {Wineland}\ \emph {et~al.}(1994)\citenamefont
  {Wineland}, \citenamefont {Bollinger}, \citenamefont {Itano},\ and\
  \citenamefont {Heinzen}}]{Wineland94}%
  \BibitemOpen
  \bibfield  {author} {\bibinfo {author} {\bibfnamefont {D.~J.}\ \bibnamefont
  {Wineland}}, \bibinfo {author} {\bibfnamefont {J.~J.}\ \bibnamefont
  {Bollinger}}, \bibinfo {author} {\bibfnamefont {W.~M.}\ \bibnamefont
  {Itano}}, \ and\ \bibinfo {author} {\bibfnamefont {D.~J.}\ \bibnamefont
  {Heinzen}},\ }\href {\doibase 10.1103/PhysRevA.50.67} {\bibfield  {journal}
  {\bibinfo  {journal} {Phys. Rev. A}\ }\textbf {\bibinfo {volume} {50}},\
  \bibinfo {pages} {67} (\bibinfo {year} {1994})}\BibitemShut {NoStop}%
\bibitem [{\citenamefont {Leroux}\ \emph {et~al.}(2010)\citenamefont {Leroux},
  \citenamefont {Schleier-Smith},\ and\ \citenamefont
  {Vuleti\ifmmode~\acute{c}\else \'{c}\fi{}}}]{Vuletic10b}%
  \BibitemOpen
  \bibfield  {author} {\bibinfo {author} {\bibfnamefont {I.~D.}\ \bibnamefont
  {Leroux}}, \bibinfo {author} {\bibfnamefont {M.~H.}\ \bibnamefont
  {Schleier-Smith}}, \ and\ \bibinfo {author} {\bibfnamefont {V.}~\bibnamefont
  {Vuleti\ifmmode~\acute{c}\else \'{c}\fi{}}},\ }\href {\doibase
  10.1103/PhysRevLett.104.073602} {\bibfield  {journal} {\bibinfo  {journal}
  {Phys. Rev. Lett.}\ }\textbf {\bibinfo {volume} {104}},\ \bibinfo {pages}
  {073602} (\bibinfo {year} {2010})}\BibitemShut {NoStop}%
\bibitem [{\citenamefont {Schleier-Smith}\ \emph
  {et~al.}(2010{\natexlab{b}})\citenamefont {Schleier-Smith}, \citenamefont
  {Leroux},\ and\ \citenamefont {Vuleti\ifmmode~\acute{c}\else
  \'{c}\fi{}}}]{Vuletic10c}%
  \BibitemOpen
  \bibfield  {author} {\bibinfo {author} {\bibfnamefont {M.~H.}\ \bibnamefont
  {Schleier-Smith}}, \bibinfo {author} {\bibfnamefont {I.~D.}\ \bibnamefont
  {Leroux}}, \ and\ \bibinfo {author} {\bibfnamefont {V.}~\bibnamefont
  {Vuleti\ifmmode~\acute{c}\else \'{c}\fi{}}},\ }\href {\doibase
  10.1103/PhysRevA.81.021804} {\bibfield  {journal} {\bibinfo  {journal} {Phys.
  Rev. A}\ }\textbf {\bibinfo {volume} {81}},\ \bibinfo {pages} {021804}
  (\bibinfo {year} {2010}{\natexlab{b}})}\BibitemShut {NoStop}%
\bibitem [{\citenamefont {Leroux}\ \emph {et~al.}(2012)\citenamefont {Leroux},
  \citenamefont {Schleier-Smith}, \citenamefont {Zhang},\ and\ \citenamefont
  {Vuleti\ifmmode~\acute{c}\else \'{c}\fi{}}}]{Vuletic12b}%
  \BibitemOpen
  \bibfield  {author} {\bibinfo {author} {\bibfnamefont {I.~D.}\ \bibnamefont
  {Leroux}}, \bibinfo {author} {\bibfnamefont {M.~H.}\ \bibnamefont
  {Schleier-Smith}}, \bibinfo {author} {\bibfnamefont {H.}~\bibnamefont
  {Zhang}}, \ and\ \bibinfo {author} {\bibfnamefont {V.}~\bibnamefont
  {Vuleti\ifmmode~\acute{c}\else \'{c}\fi{}}},\ }\href {\doibase
  10.1103/PhysRevA.85.013803} {\bibfield  {journal} {\bibinfo  {journal} {Phys.
  Rev. A}\ }\textbf {\bibinfo {volume} {85}},\ \bibinfo {pages} {013803}
  (\bibinfo {year} {2012})}\BibitemShut {NoStop}%
\bibitem [{\citenamefont {Takeuchi}\ \emph {et~al.}(2005)\citenamefont
  {Takeuchi}, \citenamefont {Ichihara}, \citenamefont {Takano}, \citenamefont
  {Kumakura}, \citenamefont {Yabuzaki},\ and\ \citenamefont
  {Takahashi}}]{Takahashi05}%
  \BibitemOpen
  \bibfield  {author} {\bibinfo {author} {\bibfnamefont {M.}~\bibnamefont
  {Takeuchi}}, \bibinfo {author} {\bibfnamefont {S.}~\bibnamefont {Ichihara}},
  \bibinfo {author} {\bibfnamefont {T.}~\bibnamefont {Takano}}, \bibinfo
  {author} {\bibfnamefont {M.}~\bibnamefont {Kumakura}}, \bibinfo {author}
  {\bibfnamefont {T.}~\bibnamefont {Yabuzaki}}, \ and\ \bibinfo {author}
  {\bibfnamefont {Y.}~\bibnamefont {Takahashi}},\ }\href {\doibase
  10.1103/PhysRevLett.94.023003} {\bibfield  {journal} {\bibinfo  {journal}
  {Phys. Rev. Lett.}\ }\textbf {\bibinfo {volume} {94}},\ \bibinfo {pages}
  {023003} (\bibinfo {year} {2005})}\BibitemShut {NoStop}%
\bibitem [{\citenamefont {Trail}\ \emph {et~al.}(2010)\citenamefont {Trail},
  \citenamefont {Jessen},\ and\ \citenamefont {Deutsch}}]{Deutsch10}%
  \BibitemOpen
  \bibfield  {author} {\bibinfo {author} {\bibfnamefont {C.~M.}\ \bibnamefont
  {Trail}}, \bibinfo {author} {\bibfnamefont {P.~S.}\ \bibnamefont {Jessen}}, \
  and\ \bibinfo {author} {\bibfnamefont {I.~H.}\ \bibnamefont {Deutsch}},\
  }\href {\doibase 10.1103/PhysRevLett.105.193602} {\bibfield  {journal}
  {\bibinfo  {journal} {Phys. Rev. Lett.}\ }\textbf {\bibinfo {volume} {105}},\
  \bibinfo {pages} {193602} (\bibinfo {year} {2010})}\BibitemShut {NoStop}%
\bibitem [{\citenamefont {Norris}\ \emph {et~al.}(2012)\citenamefont {Norris},
  \citenamefont {Trail}, \citenamefont {Jessen},\ and\ \citenamefont
  {Deutsch}}]{Deutsch12}%
  \BibitemOpen
  \bibfield  {author} {\bibinfo {author} {\bibfnamefont {L.~M.}\ \bibnamefont
  {Norris}}, \bibinfo {author} {\bibfnamefont {C.~M.}\ \bibnamefont {Trail}},
  \bibinfo {author} {\bibfnamefont {P.~S.}\ \bibnamefont {Jessen}}, \ and\
  \bibinfo {author} {\bibfnamefont {I.~H.}\ \bibnamefont {Deutsch}},\ }\href
  {\doibase 10.1103/PhysRevLett.109.173603} {\bibfield  {journal} {\bibinfo
  {journal} {Phys. Rev. Lett.}\ }\textbf {\bibinfo {volume} {109}},\ \bibinfo
  {pages} {173603} (\bibinfo {year} {2012})}\BibitemShut {NoStop}%
\bibitem [{\citenamefont {Grangier}\ \emph {et~al.}(1998)\citenamefont
  {Grangier}, \citenamefont {Levenson},\ and\ \citenamefont
  {Poizat}}]{Grangier98}%
  \BibitemOpen
  \bibfield  {author} {\bibinfo {author} {\bibfnamefont {P.}~\bibnamefont
  {Grangier}}, \bibinfo {author} {\bibfnamefont {J.~A.}\ \bibnamefont
  {Levenson}}, \ and\ \bibinfo {author} {\bibfnamefont {J.-P.}\ \bibnamefont
  {Poizat}},\ }\href {http://dx.doi.org/10.1038/25059} {\bibfield  {journal}
  {\bibinfo  {journal} {Nature}\ }\textbf {\bibinfo {volume} {396}},\ \bibinfo
  {pages} {537} (\bibinfo {year} {1998})}\BibitemShut {NoStop}%
\bibitem [{\citenamefont {Chaudhury}\ \emph {et~al.}(2006)\citenamefont
  {Chaudhury}, \citenamefont {Smith}, \citenamefont {Schulz},\ and\
  \citenamefont {Jessen}}]{Jessen06}%
  \BibitemOpen
  \bibfield  {author} {\bibinfo {author} {\bibfnamefont {S.}~\bibnamefont
  {Chaudhury}}, \bibinfo {author} {\bibfnamefont {G.~A.}\ \bibnamefont
  {Smith}}, \bibinfo {author} {\bibfnamefont {K.}~\bibnamefont {Schulz}}, \
  and\ \bibinfo {author} {\bibfnamefont {P.~S.}\ \bibnamefont {Jessen}},\
  }\href {\doibase 10.1103/PhysRevLett.96.043001} {\bibfield  {journal}
  {\bibinfo  {journal} {Phys. Rev. Lett.}\ }\textbf {\bibinfo {volume} {96}},\
  \bibinfo {pages} {043001} (\bibinfo {year} {2006})}\BibitemShut {NoStop}%
\bibitem [{\citenamefont {Takahashi}\ \emph {et~al.}(1999)\citenamefont
  {Takahashi}, \citenamefont {Honda}, \citenamefont {Tanaka}, \citenamefont
  {Toyoda}, \citenamefont {Ishikawa},\ and\ \citenamefont
  {Yabuzaki}}]{Takahashi99}%
  \BibitemOpen
  \bibfield  {author} {\bibinfo {author} {\bibfnamefont {Y.}~\bibnamefont
  {Takahashi}}, \bibinfo {author} {\bibfnamefont {K.}~\bibnamefont {Honda}},
  \bibinfo {author} {\bibfnamefont {N.}~\bibnamefont {Tanaka}}, \bibinfo
  {author} {\bibfnamefont {K.}~\bibnamefont {Toyoda}}, \bibinfo {author}
  {\bibfnamefont {K.}~\bibnamefont {Ishikawa}}, \ and\ \bibinfo {author}
  {\bibfnamefont {T.}~\bibnamefont {Yabuzaki}},\ }\href {\doibase
  10.1103/PhysRevA.60.4974} {\bibfield  {journal} {\bibinfo  {journal} {Phys.
  Rev. A}\ }\textbf {\bibinfo {volume} {60}},\ \bibinfo {pages} {4974}
  (\bibinfo {year} {1999})}\BibitemShut {NoStop}%
\bibitem [{\citenamefont {Takano}\ \emph {et~al.}(2009)\citenamefont {Takano},
  \citenamefont {Fuyama}, \citenamefont {Namiki},\ and\ \citenamefont
  {Takahashi}}]{Takano09}%
  \BibitemOpen
  \bibfield  {author} {\bibinfo {author} {\bibfnamefont {T.}~\bibnamefont
  {Takano}}, \bibinfo {author} {\bibfnamefont {M.}~\bibnamefont {Fuyama}},
  \bibinfo {author} {\bibfnamefont {R.}~\bibnamefont {Namiki}}, \ and\ \bibinfo
  {author} {\bibfnamefont {Y.}~\bibnamefont {Takahashi}},\ }\href {\doibase
  10.1103/PhysRevLett.102.033601} {\bibfield  {journal} {\bibinfo  {journal}
  {Phys. Rev. Lett.}\ }\textbf {\bibinfo {volume} {102}},\ \bibinfo {pages}
  {033601} (\bibinfo {year} {2009})}\BibitemShut {NoStop}%
\bibitem [{\citenamefont {Kuzmich}\ \emph {et~al.}(1998)\citenamefont
  {Kuzmich}, \citenamefont {Bigelow},\ and\ \citenamefont {Mandel}}]{Mandel98}%
  \BibitemOpen
  \bibfield  {author} {\bibinfo {author} {\bibfnamefont {A.}~\bibnamefont
  {Kuzmich}}, \bibinfo {author} {\bibfnamefont {N.~P.}\ \bibnamefont
  {Bigelow}}, \ and\ \bibinfo {author} {\bibfnamefont {L.}~\bibnamefont
  {Mandel}},\ }\href {http://stacks.iop.org/0295-5075/42/i=5/a=481} {\bibfield
  {journal} {\bibinfo  {journal} {Europhys. Lett.}\ }\textbf {\bibinfo {volume}
  {42}},\ \bibinfo {pages} {481} (\bibinfo {year} {1998})}\BibitemShut
  {NoStop}%
\bibitem [{\citenamefont {Thomsen}\ \emph {et~al.}(2002)\citenamefont
  {Thomsen}, \citenamefont {Mancini},\ and\ \citenamefont
  {Wiseman}}]{Wiseman02}%
  \BibitemOpen
  \bibfield  {author} {\bibinfo {author} {\bibfnamefont {L.~K.}\ \bibnamefont
  {Thomsen}}, \bibinfo {author} {\bibfnamefont {S.}~\bibnamefont {Mancini}}, \
  and\ \bibinfo {author} {\bibfnamefont {H.~M.}\ \bibnamefont {Wiseman}},\
  }\href {http://stacks.iop.org/0953-4075/35/i=23/a=316} {\bibfield  {journal}
  {\bibinfo  {journal} {J. Phys. B: At. Mo. Opt. Phys.}\ }\textbf {\bibinfo
  {volume} {35}},\ \bibinfo {pages} {4937} (\bibinfo {year}
  {2002})}\BibitemShut {NoStop}%
\bibitem [{\citenamefont {Geremia}\ \emph {et~al.}(2003)\citenamefont
  {Geremia}, \citenamefont {Stockton}, \citenamefont {Doherty},\ and\
  \citenamefont {Mabuchi}}]{Mabuchi03}%
  \BibitemOpen
  \bibfield  {author} {\bibinfo {author} {\bibfnamefont {J.}~\bibnamefont
  {Geremia}}, \bibinfo {author} {\bibfnamefont {J.~K.}\ \bibnamefont
  {Stockton}}, \bibinfo {author} {\bibfnamefont {A.~C.}\ \bibnamefont
  {Doherty}}, \ and\ \bibinfo {author} {\bibfnamefont {H.}~\bibnamefont
  {Mabuchi}},\ }\href {\doibase 10.1103/PhysRevLett.91.250801} {\bibfield
  {journal} {\bibinfo  {journal} {Phys. Rev. Lett.}\ }\textbf {\bibinfo
  {volume} {91}},\ \bibinfo {pages} {250801} (\bibinfo {year}
  {2003})}\BibitemShut {NoStop}%
\bibitem [{\citenamefont {Madsen}\ and\ \citenamefont
  {M\o{}lmer}(2004)}]{Molmer04a}%
  \BibitemOpen
  \bibfield  {author} {\bibinfo {author} {\bibfnamefont {L.~B.}\ \bibnamefont
  {Madsen}}\ and\ \bibinfo {author} {\bibfnamefont {K.}~\bibnamefont
  {M\o{}lmer}},\ }\href {\doibase 10.1103/PhysRevA.70.052324} {\bibfield
  {journal} {\bibinfo  {journal} {Phys. Rev. A}\ }\textbf {\bibinfo {volume}
  {70}},\ \bibinfo {pages} {052324} (\bibinfo {year} {2004})}\BibitemShut
  {NoStop}%
\bibitem [{\citenamefont {M\o{}lmer}\ and\ \citenamefont
  {Madsen}(2004)}]{Molmer04b}%
  \BibitemOpen
  \bibfield  {author} {\bibinfo {author} {\bibfnamefont {K.}~\bibnamefont
  {M\o{}lmer}}\ and\ \bibinfo {author} {\bibfnamefont {L.~B.}\ \bibnamefont
  {Madsen}},\ }\href {\doibase 10.1103/PhysRevA.70.052102} {\bibfield
  {journal} {\bibinfo  {journal} {Phys. Rev. A}\ }\textbf {\bibinfo {volume}
  {70}},\ \bibinfo {pages} {052102} (\bibinfo {year} {2004})}\BibitemShut
  {NoStop}%
\bibitem [{\citenamefont {Kuzmich}\ and\ \citenamefont
  {Kennedy}(2004)}]{Kennedy04}%
  \BibitemOpen
  \bibfield  {author} {\bibinfo {author} {\bibfnamefont {A.}~\bibnamefont
  {Kuzmich}}\ and\ \bibinfo {author} {\bibfnamefont {T.~A.~B.}\ \bibnamefont
  {Kennedy}},\ }\href {\doibase 10.1103/PhysRevLett.92.030407} {\bibfield
  {journal} {\bibinfo  {journal} {Phys. Rev. Lett.}\ }\textbf {\bibinfo
  {volume} {92}},\ \bibinfo {pages} {030407} (\bibinfo {year}
  {2004})}\BibitemShut {NoStop}%
\bibitem [{\citenamefont {Auzinsh}\ \emph {et~al.}(2004)\citenamefont
  {Auzinsh}, \citenamefont {Budker}, \citenamefont {Kimball}, \citenamefont
  {Rochester}, \citenamefont {Stalnaker}, \citenamefont {Sushkov},\ and\
  \citenamefont {Yashchuk}}]{Auzinsh04}%
  \BibitemOpen
  \bibfield  {author} {\bibinfo {author} {\bibfnamefont {M.}~\bibnamefont
  {Auzinsh}}, \bibinfo {author} {\bibfnamefont {D.}~\bibnamefont {Budker}},
  \bibinfo {author} {\bibfnamefont {D.~F.}\ \bibnamefont {Kimball}}, \bibinfo
  {author} {\bibfnamefont {S.~M.}\ \bibnamefont {Rochester}}, \bibinfo {author}
  {\bibfnamefont {J.~E.}\ \bibnamefont {Stalnaker}}, \bibinfo {author}
  {\bibfnamefont {A.~O.}\ \bibnamefont {Sushkov}}, \ and\ \bibinfo {author}
  {\bibfnamefont {V.~V.}\ \bibnamefont {Yashchuk}},\ }\href {\doibase
  10.1103/PhysRevLett.93.173002} {\bibfield  {journal} {\bibinfo  {journal}
  {Phys. Rev. Lett.}\ }\textbf {\bibinfo {volume} {93}},\ \bibinfo {pages}
  {173002} (\bibinfo {year} {2004})}\BibitemShut {NoStop}%
\bibitem [{\citenamefont {Hope}\ and\ \citenamefont {Close}(2004)}]{Hope04}%
  \BibitemOpen
  \bibfield  {author} {\bibinfo {author} {\bibfnamefont {J.~J.}\ \bibnamefont
  {Hope}}\ and\ \bibinfo {author} {\bibfnamefont {J.~D.}\ \bibnamefont
  {Close}},\ }\href {\doibase 10.1103/PhysRevLett.93.180402} {\bibfield
  {journal} {\bibinfo  {journal} {Phys. Rev. Lett.}\ }\textbf {\bibinfo
  {volume} {93}},\ \bibinfo {pages} {180402} (\bibinfo {year}
  {2004})}\BibitemShut {NoStop}%
\bibitem [{\citenamefont {Stockton}\ \emph {et~al.}(2004)\citenamefont
  {Stockton}, \citenamefont {van Handel},\ and\ \citenamefont
  {Mabuchi}}]{Mabuchi04}%
  \BibitemOpen
  \bibfield  {author} {\bibinfo {author} {\bibfnamefont {J.~K.}\ \bibnamefont
  {Stockton}}, \bibinfo {author} {\bibfnamefont {R.}~\bibnamefont {van
  Handel}}, \ and\ \bibinfo {author} {\bibfnamefont {H.}~\bibnamefont
  {Mabuchi}},\ }\href {\doibase 10.1103/PhysRevA.70.022106} {\bibfield
  {journal} {\bibinfo  {journal} {Phys. Rev. A}\ }\textbf {\bibinfo {volume}
  {70}},\ \bibinfo {pages} {022106} (\bibinfo {year} {2004})}\BibitemShut
  {NoStop}%
\bibitem [{\citenamefont {Hope}\ and\ \citenamefont {Close}(2005)}]{Hope05}%
  \BibitemOpen
  \bibfield  {author} {\bibinfo {author} {\bibfnamefont {J.~J.}\ \bibnamefont
  {Hope}}\ and\ \bibinfo {author} {\bibfnamefont {J.~D.}\ \bibnamefont
  {Close}},\ }\href {\doibase 10.1103/PhysRevA.71.043822} {\bibfield  {journal}
  {\bibinfo  {journal} {Phys. Rev. A}\ }\textbf {\bibinfo {volume} {71}},\
  \bibinfo {pages} {043822} (\bibinfo {year} {2005})}\BibitemShut {NoStop}%
\bibitem [{\citenamefont {de~Echaniz}\ \emph {et~al.}(2005)\citenamefont
  {de~Echaniz}, \citenamefont {Mitchell}, \citenamefont {Kubasik},
  \citenamefont {Koschorreck}, \citenamefont {Crepaz}, \citenamefont
  {Eschner},\ and\ \citenamefont {Polzik}}]{Polzik05}%
  \BibitemOpen
  \bibfield  {author} {\bibinfo {author} {\bibfnamefont {S.~R.}\ \bibnamefont
  {de~Echaniz}}, \bibinfo {author} {\bibfnamefont {M.~W.}\ \bibnamefont
  {Mitchell}}, \bibinfo {author} {\bibfnamefont {M.}~\bibnamefont {Kubasik}},
  \bibinfo {author} {\bibfnamefont {M.}~\bibnamefont {Koschorreck}}, \bibinfo
  {author} {\bibfnamefont {H.}~\bibnamefont {Crepaz}}, \bibinfo {author}
  {\bibfnamefont {J.}~\bibnamefont {Eschner}}, \ and\ \bibinfo {author}
  {\bibfnamefont {E.~S.}\ \bibnamefont {Polzik}},\ }\href
  {http://stacks.iop.org/1464-4266/7/i=12/a=016} {\bibfield  {journal}
  {\bibinfo  {journal} {J. Opt. B: Quantum Semiclass. Opt.}\ }\textbf {\bibinfo
  {volume} {7}},\ \bibinfo {pages} {S548} (\bibinfo {year} {2005})}\BibitemShut
  {NoStop}%
\bibitem [{\citenamefont {Baragiola}\ \emph {et~al.}(2013)\citenamefont
  {Baragiola}, \citenamefont {Norris}, \citenamefont {Montano}, \citenamefont
  {Mickelson}, \citenamefont {Jessen},\ and\ \citenamefont
  {Deutsch}}]{Baragiola13}%
  \BibitemOpen
  \bibfield  {author} {\bibinfo {author} {\bibfnamefont {B.~Q.}\ \bibnamefont
  {Baragiola}}, \bibinfo {author} {\bibfnamefont {L.~M.}\ \bibnamefont
  {Norris}}, \bibinfo {author} {\bibfnamefont {E.}~\bibnamefont {Montano}},
  \bibinfo {author} {\bibfnamefont {P.~G.}\ \bibnamefont {Mickelson}}, \bibinfo
  {author} {\bibfnamefont {P.~S.}\ \bibnamefont {Jessen}}, \ and\ \bibinfo
  {author} {\bibfnamefont {I.~H.}\ \bibnamefont {Deutsch}},\ }\href@noop {}
  {\bibfield  {journal} {\bibinfo  {journal} {ArXiv e-prints}\ } (\bibinfo
  {year} {2013})},\ \Eprint {http://arxiv.org/abs/1311.2328} {arXiv:1311.2328
  [quant-ph]} \BibitemShut {NoStop}%
\bibitem [{\citenamefont {Kuzmich}\ \emph {et~al.}(1999)\citenamefont
  {Kuzmich}, \citenamefont {Mandel}, \citenamefont {Janis}, \citenamefont
  {Young}, \citenamefont {Ejnisman},\ and\ \citenamefont
  {Bigelow}}]{Kuzmich99}%
  \BibitemOpen
  \bibfield  {author} {\bibinfo {author} {\bibfnamefont {A.}~\bibnamefont
  {Kuzmich}}, \bibinfo {author} {\bibfnamefont {L.}~\bibnamefont {Mandel}},
  \bibinfo {author} {\bibfnamefont {J.}~\bibnamefont {Janis}}, \bibinfo
  {author} {\bibfnamefont {Y.~E.}\ \bibnamefont {Young}}, \bibinfo {author}
  {\bibfnamefont {R.}~\bibnamefont {Ejnisman}}, \ and\ \bibinfo {author}
  {\bibfnamefont {N.~P.}\ \bibnamefont {Bigelow}},\ }\href {\doibase
  10.1103/PhysRevA.60.2346} {\bibfield  {journal} {\bibinfo  {journal} {Phys.
  Rev. A}\ }\textbf {\bibinfo {volume} {60}},\ \bibinfo {pages} {2346}
  (\bibinfo {year} {1999})}\BibitemShut {NoStop}%
\bibitem [{\citenamefont {Kuzmich}\ \emph {et~al.}(2000)\citenamefont
  {Kuzmich}, \citenamefont {Mandel},\ and\ \citenamefont
  {Bigelow}}]{Bigelow00}%
  \BibitemOpen
  \bibfield  {author} {\bibinfo {author} {\bibfnamefont {A.}~\bibnamefont
  {Kuzmich}}, \bibinfo {author} {\bibfnamefont {L.}~\bibnamefont {Mandel}}, \
  and\ \bibinfo {author} {\bibfnamefont {N.~P.}\ \bibnamefont {Bigelow}},\
  }\href {\doibase 10.1103/PhysRevLett.85.1594} {\bibfield  {journal} {\bibinfo
   {journal} {Phys. Rev. Lett.}\ }\textbf {\bibinfo {volume} {85}},\ \bibinfo
  {pages} {1594} (\bibinfo {year} {2000})}\BibitemShut {NoStop}%
\bibitem [{\citenamefont {Windpassinger}\ \emph {et~al.}(2008)\citenamefont
  {Windpassinger}, \citenamefont {Oblak}, \citenamefont {Hoff}, \citenamefont
  {Appel}, \citenamefont {Kj¾rgaard},\ and\ \citenamefont
  {Polzik}}]{Polzik08}%
  \BibitemOpen
  \bibfield  {author} {\bibinfo {author} {\bibfnamefont {P.~J.}\ \bibnamefont
  {Windpassinger}}, \bibinfo {author} {\bibfnamefont {D.}~\bibnamefont
  {Oblak}}, \bibinfo {author} {\bibfnamefont {U.~B.}\ \bibnamefont {Hoff}},
  \bibinfo {author} {\bibfnamefont {J.}~\bibnamefont {Appel}}, \bibinfo
  {author} {\bibfnamefont {N.}~\bibnamefont {Kj¾rgaard}}, \ and\ \bibinfo
  {author} {\bibfnamefont {E.~S.}\ \bibnamefont {Polzik}},\ }\href
  {http://stacks.iop.org/1367-2630/10/i=5/a=053032} {\bibfield  {journal}
  {\bibinfo  {journal} {New J. Phys.}\ }\textbf {\bibinfo {volume} {10}},\
  \bibinfo {pages} {053032} (\bibinfo {year} {2008})}\BibitemShut {NoStop}%
\bibitem [{\citenamefont {Wasilewski}\ \emph {et~al.}(2010)\citenamefont
  {Wasilewski}, \citenamefont {Jensen}, \citenamefont {Krauter}, \citenamefont
  {Renema}, \citenamefont {Balabas},\ and\ \citenamefont {Polzik}}]{Polzik10b}%
  \BibitemOpen
  \bibfield  {author} {\bibinfo {author} {\bibfnamefont {W.}~\bibnamefont
  {Wasilewski}}, \bibinfo {author} {\bibfnamefont {K.}~\bibnamefont {Jensen}},
  \bibinfo {author} {\bibfnamefont {H.}~\bibnamefont {Krauter}}, \bibinfo
  {author} {\bibfnamefont {J.~J.}\ \bibnamefont {Renema}}, \bibinfo {author}
  {\bibfnamefont {M.~V.}\ \bibnamefont {Balabas}}, \ and\ \bibinfo {author}
  {\bibfnamefont {E.~S.}\ \bibnamefont {Polzik}},\ }\href {\doibase
  10.1103/PhysRevLett.104.133601} {\bibfield  {journal} {\bibinfo  {journal}
  {Phys. Rev. Lett.}\ }\textbf {\bibinfo {volume} {104}},\ \bibinfo {pages}
  {133601} (\bibinfo {year} {2010})}\BibitemShut {NoStop}%
\bibitem [{\citenamefont {Koschorreck}\ \emph
  {et~al.}(2010{\natexlab{a}})\citenamefont {Koschorreck}, \citenamefont
  {Napolitano}, \citenamefont {Dubost},\ and\ \citenamefont
  {Mitchell}}]{Mitchell10a}%
  \BibitemOpen
  \bibfield  {author} {\bibinfo {author} {\bibfnamefont {M.}~\bibnamefont
  {Koschorreck}}, \bibinfo {author} {\bibfnamefont {M.}~\bibnamefont
  {Napolitano}}, \bibinfo {author} {\bibfnamefont {B.}~\bibnamefont {Dubost}},
  \ and\ \bibinfo {author} {\bibfnamefont {M.~W.}\ \bibnamefont {Mitchell}},\
  }\href {\doibase 10.1103/PhysRevLett.104.093602} {\bibfield  {journal}
  {\bibinfo  {journal} {Phys. Rev. Lett.}\ }\textbf {\bibinfo {volume} {104}},\
  \bibinfo {pages} {093602} (\bibinfo {year} {2010}{\natexlab{a}})}\BibitemShut
  {NoStop}%
\bibitem [{\citenamefont {Koschorreck}\ \emph
  {et~al.}(2010{\natexlab{b}})\citenamefont {Koschorreck}, \citenamefont
  {Napolitano}, \citenamefont {Dubost},\ and\ \citenamefont
  {Mitchell}}]{Mitchell10b}%
  \BibitemOpen
  \bibfield  {author} {\bibinfo {author} {\bibfnamefont {M.}~\bibnamefont
  {Koschorreck}}, \bibinfo {author} {\bibfnamefont {M.}~\bibnamefont
  {Napolitano}}, \bibinfo {author} {\bibfnamefont {B.}~\bibnamefont {Dubost}},
  \ and\ \bibinfo {author} {\bibfnamefont {M.~W.}\ \bibnamefont {Mitchell}},\
  }\href {\doibase 10.1103/PhysRevLett.105.093602} {\bibfield  {journal}
  {\bibinfo  {journal} {Phys. Rev. Lett.}\ }\textbf {\bibinfo {volume} {105}},\
  \bibinfo {pages} {093602} (\bibinfo {year} {2010}{\natexlab{b}})}\BibitemShut
  {NoStop}%
\bibitem [{\citenamefont {Shah}\ \emph {et~al.}(2010)\citenamefont {Shah},
  \citenamefont {Vasilakis},\ and\ \citenamefont {Romalis}}]{Romalis10}%
  \BibitemOpen
  \bibfield  {author} {\bibinfo {author} {\bibfnamefont {V.}~\bibnamefont
  {Shah}}, \bibinfo {author} {\bibfnamefont {G.}~\bibnamefont {Vasilakis}}, \
  and\ \bibinfo {author} {\bibfnamefont {M.~V.}\ \bibnamefont {Romalis}},\
  }\href {\doibase 10.1103/PhysRevLett.104.013601} {\bibfield  {journal}
  {\bibinfo  {journal} {Phys. Rev. Lett.}\ }\textbf {\bibinfo {volume} {104}},\
  \bibinfo {pages} {013601} (\bibinfo {year} {2010})}\BibitemShut {NoStop}%
\bibitem [{\citenamefont {Li}\ \emph {et~al.}(2011)\citenamefont {Li},
  \citenamefont {Vachaspati}, \citenamefont {Sheng}, \citenamefont {Dural},\
  and\ \citenamefont {Romalis}}]{Romalis11}%
  \BibitemOpen
  \bibfield  {author} {\bibinfo {author} {\bibfnamefont {S.}~\bibnamefont
  {Li}}, \bibinfo {author} {\bibfnamefont {P.}~\bibnamefont {Vachaspati}},
  \bibinfo {author} {\bibfnamefont {D.}~\bibnamefont {Sheng}}, \bibinfo
  {author} {\bibfnamefont {N.}~\bibnamefont {Dural}}, \ and\ \bibinfo {author}
  {\bibfnamefont {M.~V.}\ \bibnamefont {Romalis}},\ }\href {\doibase
  10.1103/PhysRevA.84.061403} {\bibfield  {journal} {\bibinfo  {journal} {Phys.
  Rev. A}\ }\textbf {\bibinfo {volume} {84}},\ \bibinfo {pages} {061403}
  (\bibinfo {year} {2011})}\BibitemShut {NoStop}%
\bibitem [{\citenamefont {Ye}\ \emph {et~al.}(1998)\citenamefont {Ye},
  \citenamefont {Ma},\ and\ \citenamefont {Hall}}]{Ye98}%
  \BibitemOpen
  \bibfield  {author} {\bibinfo {author} {\bibfnamefont {J.}~\bibnamefont
  {Ye}}, \bibinfo {author} {\bibfnamefont {L.-S.}\ \bibnamefont {Ma}}, \ and\
  \bibinfo {author} {\bibfnamefont {J.~L.}\ \bibnamefont {Hall}},\ }\href
  {\doibase 10.1364/JOSAB.15.000006} {\bibfield  {journal} {\bibinfo  {journal}
  {J. Opt. Soc. Am. B}\ }\textbf {\bibinfo {volume} {15}},\ \bibinfo {pages}
  {6} (\bibinfo {year} {1998})}\BibitemShut {NoStop}%
\bibitem [{\citenamefont {Teper}\ \emph {et~al.}(2006)\citenamefont {Teper},
  \citenamefont {Lin},\ and\ \citenamefont {Vuleti\ifmmode~\acute{c}\else
  \'{c}\fi{}}}]{Vuletic06}%
  \BibitemOpen
  \bibfield  {author} {\bibinfo {author} {\bibfnamefont {I.}~\bibnamefont
  {Teper}}, \bibinfo {author} {\bibfnamefont {Y.-J.}\ \bibnamefont {Lin}}, \
  and\ \bibinfo {author} {\bibfnamefont {V.}~\bibnamefont
  {Vuleti\ifmmode~\acute{c}\else \'{c}\fi{}}},\ }\href {\doibase
  10.1103/PhysRevLett.97.023002} {\bibfield  {journal} {\bibinfo  {journal}
  {Phys. Rev. Lett.}\ }\textbf {\bibinfo {volume} {97}},\ \bibinfo {pages}
  {023002} (\bibinfo {year} {2006})}\BibitemShut {NoStop}%
\bibitem [{\citenamefont {Nielsen}\ and\ \citenamefont
  {M\o{}lmer}(2008)}]{Molmer08}%
  \BibitemOpen
  \bibfield  {author} {\bibinfo {author} {\bibfnamefont {A.~E.~B.}\
  \bibnamefont {Nielsen}}\ and\ \bibinfo {author} {\bibfnamefont
  {K.}~\bibnamefont {M\o{}lmer}},\ }\href {\doibase 10.1103/PhysRevA.77.063811}
  {\bibfield  {journal} {\bibinfo  {journal} {Phys. Rev. A}\ }\textbf {\bibinfo
  {volume} {77}},\ \bibinfo {pages} {063811} (\bibinfo {year}
  {2008})}\BibitemShut {NoStop}%
\bibitem [{\citenamefont {Teper}\ \emph {et~al.}(2008)\citenamefont {Teper},
  \citenamefont {Vrijsen}, \citenamefont {Lee},\ and\ \citenamefont
  {Kasevich}}]{Kasevich08}%
  \BibitemOpen
  \bibfield  {author} {\bibinfo {author} {\bibfnamefont {I.}~\bibnamefont
  {Teper}}, \bibinfo {author} {\bibfnamefont {G.}~\bibnamefont {Vrijsen}},
  \bibinfo {author} {\bibfnamefont {J.}~\bibnamefont {Lee}}, \ and\ \bibinfo
  {author} {\bibfnamefont {M.~A.}\ \bibnamefont {Kasevich}},\ }\href {\doibase
  10.1103/PhysRevA.78.051803} {\bibfield  {journal} {\bibinfo  {journal} {Phys.
  Rev. A}\ }\textbf {\bibinfo {volume} {78}},\ \bibinfo {pages} {051803}
  (\bibinfo {year} {2008})}\BibitemShut {NoStop}%
\bibitem [{\citenamefont {Bernon}\ \emph {et~al.}(2011)\citenamefont {Bernon},
  \citenamefont {Vanderbruggen}, \citenamefont {Kohlhaas}, \citenamefont
  {Bertoldi}, \citenamefont {Landragin},\ and\ \citenamefont
  {Bouyer}}]{Bouyer11}%
  \BibitemOpen
  \bibfield  {author} {\bibinfo {author} {\bibfnamefont {S.}~\bibnamefont
  {Bernon}}, \bibinfo {author} {\bibfnamefont {T.}~\bibnamefont
  {Vanderbruggen}}, \bibinfo {author} {\bibfnamefont {R.}~\bibnamefont
  {Kohlhaas}}, \bibinfo {author} {\bibfnamefont {A.}~\bibnamefont {Bertoldi}},
  \bibinfo {author} {\bibfnamefont {A.}~\bibnamefont {Landragin}}, \ and\
  \bibinfo {author} {\bibfnamefont {P.}~\bibnamefont {Bouyer}},\ }\href
  {http://stacks.iop.org/1367-2630/13/i=6/a=065021} {\bibfield  {journal}
  {\bibinfo  {journal} {New J. Phys.}\ }\textbf {\bibinfo {volume} {13}},\
  \bibinfo {pages} {065021} (\bibinfo {year} {2011})}\BibitemShut {NoStop}%
\bibitem [{\citenamefont {Zhang}\ \emph {et~al.}(2012)\citenamefont {Zhang},
  \citenamefont {McConnell}, \citenamefont {\ifmmode~\acute{C}\else
  \'{C}\fi{}uk}, \citenamefont {Lin}, \citenamefont {Schleier-Smith},
  \citenamefont {Leroux},\ and\ \citenamefont {Vuleti\ifmmode~\acute{c}\else
  \'{c}\fi{}}}]{Vuletic12}%
  \BibitemOpen
  \bibfield  {author} {\bibinfo {author} {\bibfnamefont {H.}~\bibnamefont
  {Zhang}}, \bibinfo {author} {\bibfnamefont {R.}~\bibnamefont {McConnell}},
  \bibinfo {author} {\bibfnamefont {S.}~\bibnamefont {\ifmmode~\acute{C}\else
  \'{C}\fi{}uk}}, \bibinfo {author} {\bibfnamefont {Q.}~\bibnamefont {Lin}},
  \bibinfo {author} {\bibfnamefont {M.~H.}\ \bibnamefont {Schleier-Smith}},
  \bibinfo {author} {\bibfnamefont {I.~D.}\ \bibnamefont {Leroux}}, \ and\
  \bibinfo {author} {\bibfnamefont {V.}~\bibnamefont
  {Vuleti\ifmmode~\acute{c}\else \'{c}\fi{}}},\ }\href {\doibase
  10.1103/PhysRevLett.109.133603} {\bibfield  {journal} {\bibinfo  {journal}
  {Phys. Rev. Lett.}\ }\textbf {\bibinfo {volume} {109}},\ \bibinfo {pages}
  {133603} (\bibinfo {year} {2012})}\BibitemShut {NoStop}%
\bibitem [{\citenamefont {Tanji-Suzuki}\ \emph {et~al.}(2011)\citenamefont
  {Tanji-Suzuki}, \citenamefont {Leroux}, \citenamefont {Schleier-Smith},
  \citenamefont {Cetina}, \citenamefont {Grier}, \citenamefont {Simon},\ and\
  \citenamefont {Vuletic}}]{Vuletic11}%
  \BibitemOpen
  \bibfield  {author} {\bibinfo {author} {\bibfnamefont {H.}~\bibnamefont
  {Tanji-Suzuki}}, \bibinfo {author} {\bibfnamefont {I.~D.}\ \bibnamefont
  {Leroux}}, \bibinfo {author} {\bibfnamefont {M.~H.}\ \bibnamefont
  {Schleier-Smith}}, \bibinfo {author} {\bibfnamefont {M.}~\bibnamefont
  {Cetina}}, \bibinfo {author} {\bibfnamefont {A.~T.}\ \bibnamefont {Grier}},
  \bibinfo {author} {\bibfnamefont {J.}~\bibnamefont {Simon}}, \ and\ \bibinfo
  {author} {\bibfnamefont {V.}~\bibnamefont {Vuletic}},\ }in\ \href {\doibase
  10.1016/B978-0-12-385508-4.00004-8} {\emph {\bibinfo {booktitle} {Advances in
  Atomic, Molecular, and Optical Physics}}},\ Vol.~\bibinfo {volume} {60},\
  \bibinfo {editor} {edited by\ \bibinfo {editor} {\bibfnamefont
  {E.}~\bibnamefont {Arimondo}}, \bibinfo {editor} {\bibfnamefont {P.~R.}\
  \bibnamefont {Berman}}, \ and\ \bibinfo {editor} {\bibfnamefont {C.~C.}\
  \bibnamefont {Lin}}}\ (\bibinfo  {publisher} {Academic Press},\ \bibinfo
  {year} {2011})\ pp.\ \bibinfo {pages} {201 -- 237}\BibitemShut {NoStop}%
\bibitem [{\citenamefont {Wesenberg}\ \emph {et~al.}(2009)\citenamefont
  {Wesenberg}, \citenamefont {Ardavan}, \citenamefont {Briggs}, \citenamefont
  {Morton}, \citenamefont {Schoelkopf}, \citenamefont {Schuster},\ and\
  \citenamefont {M\o{}lmer}}]{Molmer09}%
  \BibitemOpen
  \bibfield  {author} {\bibinfo {author} {\bibfnamefont {J.~H.}\ \bibnamefont
  {Wesenberg}}, \bibinfo {author} {\bibfnamefont {A.}~\bibnamefont {Ardavan}},
  \bibinfo {author} {\bibfnamefont {G.~A.~D.}\ \bibnamefont {Briggs}}, \bibinfo
  {author} {\bibfnamefont {J.~J.~L.}\ \bibnamefont {Morton}}, \bibinfo {author}
  {\bibfnamefont {R.~J.}\ \bibnamefont {Schoelkopf}}, \bibinfo {author}
  {\bibfnamefont {D.~I.}\ \bibnamefont {Schuster}}, \ and\ \bibinfo {author}
  {\bibfnamefont {K.}~\bibnamefont {M\o{}lmer}},\ }\href {\doibase
  10.1103/PhysRevLett.103.070502} {\bibfield  {journal} {\bibinfo  {journal}
  {Phys. Rev. Lett.}\ }\textbf {\bibinfo {volume} {103}},\ \bibinfo {pages}
  {070502} (\bibinfo {year} {2009})}\BibitemShut {NoStop}%
\bibitem [{\citenamefont {Diniz}\ \emph {et~al.}(2011)\citenamefont {Diniz},
  \citenamefont {Portolan}, \citenamefont {Ferreira}, \citenamefont {G\'erard},
  \citenamefont {Bertet},\ and\ \citenamefont {Auff\`eves}}]{Auffeves11}%
  \BibitemOpen
  \bibfield  {author} {\bibinfo {author} {\bibfnamefont {I.}~\bibnamefont
  {Diniz}}, \bibinfo {author} {\bibfnamefont {S.}~\bibnamefont {Portolan}},
  \bibinfo {author} {\bibfnamefont {R.}~\bibnamefont {Ferreira}}, \bibinfo
  {author} {\bibfnamefont {J.~M.}\ \bibnamefont {G\'erard}}, \bibinfo {author}
  {\bibfnamefont {P.}~\bibnamefont {Bertet}}, \ and\ \bibinfo {author}
  {\bibfnamefont {A.}~\bibnamefont {Auff\`eves}},\ }\href {\doibase
  10.1103/PhysRevA.84.063810} {\bibfield  {journal} {\bibinfo  {journal} {Phys.
  Rev. A}\ }\textbf {\bibinfo {volume} {84}},\ \bibinfo {pages} {063810}
  (\bibinfo {year} {2011})}\BibitemShut {NoStop}%
\bibitem [{\citenamefont {Albert}\ \emph {et~al.}(2012)\citenamefont {Albert},
  \citenamefont {Marler}, \citenamefont {Herskind}, \citenamefont {Dantan},\
  and\ \citenamefont {Drewsen}}]{Albert12}%
  \BibitemOpen
  \bibfield  {author} {\bibinfo {author} {\bibfnamefont {M.}~\bibnamefont
  {Albert}}, \bibinfo {author} {\bibfnamefont {J.~P.}\ \bibnamefont {Marler}},
  \bibinfo {author} {\bibfnamefont {P.~F.}\ \bibnamefont {Herskind}}, \bibinfo
  {author} {\bibfnamefont {A.}~\bibnamefont {Dantan}}, \ and\ \bibinfo {author}
  {\bibfnamefont {M.}~\bibnamefont {Drewsen}},\ }\href {\doibase
  10.1103/PhysRevA.85.023818} {\bibfield  {journal} {\bibinfo  {journal} {Phys.
  Rev. A}\ }\textbf {\bibinfo {volume} {85}},\ \bibinfo {pages} {023818}
  (\bibinfo {year} {2012})}\BibitemShut {NoStop}%
\bibitem [{\citenamefont {Holstein}\ and\ \citenamefont
  {Primakoff}(1940)}]{HP40}%
  \BibitemOpen
  \bibfield  {author} {\bibinfo {author} {\bibfnamefont {T.}~\bibnamefont
  {Holstein}}\ and\ \bibinfo {author} {\bibfnamefont {H.}~\bibnamefont
  {Primakoff}},\ }\href {\doibase 10.1103/PhysRev.58.1098} {\bibfield
  {journal} {\bibinfo  {journal} {Phys. Rev.}\ }\textbf {\bibinfo {volume}
  {58}},\ \bibinfo {pages} {1098} (\bibinfo {year} {1940})}\BibitemShut
  {NoStop}%
\bibitem [{\citenamefont {Gardiner}\ and\ \citenamefont
  {Collett}(1985)}]{Gardiner85}%
  \BibitemOpen
  \bibfield  {author} {\bibinfo {author} {\bibfnamefont {C.~W.}\ \bibnamefont
  {Gardiner}}\ and\ \bibinfo {author} {\bibfnamefont {M.~J.}\ \bibnamefont
  {Collett}},\ }\href {\doibase 10.1103/PhysRevA.31.3761} {\bibfield  {journal}
  {\bibinfo  {journal} {Phys. Rev. A}\ }\textbf {\bibinfo {volume} {31}},\
  \bibinfo {pages} {3761} (\bibinfo {year} {1985})}\BibitemShut {NoStop}%
\bibitem [{Note1()}]{Note1}%
  \BibitemOpen
  \bibinfo {note} {The approximation that the single-particle decay rate
  $\Gamma $ into all modes other than the cavity mode holds true in the limit
  that the cavity subtends a small fraction of the total solid angle as seen by
  the atom~\cite {Kimble98}.}\BibitemShut {Stop}%
\bibitem [{\citenamefont {Zhu}\ \emph {et~al.}(1990)\citenamefont {Zhu},
  \citenamefont {Gauthier}, \citenamefont {Morin}, \citenamefont {Wu},
  \citenamefont {Carmichael},\ and\ \citenamefont {Mossberg}}]{ZGM90}%
  \BibitemOpen
  \bibfield  {author} {\bibinfo {author} {\bibfnamefont {Y.}~\bibnamefont
  {Zhu}}, \bibinfo {author} {\bibfnamefont {D.~J.}\ \bibnamefont {Gauthier}},
  \bibinfo {author} {\bibfnamefont {S.~E.}\ \bibnamefont {Morin}}, \bibinfo
  {author} {\bibfnamefont {Q.}~\bibnamefont {Wu}}, \bibinfo {author}
  {\bibfnamefont {H.~J.}\ \bibnamefont {Carmichael}}, \ and\ \bibinfo {author}
  {\bibfnamefont {T.~W.}\ \bibnamefont {Mossberg}},\ }\href {\doibase
  10.1103/PhysRevLett.64.2499} {\bibfield  {journal} {\bibinfo  {journal}
  {Phys. Rev. Lett.}\ }\textbf {\bibinfo {volume} {64}},\ \bibinfo {pages}
  {2499} (\bibinfo {year} {1990})}\BibitemShut {NoStop}%
\bibitem [{\citenamefont {Bollinger}\ \emph {et~al.}(1996)\citenamefont
  {Bollinger}, \citenamefont {Itano}, \citenamefont {Wineland},\ and\
  \citenamefont {Heinzen}}]{Bollinger96}%
  \BibitemOpen
  \bibfield  {author} {\bibinfo {author} {\bibfnamefont {J.~J.~.}\ \bibnamefont
  {Bollinger}}, \bibinfo {author} {\bibfnamefont {W.~M.}\ \bibnamefont
  {Itano}}, \bibinfo {author} {\bibfnamefont {D.~J.}\ \bibnamefont {Wineland}},
  \ and\ \bibinfo {author} {\bibfnamefont {D.~J.}\ \bibnamefont {Heinzen}},\
  }\href {\doibase 10.1103/PhysRevA.54.R4649} {\bibfield  {journal} {\bibinfo
  {journal} {Phys. Rev. A}\ }\textbf {\bibinfo {volume} {54}},\ \bibinfo
  {pages} {R4649} (\bibinfo {year} {1996})}\BibitemShut {NoStop}%
\bibitem [{\citenamefont {Holland}\ and\ \citenamefont
  {Burnett}(1993)}]{Holland93}%
  \BibitemOpen
  \bibfield  {author} {\bibinfo {author} {\bibfnamefont {M.~J.}\ \bibnamefont
  {Holland}}\ and\ \bibinfo {author} {\bibfnamefont {K.}~\bibnamefont
  {Burnett}},\ }\href {\doibase 10.1103/PhysRevLett.71.1355} {\bibfield
  {journal} {\bibinfo  {journal} {Phys. Rev. Lett.}\ }\textbf {\bibinfo
  {volume} {71}},\ \bibinfo {pages} {1355} (\bibinfo {year}
  {1993})}\BibitemShut {NoStop}%
\bibitem [{\citenamefont {Uys}\ \emph {et~al.}(2010)\citenamefont {Uys},
  \citenamefont {Biercuk}, \citenamefont {VanDevender}, \citenamefont
  {Ospelkaus}, \citenamefont {Meiser}, \citenamefont {Ozeri},\ and\
  \citenamefont {Bollinger}}]{Uys2010}%
  \BibitemOpen
  \bibfield  {author} {\bibinfo {author} {\bibfnamefont {H.}~\bibnamefont
  {Uys}}, \bibinfo {author} {\bibfnamefont {M.~J.}\ \bibnamefont {Biercuk}},
  \bibinfo {author} {\bibfnamefont {A.~P.}\ \bibnamefont {VanDevender}},
  \bibinfo {author} {\bibfnamefont {C.}~\bibnamefont {Ospelkaus}}, \bibinfo
  {author} {\bibfnamefont {D.}~\bibnamefont {Meiser}}, \bibinfo {author}
  {\bibfnamefont {R.}~\bibnamefont {Ozeri}}, \ and\ \bibinfo {author}
  {\bibfnamefont {J.~J.}\ \bibnamefont {Bollinger}},\ }\href {\doibase
  10.1103/PhysRevLett.105.200401} {\bibfield  {journal} {\bibinfo  {journal}
  {Phys. Rev. Lett.}\ }\textbf {\bibinfo {volume} {105}},\ \bibinfo {pages}
  {200401} (\bibinfo {year} {2010})}\BibitemShut {NoStop}%
\bibitem [{\citenamefont {Saffman}\ \emph {et~al.}(2009)\citenamefont
  {Saffman}, \citenamefont {Oblak}, \citenamefont {Appel},\ and\ \citenamefont
  {Polzik}}]{Saffman09}%
  \BibitemOpen
  \bibfield  {author} {\bibinfo {author} {\bibfnamefont {M.}~\bibnamefont
  {Saffman}}, \bibinfo {author} {\bibfnamefont {D.}~\bibnamefont {Oblak}},
  \bibinfo {author} {\bibfnamefont {J.}~\bibnamefont {Appel}}, \ and\ \bibinfo
  {author} {\bibfnamefont {E.~S.}\ \bibnamefont {Polzik}},\ }\href {\doibase
  10.1103/PhysRevA.79.023831} {\bibfield  {journal} {\bibinfo  {journal} {Phys.
  Rev. A}\ }\textbf {\bibinfo {volume} {79}},\ \bibinfo {pages} {023831}
  (\bibinfo {year} {2009})}\BibitemShut {NoStop}%
\bibitem [{\citenamefont {Hinkley}\ \emph {et~al.}(2013)\citenamefont
  {Hinkley}, \citenamefont {Sherman}, \citenamefont {Phillips}, \citenamefont
  {Schioppo}, \citenamefont {Lemke}, \citenamefont {Beloy}, \citenamefont
  {Pizzocaro}, \citenamefont {Oates},\ and\ \citenamefont
  {Ludlow}}]{Hinkley2013}%
  \BibitemOpen
  \bibfield  {author} {\bibinfo {author} {\bibfnamefont {N.}~\bibnamefont
  {Hinkley}}, \bibinfo {author} {\bibfnamefont {J.~A.}\ \bibnamefont
  {Sherman}}, \bibinfo {author} {\bibfnamefont {N.~B.}\ \bibnamefont
  {Phillips}}, \bibinfo {author} {\bibfnamefont {M.}~\bibnamefont {Schioppo}},
  \bibinfo {author} {\bibfnamefont {N.~D.}\ \bibnamefont {Lemke}}, \bibinfo
  {author} {\bibfnamefont {K.}~\bibnamefont {Beloy}}, \bibinfo {author}
  {\bibfnamefont {M.}~\bibnamefont {Pizzocaro}}, \bibinfo {author}
  {\bibfnamefont {C.~W.}\ \bibnamefont {Oates}}, \ and\ \bibinfo {author}
  {\bibfnamefont {A.~D.}\ \bibnamefont {Ludlow}},\ }\href {\doibase
  10.1126/science.1240420} {\bibfield  {journal} {\bibinfo  {journal}
  {Science}\ } (\bibinfo {year} {2013}),\ 10.1126/science.1240420}\BibitemShut
  {NoStop}%
\bibitem [{\citenamefont {Bloom}\ \emph {et~al.}(2014)\citenamefont {Bloom},
  \citenamefont {Nicholson}, \citenamefont {Williams}, \citenamefont
  {Campbell}, \citenamefont {Bishof}, \citenamefont {Zhang}, \citenamefont
  {Zhang}, \citenamefont {Bromley},\ and\ \citenamefont {Ye}}]{Bloom2013}%
  \BibitemOpen
  \bibfield  {author} {\bibinfo {author} {\bibfnamefont {B.~J.}\ \bibnamefont
  {Bloom}}, \bibinfo {author} {\bibfnamefont {T.~L.}\ \bibnamefont
  {Nicholson}}, \bibinfo {author} {\bibfnamefont {J.~R.}\ \bibnamefont
  {Williams}}, \bibinfo {author} {\bibfnamefont {S.~L.}\ \bibnamefont
  {Campbell}}, \bibinfo {author} {\bibfnamefont {M.}~\bibnamefont {Bishof}},
  \bibinfo {author} {\bibfnamefont {X.}~\bibnamefont {Zhang}}, \bibinfo
  {author} {\bibfnamefont {W.}~\bibnamefont {Zhang}}, \bibinfo {author}
  {\bibfnamefont {S.~L.}\ \bibnamefont {Bromley}}, \ and\ \bibinfo {author}
  {\bibfnamefont {J.}~\bibnamefont {Ye}},\ }\href {\doibase
  10.1038/nature12941} {\bibfield  {journal} {\bibinfo  {journal} {Nature}\ }
  (\bibinfo {year} {2014}),\ 10.1038/nature12941}\BibitemShut {NoStop}%
\bibitem [{\citenamefont {Kimble}(1998)}]{Kimble98}%
  \BibitemOpen
  \bibfield  {author} {\bibinfo {author} {\bibfnamefont {H.~J.}\ \bibnamefont
  {Kimble}},\ }\href {http://stacks.iop.org/1402-4896/1998/i=T76/a=019}
  {\bibfield  {journal} {\bibinfo  {journal} {Physica Scripta}\ }\textbf
  {\bibinfo {volume} {1998}},\ \bibinfo {pages} {127} (\bibinfo {year}
  {1998})}\BibitemShut {NoStop}%
\end{thebibliography}
%merlin.mbs apsrev4-1.bst 2010-07-25 4.21a (PWD, AO, DPC) hacked
%Control: key (0)
%Control: author (8) initials jnrlst
%Control: editor formatted (1) identically to author
%Control: production of article title (-1) disabled
%Control: page (0) single
%Control: year (1) truncated
%Control: production of eprint (0) enabled
%

\end{document}